# Adaptive music:
# Automated music composition
# and distribution

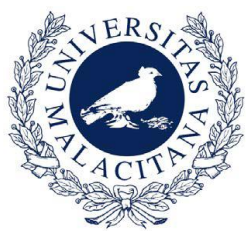

UNIVERSIDAD
DE MÁLAGA

## TESIS DOCTORAL

David Daniel Albarracín Molina

Director: Francisco J. Vico Vela

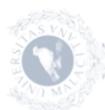

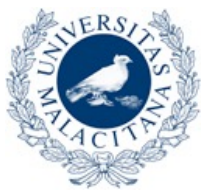

UNIVERSIDAD
DE MÁLAGA

AUTOR: David Daniel Albarracín Molina

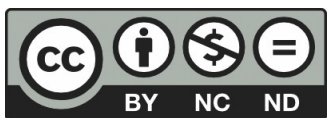 https://orcid.org/0000-0001-6311-1304

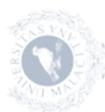

**Francisco J. Vico, Ph.D.**, Full Professor of Computer Science, Department of Lenguajes y Ciencias de la Computación, Universidad de Málaga, Spain.

Hereby CERTIFIES:

That **David Daniel Albarracín Molina**, M.Sc. in Computer Science, has completed his doctoral thesis:

**Adaptive music: Automated music composition and distribution**

in the Department of Lenguajes y Ciencias de la Computación at Universidad de Málaga, under my supervision. I have directed and approved this thesis to be presented to the committee of the Universidad de Málaga.

Málaga, July 2021.

Ph.D. Francisco J. Vico
Full Professor
Universidad de Málaga
Spain

# DECLARACIÓN DE AUTORÍA Y ORIGINALIDAD DE LA TESIS PRESENTADA PARA OBTENER EL TÍTULO DE DOCTOR

D. DAVID DANIEL ALBARRACÍN MOLINA

Estudiante del programa de doctorado TECNOLOGÍAS INFORMÁTICAS de la Universidad de Málaga, autor de la tesis, presentada para la obtención del título de doctor por la Universidad de Málaga, titulada: ADAPTIVE MUSIC: AUTOMATED MUSIC COMPOSITION AND DISTRIBUTION.

Realizada bajo la tutorización de FRANCISCO J. VICO y dirección de FRANCISCO J. VICO.

DECLARO QUE:

La tesis presentada es una obra original que no infringe los derechos de propiedad intelectual ni los derechos de propiedad industrial u otros, conforme al ordenamiento jurídico vigente (Real Decreto Legislativo 1/1996, de 12 de abril, por el que se aprueba el texto refundido de la Ley de Propiedad Intelectual, regularizando, aclarando y armonizando las disposiciones legales vigentes sobre la materia), modificado por la Ley 2/2019, de 1 de marzo.

Igualmente asumo, ante a la Universidad de Málaga y ante cualquier otra instancia, la responsabilidad que pudiera derivarse en caso de plagio de contenidos en la tesis presentada, conforme al ordenamiento jurídico vigente.

En Málaga, a 01 de JULIO de 2021

| | |
|---|---|
| **Fdo.: DAVID D. ALBARRACÍN MOLINA** <br> **Doctorando** | **Fdo.: FRANCISCO J. VICO** <br> **Tutor** |
| **Fdo.: FRANCISCO J. VICO** <br> **Director de tesis** | |

*A mi familia,*

*por la educación, cariño y apoyo.*

*Y a todo aquel que no deje de cuestionarse lo que cree saber.*

# Abstract

Creativity, or the ability to produce new useful ideas, is commonly associated to the human being; but there are many other examples in nature where this phenomenon can be observed. Inspired by this fact, in engineering and particularly in computational sciences, many different models have been developed to tackle a number of problems.

Composing music, a form of art broadly present along the human history, is the main topic addressed in this thesis. Taking advantage of the kind of ideas that bring diversity and creativity to nature and computation, we present Melomics: an algorithmic composition method based on evolutionary search. The solutions have a genetic encoding based on formal grammars and these are interpreted in a complex developmental process followed by a fitness assessment, to produce valid music compositions in standard formats.

The system has exhibited a high creative power and versatility to produce music of different types and it has been tested, proving on many occasions the outcome to be indistinguishable from the music made by human composers. The system has also enabled the emergence of a set of completely novel applications: from effective tools to help anyone to easily obtain the precise music that they need, to radically new uses, such as adaptive music for therapy, exercise, amusement and many others. It seems clear that automated composition is an active research area and that countless new uses will be discovered.

# Acknowledgements

I have had the privilege to work in a laboratory provided by Universidad de Málaga in the PTA, such a nice environment, and to have been funded by the spin-off program of this university and by the national programs INNPACTO and Plan Avanza2. These are mechanisms of this country to enhance the so essential research and technological development activities, which I believe highly determine a society's future wellbeing.

I thank my advisor Fransico Vico, for introducing me to research and for letting me be part of the GEB, a surprisingly heterogeneous place that has allowed me to work along with many singular and interesting people who all have influenced me and whom I consider my friends. Jose, a natural inquirer, taught me music for the first time after high school. Then, with Gustavo, I explored music from a more theoretical perspective and I thank him for his patience, especially at the beginning of my re-learning process. Carlos and I helped each other since the beginning, during those nights of writing project proposals; and Fran Moreno, during those nights coding. Thanks to Juanca and Guti for sharing their wisdom and for those intense discussions about both relevant and trivial matters. Miguel, the ultimate IT engineer, also with a charming personality. Jose David, a broad and probably endless source of relevant information. He showed me, together with Dani and Gema, how beautiful a research environment can be. Julio, providing his experienced help in technical situations and his peculiar sense of humor. Curro, always helpful in any bureaucratic process, and many others with whom I have shared interesting thoughts and established a good relationship: Ellen, Vicente, Rafa, Mario, Enrique, Héctor, Cristóbal, Pablo, Pedro, Fran Palma, Oliver, Juan Arenas, Juan Travesedo, Sirenia and Chelo. Let's not forget the people outside the group that I have had the opportunity to work with, like Cory McKay, always disposed to help with MIR matters; the people at the UGR department of psychology and the people at the Montecanal hospital neonatal unit.

I especially thank my parents Ana María and Andrés José, whose education and encouragement have undoubtedly allowed me to choose my path and make this humble contribution. I also thank my brother Andrés and my sister Mariam, for all the great time and support; and Hadjira for her company and for helping me during the last part of this endeavor. Lastly, I thank my grandparents and the rest of my family for their nice company and support.

Thank you all.

# Credits

This thesis is largely a consequence of a research project commonly known as Melomics, driven by the GEB, Universidad de Málaga and mainly funded and promoted by two national research projects granted in 2010 and 2011, and a university spin-off award in 2010. Many people have been involved in one way of another. Below we give credit to the people whose work has contributed to this thesis:

- The present compositional system, starting from a collection of scripts and ideas provided by my advisor Francisco Vico, has continuously been improving with the addition of countless functionalities in the synthesis, score-writing and core systems, which have been coded with the help of Francisco Moreno Arcas, Rafael De Vicente Milans, Pedro Guillén Aroca and José David Gutiérrez Gutiérrez; and specified with the help of experts musicians: Gustavo Díaz Jerez for contemporary music and José F. Serrano García, Óliver Moya Bueno, Francisco J. Palma Olmos and, especially, Juan Carlos Moya Olea, co-author in some of the published work (Albarracin-Molina, Moya, & Vico, 2016).

- To design and build the mobile applications and web tools, essential parts of the adaptive music system described in Chapter 4, we counted on: Miguel Carmona Bermúdez, Cristóbal Carnero Liñán, Héctor Mesa Jiménez, Pablo Moreno Olalla, Salvador Burrezo González, José David Gutiérrez Gutiérrez, Enrique Morris Cordón and Mario A. Ureña Martínez.

- The perceptive study described in section 3.2 and published in a scientific journal (Albarracín-Molina, Raglio, Rivas-Ruiz, & Vico, 2021) was designed and executed with the collaboration of Francisco Vico, Alfredo Raglio, Francisco Rivas-Ruiz, Juan C. Moya Olea and Carlos A. Sánchez Quintana.

# Contents

# Contents

# List of figures

# List of tables

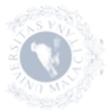

# List of musical content

# Chapter 1

# Introduction

Specifying what music is and distinguishing it from other kind of sounds is a complex issue and depends on many cultural factors. Music compositions, as a form of art, do influence the audience; they raise on them some kind of emotions or psychological effects (Juslin & Sloboda, 2011), sometimes in an intended way by the composer. As we may think, the composition process, the notation methods, the instruments, the styles, etc. have evolved since their emergence in the prehistoric eras, always linked to the social progression and the technological advancements (Crocker, 1966) (Meyer L. B., 1989) (Christensen, 2002) (Sachs, 2012). Nowadays music is taking advantage from the current digital means, such as new musical instruments or computer programs. These tools ease and improve the processes of composing and performing music, also leading to the appearance of new genres and, especially, new ways of making music.

In this thesis we present a computational system aimed to create original full-scale musical compositions. We describe the methods that we use from the field of artificial intelligence, in order to achieve the intended creativity on its creations. Furthermore, some innovative applications that take advantage of automatically generated music are presented.

Biomimetics is a branch of the engineering, consisting of mimicking solutions that are observed in nature, such as structures or behaviors. Over the years this approach has shown its potential to help solve complex problems in computer science. Some of the most known and relevant methods are artificial neural networks, fuzzy logic or evolutionary algorithms. Biomimetic algorithms exist because they can bring very creative and almost unpredictable solutions. These results are not always optimal, but they can answer the problem reasonably well; besides, sometimes the problem is not supposed to comprise a unique optimal solution. Here we describe how we leverage this less orthodox way of computing as the basis for a music composition system.

A tool allowing input specifications and capable of automatically generating pieces with similar characteristics to music created by humans can bring a number of advantages, for instance: producing compositions very quickly, making easier to explore new musical genres or styles, enabling the possibility to obtain music tailored to anyone's particular taste or producing multiple variations of a specific composition or musical idea. This thesis explores some of these new paths arisen from automated composition.

Before attending to the main topics of this dissertation, some relevant concepts are introduced: In section 1.1 we focus on what we understand as *computational creativity* and provide some examples. In section 1.2 we describe how the music field in general is benefiting from computers and mention different existing methods for composing in an automated way. In section 1.3, we outline how computer science takes advantage from mimicking nature. Finally, section 1.4 summarizes the work presented in this dissertation.

## 1.1  Computational creativity

Creativity is an idea still difficult to define. In fact, the current concept has been forged during the modern eras, the Renaissance, the Enlightenment and so on. In previous ages, especially the middle ages, what we understand now as creating something was seen more like a manner of discovery, imitation or even a divine inspiration; Gods could have this ability, but not humans (Runco & Albert, 2010) (Tatarkiewicz, 1980). Then, creativity was started to be perceived as a capability of great men, in contrast to one just being productive (Dacey, 1999); and it was clear, as it is now, that it was linked to the capability of imagination (Tatarkiewicz, 1980). There is a multitude of definitions, but a suitable one to us can be creativity as *the ability to generate novel and valuable ideas* (Boden, 2009).

Creativity has generally been seen as a characteristic of the human being, distinguishing them from the rest of beings, in the same way it happens with intelligence. So, it stands to reason that these two qualities may be related (Sternberg & O'Hara, 1999). Being a distinctive feature might carry certain skepticism when attributed to other entities, but building machines that exhibit this capability certainly represents a challenge for scientists and engineers. From the point of view of computer science, this is a cutting-edge area in artificial intelligence, usually addressing complex and vague problems, with ill-defined spaces of search (Gero & Maher, 1993) (De Smedt, 2013), so that using conventional algorithms is not feasible in this case.

As we have suggested, creativity is about combining pre-existing ideas or things to produce a new and valuable one. We can summarize the process in four stages (Wallas, 1926): (1) gathering inputs, like aims, pre-existing solutions, limitations, etc.; (2) exploring ideas, by combining the initial products, trying new approaches, etc.; (3) generate effective results and (4) testing the results. Hence, it seems a process that could be modeled into a current machine, at least for specific purposes.

Along the development of computer science, artificial creativity has regularly been approached in arts, the main area associated to this ability. For example there are programs that can make original paintings (Cook & Colton, 2011), videos (Chambel, et al., 2007) or writings (Gervás, 2009) (Turner, 2014). But creativity has also been treated in some other fields, for example in AI to play board games having a large search space. For example, Selfo (Vico F. J., 2007) is a connection game for which we implemented strategies based on self-organization observed in living organisms (Albarracín-Molina, Diseño e implementación de una estrategia para un juego de conexión. M.A. Thesis, 2010). A better known game, GO, with more states than chess, required flexible algorithms to be played by an AI (Müller M. , 2002) and it has not been until recently that they defeated professional players, using reinforcement learning (Silver, et al., 2016). Another field where creativity is explored is in AI for video games. There are, for example, non-player characters in video games that can cooperate in groups, share tasks and exhibit unpredictable reactions to different situations, such as new visual information, events in the environment, perceived sounds, human player presence or other non-player character actions (Yannakakis, 2012). One last specific case of what can be considered computational creativity is Watson, the IBM system that includes natural language processing, information retrieval, knowledge representation, automated reasoning and machine learning technologies (Ferrucci, et al., 2010). The system proved its capability of producing original output by competing and defeating the human players in the TV game show Jeopardy! (Markoff, 2011). Since then, Watson has been used for applications involving understanding questions asked by humans and providing valuable answers, including purposes as diverse as financial applications, medical problems, legal consultancy or even cooking (Ferrucci, Levas, Bagchi, Gondek, & Mueller, 2013).

In research there always has been an interest about creativity. It has been studied in nature (Bentley, Is evolution creative, 1999), in computer science (Colton, López de Mantaras, & Stock, 2009), it has been discussed how to measure it (Ritchie, 2007) (Bringsjord, Bello, & Ferrucci, 2001) (Pease & Colton, 2011) and how to distinguish creative systems from automated systems that just pretend to be creative (Colton, 2008). In artificial intelligence, many paths have been followed to achieve creativity and evolutionary computation is one of the most frequent approaches (Bentley & Corne, An Introduction to Creative Evolutionary Systems, 2001) (Boden, 2009).

The system that we propose incorporates techniques mainly from three areas of artificial intelligence: (a) Knowledge-based systems, which can be implemented in the form of rules for a specific domain, and have been used before in computer music (Schaffer & McGee, 1997). We make use of a knowledge-based approach to build the general rules for composing (e.g., physical limitations of musical instruments) or to specify musical styles (see section 2.2.3.1 Musical styles), where the pursued novelty is achieved through exploration guided by these rules. (b) Formal grammars, which can be roughly defined as a vocabulary of symbols plus a set of rewriting rules that guide the development of an initial character into possible strings of symbols. These methods have been used for example in computational design (Stiny, 1980) (Shea & Fenves, 1997). In our system we use two different indirect encoding models based on grammars to represent musical genomes which are developed and interpreted into legible compositions. (c) Search methods, which includes a multitude of algorithms of different nature in order to explore search spaces of different kinds, in many cases including the use of heuristics (Russell & Norvig, 2009) (Campbell, Hoane, & Hsu, 2002). Our approach is based on evolutionary search, a genetic algorithm in particular. Compositions, encoded as genomes in the form of a formal grammar, are developed and then assessed using both general and style-based musical rules. Genomes are discarded if they fail the test, but they are kept and reused in the future if they pass.

There is evidence that suggests that, using the mentioned approaches, Melomics output is indeed creative (Ball, Iamus, classical music's computer composer, live from Malaga, 2012). Very frequently people, including professional musicians, have mistakenly attributed Melomics works to human styles (Ball, Artificial music: The computers that create melodies, 2014) or even to specific composers; a number of musical pieces are provided and studied in this thesis. Finally, there is a curious incident that occurred in December 2014, related to the fact that an autonomous system was producing genuine content. On that occasion, Melomics entries in Wikipedia experienced massive deletion of its media content due to copyright-related issues[1]. All this content was after restored, since it was established that there was "no creative input from a human".

---

[1] https://commons.wikimedia.org/w/index.php?title=Commons:Village_pump/Copyright&oldid=143770481#melomics.com (accessed on May 16, 2021)

## 1.2 Computer music and algorithmic composition

Music, as every other field in human knowledge, has benefited from the appearance and development of computers and computer science. New tools to help in the process of creating music have appeared: for recording and reproducing sounds, mixing and editing audio, writing scores, synthesizing; and digital audio workstations (DAW) have been developed, integrating most of these functionalities. There is also a kind of tool known as Computer-Aided Algorithmic Composition (CAAC) to help in the composition process, some examples are Csound (Boulanger, 2000), SuperCollider (McCartney, 2002) or MAX/MSP (Puckette, 2002). New musical instruments have been invented, with a special mention to the highly influential music synthesizer (Pinch & Trocco, 1998) and others like the electric violin (Washington, DC: U.S. Patent and Trademark Office Patent No. D403,013, 1998). Virtual instrument technologies have been developed (SoundFont Technical Specification, 2006) (Steinberg, 1999) and digital standards to store music, either in a symbolic way (MIDI,[2] SIB,[3] XML[4]) or as a wave (WAV, FLAC, MP3). The digital era has brought yet another step in the evolution of musical genres (Pachet & Cazaly, 2000) as well as the development of algorithmic composition, a discipline at the intersection of music theory and computational intelligence.

Along with the invention of transistor computers in 1950s, the earliest works of algorithmic composition came out, roughly at the same time as the concept of artificial intelligence, although the two fields did not converge until some time later. As an overview, some works in formal composition at the early age were: an unpublished work by Caplin and Prinz in 1955 (Ariza, 2011), with an implementation of Mozart's dice dame and a generator of melodic lines using stochastic transitional probabilities; in 1956, the pioneering work Hiller and Isaacson's Illiac Suite (Hiller Jr & Isaacson, 1958), based on Markov chains and rule systems; MUSICOMP by Baker (Ames, 1987), implementing some methods used by Hillers; Xenakis's stochastic algorithms in early 1960s, working as CAAC (Ames, 1987); in 1961, a dedicated computer able to compose new melodies related to previous ones, by using Markov processes (Olson & Belar, 1961); in 1963, Gill's algorithm applying classical AI techniques, hierarchical search with backtracking (Gill, 1963); PROJECT1, created by Koenig in 1964 (Ames, 1987), using serial composition and other techniques; and also in 1964 Padberg implementing a compositional framework based on procedural techniques (Padberg, 1964). Machines were becoming less expensive and more powerful, and the field of algorithmic composition grew along intersecting the field of Artificial Intelligence. However, it has been a slow process, with difficult communication between scientists and artists and with little continuity in research.

---

[2] http://www.midi.org/ (accessed on May 16, 2021)
[3] http://www.sibelius.com/ (accessed on May 16, 2021)
[4] http://www.musicxml.com/ (accessed on May 16, 2021)

Algorithmic composition aims the automation of tasks in the creative process, such as the definition of melodies, rhythms and harmonies, or the introduction of counterpoints, voice leading, articulations or effects. It can be approached from mimicking a specific style from a set of musical examples or by modeling the composer's methodology. In both cases we can obtain tools to support musicians in their creative work or fully autonomous systems capable of producing complete compositions.

The algorithmic techniques in the compositional systems observed in literature can be classified in four general categories: (a) Symbolic AI, whose main representatives in algorithmic composition are grammars and rule-based systems. They have been used both for imitating musical styles or specific composers and for automating the composition process. These are usually very labor-intensive methods since they require the musical knowledge to be encoded and maintained in the symbolic framework. (b) Machine learning, comprising Markov chains and artificial neural networks, whose nature have made them generally more suitable for imitation purposes. (c) Optimization techniques, having been used for imitation and for the automation of compositional tasks. They count on evolutionary algorithms as the most representative example. (d) Complex systems, like self-similarity or cellular automata, used to generate novel material without being restricted to the human knowledge. For a deep review see (Fernandez & Vico, 2013).

Composing music imitating an existing corpus has been approached quite successfully. For example, Cope's EMI (Cope, 1992) was able to analyze a musical style, extract patterns of short musical sequences and determine how to use them when composing in that style; and Pachet's Continuator (Pachet F. , 2002), using variable-order Markov chains, implemented a real-time improvisation system. More recently, with the rise of neural networks, while not being as widespread as other forms of arts that use Generative Adversarial Networks (Goodfellow, et al., 2014) and other methods, there are many examples of systems following this imitating approach. For example, using long short-term memory (LSTM) recurrent networks to learn complete musical structures and produce music in that style (Eck & Schmidhuber, 2002) or networks based on gate recurrent units (GRU) to produce monophonic melodies by learning from a corpus of different styles (Colombo, Seeholzer, & Gerstner, 2017). There are different works in algorithmic music based on neural networks, typically using a variation of recurrent networks. Apart from the ones mentioned, Boulanger-Lewandowski proposed a system based on recurrent temporal restricted Boltzmann machine that learns complex polyphonic structures (Boulanger-Lewandowski, Bengio, & Vincent, 2012) and Liu and Ramakrishnan another method to deal with structure and other characteristics, using LSTM with resilient propagation (Lui & Ramakrishnan, 2014); Huang and Wu created a system that takes datasets in MIDI format to train a neural network based on 2-layer LSTM and is able to build sequences of notes (Huang & Wu, 2016) and Ianigro and Bown use Continuous Time Recurrent Neural Networks combined with evolutionary algorithms to generate and manipulate audio (Ianigro & Bown, 2019). For a further review, see (Briot, Hadjeres, & Pachet, 2017). Additionally, neural networks have been used to perform other music related tasks, such as audio synthesis (Huzaifah & Wyse, 2017), genre classification (Oramas, Barbieri, Nieto Caballero, & Serra, 2018) and other kinds of analysis and tagging (Nam, Choi, Lee, Chou, & Yang, 2018). However, despite the rise of Machine Learning and Deep Learning in particular, there has not been a major explosion or success in music composition using these methods. This is maybe because, unlike natural neural networks, they still can only deal with narrow problems and domains, relying heavily on some sort of starting dataset, something very specific to imitate or to adapt to, which is almost the opposite of the goal in any form of art, like music.

Composing with no creative input has been a harder problem for artificial intelligence, partly because of the difficulty in assessing the results, although some frameworks have been proposed (Gero J. S., 2000) (Pearce & Wiggins, 2001) (Ritchie, 2007) (Boden, 2009). Genetic algorithms have proven to be useful for producing novel musical content. In 1994, GenJam (Biles, 1994) was an interactive method to create jazz melodies. More examples of interactive genetic algorithms are Vox Populi (Moroni, Manzolli, von Zuben, & Gudwin, 2000), based on evolving chords, and Darwin Tunes (MacCallum, Mauch, Burt, & Leroi, 2012) that studied the evolution of short audio loops. There has been a wide variety of automatic musical fitness assessments. Nam and Kim (Nam & Kim, 2019) recently proposed a method to create jazz melodies, encoding explicitly the musical variables in the chromosomes and using a fitness function based on the theory of harmony and note patterns. However, evolutionary and developmental methods (evo-devo) using indirect encoding, although more complex to design, are meant to provide more unpredictable and diverse results. Musical Organisms (Lindemann & Lindemann, 2018) defined a gene regulatory network and focused on structural modularity, hierarchy and repetition as a common principle of living beings and music to produce musical scores.

Despite the number of existing attempts to build automatic composition systems, most of them seem to be scattered works, with only a few of them having achieved mainstream awareness or widespread use, like CHORAL (Ebcioglu, 1988), Cope's EMI or Pachet's Continuator. The system developed and studied in this thesis, Melomics can be considered one of them (La UMA diseña un ordenador que compone música clásica, 2012) (Peckham, 2013). It has not only been used to create a web repository of music of different genres,[5] to produce two original albums published[6] [7] and served as a source of creative material for professional composers (Diaz-Jerez, 2011), but it is also the core of some innovative applications, such as adaptive music that is detailed in further chapters.

---

[5] http://geb.uma.es/melomics (accessed on May 16, 2021)
[6] https://en.wikipedia.org/wiki/Iamus_(album) (accessed on May 16, 2021)
[7] https://en.wikipedia.org/wiki/0music (accessed on May 16, 2021)

## 1.3 Biological inspiration

Nature can be a powerful source of inspiration to solve different kinds of problems and we saw some examples in the previous section. In engineering and sciences this approach, called biomimetics, has made possible great inventions, such as the well-known Velcro, copying the tiny hooks on the surface of burs; a multitude of designs of functional structures, like the wings of an airplane, the diving fins or climbing tools mimicking the anatomy of geckos (You, 2014); the radar and sonar; self-cleaning and self-repairing materials (Yu, et al., 2020) (Yin, et al., 2015); and countless other examples. Particularly, in computer science, biomimetics plays a core role in the field of computational intelligence, being artificial neural networks, fuzzy logic, evolutionary algorithms or swarm intelligence, some of its most representative methods. They are applied to solve a multitude of computational problems, such as financial trade, computer vision, networks traffic, road traffic, control systems in appliances, creative industrial design or arts.

Evolution is one of the main mechanisms that have contributed to the diversity and complexity found in nature. These mechanisms have inspired in computer science evolutionary algorithms, which have been applied in many diverse domains. This methodology is based on having an initial population of solutions that undergoes a cyclic process of reproduction (with possible variation), evaluation and selection. Classical evolutionary algorithms use direct encoding, that is, an explicit map between representations and solutions (genotypes and phenotypes). This approach has traditionally suffered from some problems (Hornby & Pollack, 2002), such as difficult scalability due to the size of the solutions and the search spaces; or unstructured solutions, for example different parts of the genotype growing uncoordinatedly. In biology, the process that transforms a zygote into a full multicellular organism, with the cells dividing and arranging themselves forming tissues and complex shapes in a development process regulated by the selective expression of genes (Carroll, Grenier, & Weatherbee, 2004) (Forgács & Newman, 2005) (Mayr, 1961), is known to play a key role in the evolutionary processes (Marcus, 2003) (Lewontin, 2000) (Müller G. B., 2007) (Bateson, 2001) (Borenstein & Krakauer, 2008). Evolutionary developmental biology (known as evo-devo) (Carroll S. B., 2005) is the branch studying these mechanisms. It considers that the evolution of diversity is related to the evolution of the developmental processes (Borenstein & Krakauer, 2008). In fact, there is evidence that the sole variation in the regulatory programs may better explain the evolutionary emergence of novelty and diversity (Levine & Tjian, 2003), (Davidson, 2010), (Hood & Galas, 2003), (Davidson & Erwin, 2006). In computer science, this school has inspired analogous methods. With evolutionary algorithms, if using an adequate indirect encoding (Stanley & Miikkulainen, 2003), a small genotype can result in a large and complex phenotype, making the approach much more scalable. Moreover, small variations in the genotype could produce greatly different phenotypes, since the changes would be propagated in a structured way during the developmental process, which would cause this method to produce more robust, structured and potentially original solutions. Hence, these techniques have been used in fields that demanded exploring a huge space of options or certain dose of creativity. For example, they have been applied to alter existing solutions toward desired targets or as an automated tool for brainstorming, traditionally only carried out by humans. Some real-life examples are the design of NASA satellites antennas (Hornby, Lohn, & Linden, 2011), the design of micro structured optical fibers (Manos, Large, & Poladian, 2007), the automatic generation of board games (Browne, 2008) or techniques for automatic character animation (Lobo, Fernández, & Vico, 2012).

In biology, the embryological development is not an isolated process solely directed by the genome, it is also studied as a growth influenced by the physical nature of the environment (Forgács & Newman, 2005). From this point of view, to understand the emergence of diversity, physical interactions are as important as genes in the developmental processes and they can have a great impact in the evolutionary dynamics in some contexts like early stages of multicellular evolution (Newman, 2003) or in some unicellular organisms (Goodwin, 2001). In computer science these ideas have also inspired a number of works, such as Lobo's thesis (Lobo D. , 2010), Fernández's (Fernández, 2012) or Sánchez-Quintana's (Sánchez-Quintana, 2014). Our system Melomics is another example; the development of a genome into a final composition, score or audio sample, is conditioned by the predefined musical context, as detailed in further sections.

As a summary, we take two main ideas from the biomimetic discipline: (1) an indirect genetic representation for the music compositions and (2) a complex developmental process affected by a sort of physical context. This approach enables us to exploit a set of advantages, among others: a great potential to produce original material with no creative input, the possibility to obtain relevant or useful variations from an existing solution and the ability to explore a huge search space, providing solutions or great variability.

## 1.4 Outline

The rest of this thesis is organized as follows: Chapter 2 introduces Melomics, a computer system that produces music, both in score and audio formats. It makes use of evolutionary computation, in particular two models of genome and their associated development procedures, to give place to the final compositions. In Chapter 3, some tools to analyze music are presented; they study music from different perspectives: dealing with symbolic notation, wave formats or even cultural information. These tools are used to analyze Melomics music properties compared with the traditional human-made music. This chapter also describes an experiment where 251 subjects assess our computer-generated music from a perceptive approach. Chapter 4 describes how the automated method is exploited to create a new way of composing and providing with music, being able to modulate it in real time according to a given context; then some working applications are detailed. In Chapter 5 the main results, conclusions and contributions of this work are presented and discussed, as well as some extended lines of work. Appendix A gives a description of each symbol of the representation models. Appendix B shows the main hardware used for developing and testing. Appendix C provides some execution statistics for both the atonal and the tonal system, using different hardware and parameter configurations. Finally, Appendix D consists of an extensive summary and conclusions of this thesis in Spanish.

# Chapter 2

# Melodies and genomics

In this chapter, we present the computational tool called Melomics, which applies bioinspired approaches to compose music. The software includes two similar genetic-based ways for representing music compositions, in both cases mimicking a developmental process inspired by biological ontogeny, after which the music pieces appear written in a more explicit notation. There are also implemented different mechanisms at different stages of the process, in order to assess the goodness of each composition and feed the system back. A fully developed composition, stored in the system in its own internal format, can then be prepared and converted to many standard file types, either musical score formats (MusicXML, MIDI) or audio formats (FLAC, MP3).

This chapter is organized as follows: In section 2.1 we introduce the basis and motivation for using the chosen genome model and we put it in context with similar existing methods. In section 2.2 we detail the two formal models which constitute the foundations of our compositional algorithm along with their respective development process and some implementation aspects. In section 2.3 we describe our internal representation of the music and how it is translated into the preferred output formats. In section 2.4 we break the automated composition process down. In section 2.5 we cite some of the most relevant experiments carried out with each sub-system. In section 2.6 we list the main technologies used during the development. Finally, in section 2.7 we provide an overview and discuss the results and next goals.

## 2.1 Motivation

Automatic music composition is challenging from both a technical and a theoretical perspective. From a technical point of view, one important matter is the design of the musical formats, which should allow the representation and management of the information in a suitable way, according to the specific purposes. For example, we could require supporting basic operations like raising the musical pitch to one instrument or changing the tempo of the whole composition; but we could also require more advanced operations, like dynamic adjustment of the whole composition (tessitura, melodic steps, dynamics, etc.) responding to the change of a performing instrument, for example. There are other technical concerns, like managing the directions that can usually be seen in musical scores as expressions in natural language. The system must be able to produce and handle this kind of instructions and incorporate them into the different musical format in a logical manner, with the least possible loss of information. From a more formal perspective, we require a mechanism to specify, describe and deal with music in more abstract terms. Section 2.2.3.1 Musical styles shows our first proposal to specify music and the whole section 2.2 describes how the system deals with these concepts of music theory in each particular instance or music composition.

One of the main properties of our genetic models is that they are based on formal grammars. A composition is described using a set of rules that expand high-level symbols into a more detailed sequence. Formal grammars have been suitable to express the inner structured nature of music, for example to study melody (Gilbert & Conklin, 2007), musical structure (Abdallah, Gold, & Marsden, 2016) or harmony in tonal music, even with a generative approach (Rohrmeier, 2011). Lindenmayer Systems (abbreviated L-systems) (Lindenmayer, 1968), a particular variant that served us to implement the first genetic model, works by performing a parallel rewriting of all the applicable rules in each iteration. They are characterized by the capability to encode relatively complex structures with simple sets of rules, and they have been especially used to model the anatomy, physiology and morphogenesis of plants (Prusinkiewicz & Lindenmayer, 1990). Implementing a version of the so called turtle interpretation, using musical elements instead of graphical information, has brought many applications to algorithmic composition (Prusinkiewicz, 1986) (Mason & Saffle, 1994) (Wilson, 2009). Using grammars as a genomic basis in a genetic algorithm for automated composition, has also been tried from different approaches, like the system described Reddin (Reddin, McDermott, & O'Neill, 2009) or GeNotator (Thywissen, 1999), with an interactive fitness function. In particular, L-systems have been frequently used as representational models in evo-devo processes due to their simplicity, sometimes with an interactive fitness as well and studying ways of visualizing music (McCormack, 2005) (Rodrigues, Costa, Cardoso, Machado, & Cruz, 2016).

## 2.2 Representation models

In order to produce full compositions through evolutionary processes, we need to encode the whole structure of the music, not only the single pitches and other music variables, but also the relations among them. That meaning, there must be a higher-level structure more compact than the entire composition. In a sense, it is a compressed version of the composition, which implies the presence of some kind of repetitions and structured behavior, since truly random data would not be susceptible to compression.

While computers are capable of sound synthesis and procedural generation of sound, we are interested in the production of a music score in the traditional staff notation. Some systems aim to reproduce a particular style of a specific artist, period, or genre by using a corpus of compositions from which recurring structures can be extracted. We pursue music composition from scratch, modeling a creative process able to generate music without imitation of any given corpus. The ability to create music in a specific style is enforced by a combination of two important constraints:

- The encoding, which generates a bias in the search space. That is, by changing the way we represent music we also change what is easy to write and what is difficult -or even impossible- to express. By forcing music to be expressed as a deterministic L-system or a deterministic context-free grammar, regular structures are easier to represent and will appear more often than completely unrelated fragments of music. A set of rules can appear not to provide any repeated structure, but its product might contain a high number of repetitions and self-similar structures. This is, indeed, an essential aspect in both biology and music (Ohno, 1987). The structure of an entire piece of music is unequivocally encoded, including performing directions, and deploys from a series of derivations. This provides a way to generate repetitions in the product from a simple axiom, without being repetitive.

• The fitness function, which associates a measure of quality to each composition, determined by looking at some of its high-level features, such as duration, amount of dissonance, etc.

The dual effect of having an encoding that restricts the search into a more structured space, where individual changes in the genome can produce a set of organized alterations on the phenotype, combined with the filtering of a fitness function, helps the generation process to "converge" to music respecting a particular style or set of conditions, without any imitation.

We have developed two different compositional methods. One is aimed to produce atonal music, which does not follow the standard rules of tonality based on the existence of a central tone or key, and the second is intended to create tonal compositions, prioritizing harmony rules, that is, dealing with specific relationships between note pitches along time and the participant instruments. We have designed an approach of encoding music by means of L-systems and context-free grammars and then compose by evolution of these formal models. The two representational systems are actually part of different programs developed from the same principles but adapted for the composition of both atonal and tonal music. These mathematical systems, able to represent self-similarity encoding compositional and performance properties, can evolve as to produce valid and esthetic music scores and then synthesize them in multiple audio formats. The two workflows slightly vary, mainly due to their different genetic formats and the different ways to evaluate the goodness of the results. However, apart from the harmony, the treatment of the rest of the musical variables is very similar in both types of music: some kind of order is sought in the compositional structure, in the durations (rhythms), in the volumes (dynamics) or the timber (ensemble of instruments); conventions in music notation or formats are equal; and, from a general point of view, what is expected from both systems is the same as well: solutions that are original, valuable, mutable, etc. Hence, the systems have been designed following similar principles.

## 2.2.1 Music variables

In the design of the representation models, two of the main problems that have been addressed are (1) how to map the structure of the genetic representation with the structure of a composition and (2) how to assign the operators and the rest of the genes to music behaviors and parameters, thus converting an arbitrary string of symbols into musical data.

- **The structure** of a composition can be described by the arrangement of certain identifiable sections or musical units. Some of these entities can be at the higher level of the musical hierarchy, such as the verses, the choruses or the bridges; and some other at the lower levels, like for example musical motifs or specific patterns. These units can be usually recognized for their particular musical features (rhythm, melodic contour, tone, performing instruments...). A complete composition shall be formed by some of these pieces of original material, which are reused along time, possibly introducing variations. In our system, the musical structure is handled by the grammar structure, providing high compactness, readability, scalability, expressivity and flexibility.

- **The instruments** define the timbres or tone colors in the composition, but they also shape the musical functions or behaviors to be performed. These roles can be of many types and may define the relationship among the participant instruments. Some examples are melodies, counterpoints, harmonizations, harmonic accompaniments, rhythmic accompaniments or effects (FX). In our system, the roles and the concrete instruments for a composition can be specified through predesigned templates and then handled by a set of rules based on expert knowledge, which influence the process of creating genetic material. Regarding the audio file that might be required to be produced, the actual virtual instruments to use during the synthesis are also managed by the expert knowledge coded in the genotype.

- **The textural evolution** is referred to the presence or absence of each instrument in the composition over time. The associated parameters settle instrumental densities, persistent musical roles or transitions among them during the composition. This feature is handled at the higher level of the grammar, where the proper regulatory genes introduced by expert knowledge rules are in charge of enabling or inhibiting the presence of roles or instruments.

- **The tempo** marks the pace of the composition, usually expressed as the number of beats (a note value of reference) per units of time (e.g. 60 beats per minute). As in musical compositions, in a genome the tempo can be changed at any place, but it is typically set globally or in the highest level of the grammar structure. There is an alternate or complementary way of specifying the tempo, based on subjective directions given by the composer (e.g. allegro, andante, molto vivace...), which is especially used in classical music. Our system handles these expressions, through rules relating explicit tempo and directions, applied at the final stage of development, while building the score or synthesizing the audio.

- **The measure** (or bar) is a segment of time containing a certain number of beats, indicated by the measure number, normally implying some kind of grouping among these beats and conditioning the resulting rhythm. For example, in a section with a measure number 3/4, the rhythmic patterns that are expected to emerge consist of sequences of 3 beats, each one filled with a quarter note[8] or a cluster of notes of equivalent duration. Additionally, the measure number may influence the internal accentuation or the way the notes are performed in terms of intensity according to their position in the subdivision of one beat.

- **The rhythm** refers to a sequence of notes and rests with specific durations, which can appear repeatedly over time. These sequences might be influenced by the chosen measures and other variables, like the role executing the rhythm, the actual chosen instrument to play the role, the musical genre or style, etc. In Western music[9] the rhythm can be typically built using a duration of reference and a set of note values representing durations obtained by multiplying the reference by a power of two (positive or negative). These note values can be used as they are, tied among them or be split or grouped into some other duration patterns or more irregular configurations. This method makes all kind of sequences of durations possible, but more frequent those that are simpler. In the genetic models we have tried to map this behavior with corresponding structural genes.

---

[8] https://en.wikipedia.org/wiki/Quarter_note (accessed on May 16, 2021)

- **The pitch** in music is the property of sound related to the frequency of vibration, letting us to distinguish between bass and treble sounds. Music harmony consists of the way of using the different pitches simultaneously and along the time. It considers the proportions between the fundamental frequencies of the involved notes, because it affects the way the sounds are perceived inside the listener's auditory system. For example, the ratio 11/12 is perceived as tense or dissonant, while 4/3, 3/2, 2/1 or 1/1 is usually more pleasant. In music composition and, therefore, in the proposed system, managing harmony is a complex issue. In the lowest level, the genetic representations include structural genes that help to represent the pitches as shifts inside the chosen scale and also to build chords or arrangements of notes with specific ratios among their pitches. In the higher levels, tools based on expert knowledge have been designed to consider harmony, especially in the system for tonal music. There is a mechanism to manage chord progressions, establish musical scales, introduce different kinds of harmonic accompaniments or locate and filter certain types of dissonances, once the current composition has been developed. One last consideration regarding the pitch property is about the boundaries. Actual instruments usually have upper and lower pitch limits, commonly known as tessitura, beyond which it is impossible for an instrument to produce a valid sound. Additionally, they usually have what is called sweet spot, a region of pitches where they make the clearest sounds. In other cases, the limits may be imposed by the style of music that is being created.

---

[9] https://en.wikipedia.org/wiki/Classical_music (accessed on May 16, 2021)

- **The scale** is a parameter related to pitch and harmony. In this compositional system, as usually in music composition, managing pitches is done in an abstract way: shifts among pitches (simultaneous and over time) are considered in relation to a concrete mode or collection of allowed pitches, plus an initial tone of reference called key (also root or tonic) giving place to a scale. The mode can be settled globally or can be changed among different sections of a composition. The keynote usually changes more often, producing different harmonic progressions, which causes, for example, that a musical fragment repeated with a shift on the root note could be identified by the listener as a variation of the same musical material. In our model for atonal music, handling the scale is a relatively flexible and dynamic process, while the model for tonal music counts on a much more controlled way to do so.

- **The dynamics** are related to the way the notes are produced with the instruments in terms of strength, that is, the resulting volume, but also its progression over time. This parameter can be considered in two different levels. In the high level there are the dynamic directions, to indicate musicians how the general volume should be, for example "*mf*" for *mezzo-forte* or moderately loud, "*pp*" for *pianissimo* or very soft and "*cresc.*" (it can also be represented with a special symbol) for an increasing volume. The low level is referred to the attack and the decay times of the notes. Musical articulations tell musicians how notes should be executed, for example, *legato* is used to indicate that notes should be played in a continuous way or tied while performing them and *staccato* to indicate short attack and decay. In this level we also consider the internal accentuation, namely, the way the subdivisions of each beat are accentuated. It depends on different variables like the current measure number, the chosen rhythm or the given genre or style. We have implemented structural genes to express concrete dynamics or modulations that can appear in any level of the musical hierarchy, although they are typically used to configure the global shape or at most to make an instrument come in or leave with a fade effect. Additionally, we have implemented structural genes to express the mentioned articulations and to specify internal accentuations.

- **The effects**, in our system, comprise the different techniques to play musical instruments (e.g. *glissando* or *flutter-tonguing*) and other kinds of alterations to the sound. These effects, implemented through a set of genes, can get translated into directions to the musicians on the score (e.g. the *pizzicato* direction or the *flutter-tonguing* symbol) and they define the way the musical pieces are synthesized into audio, by selecting the adequate virtual instrument with the required execution (*pizzicato*, *glissando*, etc.) or by enabling a certain sound effect (e.g. reverb, volume normalization, panning, etc.).

## 2.2.2 Model for atonal music

To build the first compositional method we considered and picked some of the element present in the modern Western music theory. In general, this was done to favor the sequences of sounds and rests to appear organized, which is somehow what we understand as music. These elements include note values, to enhance order in rhythm; keys and scales, to enable order in pitches; or structural components, such as measures, phrases, etc., to promote order in a higher level of a composition. However, we avoided using specific constructions or following the usual conventions of Western music, such as rhythmic patterns or modes and chords, in order to open the range of musical constructions that would be produced.

- Sample 2.1. Demo of themes with a wide configuration of parameters

Since the basic ingredients to generate music were implemented but no specific conventions were hardcoded in the genetic model, the system was suitable to aim the contemporary classical music, a relevant style in nowadays professional composition. This music comprehends several kinds of artistic tendencies present in the late 20th century, with a special mention to the modernism, and it is mainly characterized for the lack of strong rules to manage the musical elements, granting the author with freedom to explore more combinations. This is especially noticeable in the pitch dimension, generally giving place to music in the chromatic scale, with little restrictions, thus being perceived with an undefined key or tonal center, which is called atonality.

- Sample 2.2. Example of atonal music produced by Melomics

The core of the atonal system is based on a grammar model that constitutes the primary genome representation of one composition. This genome develops, i.e., the grammar is unfolded into a resulting string, which is then interpreted to musical elements, sequentially from left to right. During the unfolding process, some input parameters outside the genotype, through different mechanisms can affect and help give the final shape to the resulting composition.

For the atonal system we designed a first high-level input interface of musical parameters that can help anyone to specify the type of music to be produced. Part of this information is used to guide the process of building genomes, making them more probable to develop into songs that fit the specifications. Some examples are textural density curve, repetitiveness of pitches or repetitiveness of rhythms. Although we use indirect encoding, the genome design allows us to intervene in this phase and make the search faster. On the other hand, some other features like the amount of dissonance or the shape of melodic curves can only be assessed at the end of the process, when the genome has completely developed and translated into music. This constitutes the fitness function.

In the next sections we describe the formal basis of the genome model for atonal music: (a) the genetic representation of songs and (b) the process to develop a genome into an explicit composition. In section 2.4.2     Fitness  we show how the set of rules that comprises the fitness function are organized and in section 2.5 we present some relevant use-cases or experiments carried out with the system.

## 2.2.2.1 Genome

The atonal system uses an encoding based on a deterministic L-system (genetic part) and some global parameters used during the development phase (epigenetic part). An L-system is defined as a triple $(V, S, P)$ where $V$ is the alphabet (a non-empty, finite set of symbols), $S \in V$ is the axiom or starting symbol and $P \subseteq V \times V^*$ is the set of production rules in the form $A \to x, A \in V, x \in V^*$. Each production rule rewrites every appearance of the symbol $A$ in the current string into $x$. Because it is deterministic, there is only one rule $A \to x$ for each symbol $A \in V$. In our model, we divide symbols of the alphabet $V$ into two types:

- Operators, which are reserved symbols represented by a sequence of characters with the form $@i, i \in \mathbb{Z}$ or $\$j, j \in \{1 \ldots n\}$. There are no rules explicitly written for them, the rule $A \to A$ is assumed instead. They control the current state of the abstract machine, modulating the musical variables: pitch, duration, onset time, effects, volume, tempo, etc.
- Instruments, with the form $\#k, k \in \{1 \ldots n\}$. These symbols have an explicit production rule with them on the left side and can represent either a musical structural unit (composition, phrase, idea, etc.) or, if they appear in the final string, a note played by the instrument that has been associated to that symbol.

Since the rewriting process could potentially be infinite, in order to stop it, we introduce the following mechanism: each of the production rules have an associated real value $r_i$, indicating the possibility of that rule to be applied in the next rewriting iteration. There is a global parameter $T$ which serves as the initial value for all $r_i$ and there is a weight $I_i$ for each production rule, to compute the new value of $r_i$ from the previous time step, $r_{i_t} = f(r_{i_{t-1}}, I_i)$. All $r_i$ will progressively get a lower value until reaching 0, through an arbitrary function $f$ monotonically decreasing. At that point, the associated rule will not be applied anymore. The formula is applied if that rule is used at least once in the current iteration.

### 2.2.2.2 Development

To illustrate the process, let us introduce some of the operators that we use:

$1 increase pitch value one step in the scale.

$2 decrease pitch value one step in the scale.

$5 push current pitch and duration values in the *pitch stack*.

$6 pop pitch and duration stored in the *pitch stack*.

$7 push current time position in the *time stack*.

$8 pop time position saved in the *time stack*.

$96 apply dynamic *mezzo-forte*.

@60$102 apply tempo: quarter equals 60.

For an extensive explanation of the reserved symbols see Appendix A.1 Genome in the atonal model.

Now, let us consider the following simple and handmade L-system:

$G_a = (V, S, P)$

$V = \{\#0, \#1, \#2, \#3, \#4, \$1, \$2, \$5, \$6, \$7, \$8, \$96, @60\$102\}$

$S = \{\#0\}$

and $P$ consisting of the following rules:

$P = \{$

 $\#0 \rightarrow @60\$102\$96\#2\$1\$1\$1\$1\$1\$1\#2$

 $\#1 \rightarrow \#1$

 $\#2 \rightarrow \$7\#3\$8\#4$

 $\#3 \rightarrow \#3\$2\#3\$1\#1\#3$

 $\#4 \rightarrow \#4\#1\$5\$1\$1\#4\$6\#4$

 $\}$

Below we illustrate the developmental process to get the composition generated by $G_a$, considering $T = 1$, $I_i = 1$ for all the rules, $f : r_{i_t} = r_{i_{t-1}} - I_i$ and the following global parameters:

> Scale: *C major*
> Tempo: 80 BPM
> Default duration: *quarter note*
> Default Dynamic: *mezzo-piano*
> Initial pitch: *middle C*
> Instruments: piano (0), musical rest (1), church organ (2 and 3), cello (4)

The vector $r$ shows the current values of $r_i$ for all the production rules.

**Iteration 0**

With no iterations, the resulting string is the axiom, which is interpreted as a single note, played by its associated instrument, the piano, with all the musical parameters being on their default values still.

String: #0
$r = [11111]$

To better illustrate the interpretation procedure, Figure 2.1 shows the resulting score from the current string, if we supposed that the rewriting counter is 0 for the first rule ($r_0 = 0$) at this point. Since it is not the case, the rewriting process continues.

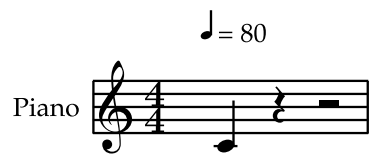

**Figure 2.1.** Resulting score at iteration 0.

- Sample 2.3. Atonal example. Iteration 0

**Iteration 1**

For the first iteration, the rule associated to the only symbol in the string needs to be applied and then, applying the formula, $r_0$ is set to 0.

String: @60$102$96#2$1$1$1$1$1$1#2
$r = [01111]$

This is interpreted from left to right as: change the tempo to "quarter equals 60", apply the dynamic *mezzo-forte*, play the current note on instrument 2, rise the pitch seven steps on the given scale and finally play the resulting note on instrument 2. As before, Figure 2.2 shows the resulting score.

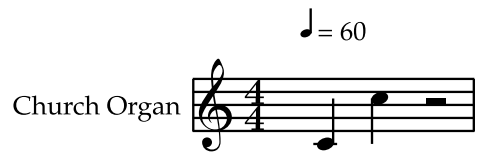

**Figure 2.2.** Resulting score at iteration 1.

- Sample 2.4. Atonal example. Iteration 1

**Iteration 2**

The previous score showed the hypothetical outcome if we supposed $r_2 = 0$ already. Since that symbol is in the string and $r_2 = 1$, the rule has to be applied once.

String: @60$102$96$7#3$8#4$1$1$1$1$1$1$7#3$8#4
$r = [01011]$

Figure 2.3 shows the hypothetical score for Iteration 2. At this point, symbol #2, as opposed to what would have happened if it had stopped at Iteration 1, no longer acts as a playing instrument, but rather as a compositional block that gives place to two instruments. These instruments play at the same time due to the use of the *time stack* operators $7 and $8. The "synchronization" between the two instruments is an emerging property from the indirect encoding.

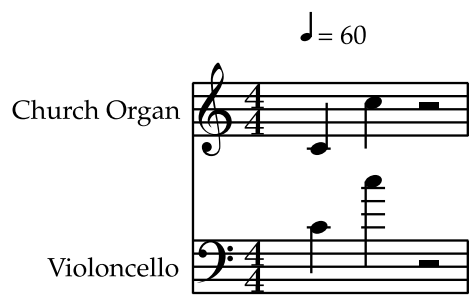

**Figure 2.3.** Resulting score at iteration 2.

- Sample 2.5. Atonal example. Iteration 2

The development has passed the second level of the genomic structure, giving place to the two final instruments, church organ and cello, playing simultaneously the same notes.

**Iteration 3**

The rules with #3 and #4 need to be applied, resulting:

String:

@60$102$96$7#3$2#3$1$1#1#3$8#4#1$5$1$1#4$6#4$1$1$1$1$1$1

$7#3$2#3$1$1#1#3$8#4#1$5$1$1#4$6#4

$r = [01000]$

The music has acquired a structure, the initial rewriting resulted in a change of tempo and dynamic and a shift up in the pitch one octave for the second part of the composition. The following rewriting developed two instruments playing in polyphony and the latest one provided the final melody to each instrument.

There is a fourth iteration that rewrites the symbol #1 into itself and sets $r_1 = 0$. The score shown in Figure 2.4 corresponds to Iteration 3 and Iteration 4, the final score in this case.

**Figure 2.4.** Resulting score at iteration 4.

- ▪ Sample 2.6. Atonal example. Iteration 4

## 2.2.3 Model for tonal music

While the system used to produce atonal music can also generate tonal music, in order to favor the emergence of structures and elements that are usually present by convention in popular Western music, we introduced different changes.

As before, the new genotype is based on a formal grammar model that undergoes a rewriting process leading to a string of symbols, which is interpreted into a matrix-based music structure and the whole process is driven by a set of initial parameters. To enhance management of harmony, texture, dynamics, rhythms, polyphony and many other relevant musical elements, we introduced a new set of operational genes, we settle a more restrictive structure for the formal grammar, we implemented additional functions to make adjustments to the resulting string and we extended the mechanism to interpret the string, which now contains more abstract data. In summary, the new model manages the exact same musical parameters, but with a higher control over them. In addition, we developed a new tool that let us specify musical styles in a much more abstract, easy and controlled way (detailed in section 2.2.3.1 Musical styles). This tool can be used as loosely as to let the system search in a space of multiple styles or as restrictively as to let the system only be able to find only one specific valid composition.

In the next sections we first describe the tool to specify musical styles (style-tool). Then we explain the formal basis of the model for tonal music: (a) the genetic representation of songs and (b) the development process. In section 2.4.2  Fitness we show how the set of rules that comprises the fitness function are organized and in section 2.5 we present some relevant use-case or experiments carried out with the system.

### 2.2.3.1 Musical styles

To make easy and standard the way to specify the music desired to produce, we designed a tool based on filling a sort of a questionnaire with a number of high-level and low-level musical parameters. The information is used to assess the developed pieces (mainly the parameters related to rhythm and harmony), but also to build genomes, whose phenotypes is likely to fit most of the restrictions.

The parameters included and the way of using them to produce genomes was studied and designed with the help of expert musicians. The nature of these inputs makes possible the interface to be handled by people with no knowledge at all about the way the software operates, but with just a bit of background in music composition; in fact, most of the times these questionnaires were completely filled in by the musicians alone. Next, we describe the categories present in the template.

## Musical structure

In order to establish a compositional order as it is common in a traditional composition process, inspired by Bent's Analysis (Bent & Pople, 2010), the production rules are explicitly structured in five hierarchical levels:

- **Composition**. This is the most abstract level and it is formed by a sequence of similar or different kinds of periods, possibly with music operators (alterations in tone, harmony, tempo, macro-dynamics...) between each of them.
- **Period**. This is the highest structural subdivision of a composition. There can be more than one type of period, built independently, becoming separate musical units recognizable in the composition.
- **Phrase**. This is the third structural level, the constituent material of the periods.
- **Idea**. Constitutes the lowest abstract level in the structure of a composition. A phrase can be composed by different ideas that can be repeated in time, with music operators in the middle. A musical idea is a short sequence of notes generated independently for each role, using many different criteria (harmony, rhythm, pitch intervals, relationship with other roles...).
- **Notes**. This level is the most concrete level, composed only by operators and notes played by instruments.

## Global parameters

Global parameters are introduced to force the occurrence of tonal music. Some of these are applied straight away when creating a new genome, by filtering the use of certain strings as the right-hand side of a production rule and some others are used as part of the fitness function (see section 2.4.2   Fitness). These parameters can be grouped in the following categories:

- **Structure**. Set of boundaries to define the global hierarchy of the composition, comprising (a) number of child compositional units that a compositional unit of the immediate higher level can contain (i.e., number of periods in a composition, phrases in a period and ideas in a phrase), expressed as a vector of integers with the valid options, and (b) similar definitions for the number of different compositional units (i.e., number of different periods in a composition, different phrases in a period and different ideas in a phrase).

- **Duration**. Boundaries or specific values, provided in seconds, to define the length of the composition or any of the structural levels.

- **Tempo**. Boundaries or vector with specific valid options to establish tempos along a composition.

- **Dynamics**. Boundaries or vector with specific valid options to establish macro-dynamics along a composition.

- **Role types**. List of roles or behaviors that can appear in a composition. There are currently 89 different roles implemented that can be classified in melody, accompaniments, homophony (harmonized melody) and counterpoint. The harmonic and rhythmic accompaniments can appear in different forms: Chords, Bass, Pads, Arpeggios, Ostinati, Drums and Percussion. These roles have presets of parameters associated, such us instruments, rhythmic patterns, etc. (see parameters below).

- **Arpeggio configuration**. If the role arpeggio is enabled, some additional parameters must be given, such as mode or contour type ("chord", "up", "down", "up-down" and others), durations or time segments to divide the primary rhythmic cell, scope, expressed in number of octaves, or tie notes, indicating whether consecutive similar notes shall be tied or not.

- **Instruments**. For each role there is a list of allowed instruments and their probability to be chosen, plus associated properties, such as tessitura, sweet spot or possible virtual instruments to be used during the synthesis.

- **Rhythmic incompatibilities**. Defines the incompatibilities of instruments to play notes at the exact same moment. Specified as a list of pairs of instruments.
- **Scales**. A set of music modes and notes of reference valid to build scales for the compositions, for example C major, C minor, D Dorian or C Mixolydian.
- **Harmony**. This category includes different parameters to describe how harmony must work specified in the different compositional levels. Some parameters are a list of valid chords; a list of valid roots; a list of chord progressions and dissonance rules, specifying group of roles to compare and intervals to check in each case.
- **Rhythmic modes**. Includes the allowed measure types and the accents to perform the notes, given in a vector of size dependent on the measure type (e.g. (1, 3, 2, 4) for measure 4-4 means that the first time is stronger, then the third, followed by the second and the fourth).
- **Rhythmic patterns**. For each role, a list of valid note values or patterns can be specified, each one with a probability of occurrence in the composition.
- **Melodic pitch intervals**. For each role of the melody kind, a weighted list of pitch intervals can be specified to constrain the way the melodies can be.
- **Texture**. This sub-section defines rules for the inclusion of roles; dependencies between them; the compositional units where they are forced, allowed or prohibited to appear in; and general evolution of the presence of instruments (see more section 2.4.2Fitness).
- **Alterations**. For each compositional level, a list of music operations (associated to genes, see section 2.2.3.2 Genome) and corresponding intensity of occurrence between 2 different compositional units, given as a list of integers. For example, increase of pitch can occur 0 to 4 times between periods.

### Special periods

It is possible to define periods with special behaviors. This enables to define for example musical fills or cadences. The parameters that can be set up for each type of special period are the same as for the regular ones plus a set of simple rules to define how they should be placed in the composition (e.g., after the first or second period or never at the end of the composition).

**Introducing the hypersong**

In order to handle the needs presented by adaptive music (see Chapter 4), we implemented a new use of the style-tool comprising two phases: (1) most of the parameters are randomly instantiated normally, giving place to a very narrow definition of the music to be produced, a meta-composition; (2) the reduced set of parameters that were left untouched in the first stage are specified as a list of combinations and the system generates sequentially a composition for each of them. For example, this list can be just formed by sets of instruments; hence the system would produce the same composition, with a different set of instruments each time. The resulting bundle of compositions obtained with this method is grouped as a single musical entity called a *hypersong* (see use in Chapter 4).

## 2.2.3.2 Genome

In the tonal system a few new operators are introduced to improve harmony management, some of them with a more complex interpretation process and dependencies with other operators and global parameters, like the operator $M$ to create chords (see development example below).

Since we enforce a strict hierarchical structure, it is useful to distinguish the symbols that represent compositional elements that are rewritten during the development from those that keep the same specific meaning at any given moment. For this reason, we use an encoding based on deterministic context-free grammar, defined as $(V, \Sigma, S, P)$, where $V$ is the set of non-terminal symbols, $\Sigma$ is the set of terminal symbols (the alphabet), $S \in V$ is the axiom or starting symbol, $P \subseteq V \times \{V \cup \Sigma\}^*$ is the set of production rules. The symbols in the set $V$ can be identified with the structural units: composition, periods, phrases and ideas, while the symbols in $\Sigma$ represent notes and music operators, such as modulators of pitch, duration, current harmonic root or current chord. In our implementation, on the right-hand side of the production rules, there are only terminal symbols or non-terminal symbols from the following level of the hierarchy in a decreasing order.

### 2.2.3.3 Development

To illustrate the rewriting process, let us introduce some of the reserved terminal symbols and their interpretation:

$N$ increase the counters pitch and harmonic root in one unit.

$n$ decrease the counters pitch and harmonic root in one unit.

[ push in a stack the current value of pitch, harmonic root and duration.

] pop from the stack the last value of pitch, harmonic root and duration.

< push in a stack the current time position, value of pitch, harmonic root and duration.

> pop from the stack the last saved time position, value of pitch, harmonic root and duration.

$W4.0$ apply the macro-dynamic *mezzo-forte*.

$M0.0.0.0$ make the next symbol linked to an instrument play the root note of the current chord, instead of the current pitch.

For an extensive description of the reserved symbols see Appendix A.2 Genome in the tonal model.

Let us define $G_t$, a simple and handmade grammar to illustrate the development using this new model:

$G_t = (V, \Sigma, S, P)$

$V = \{Z, A, B, C, D, E, F\}$

$\Sigma = \{N, n, [, ], <, >, W4.0, a, b, s, M0.0.0.0\}$

$S = \{Z\}$

and $P$ consisting of the following rules:

$P = \{$

$Z \rightarrow W4.0[ANNNNNNNA]B$

$A \rightarrow CC$

$B \rightarrow D$

$C \rightarrow ENEnE$

$D \rightarrow FFNF$

$E \rightarrow < anaNNsa > M0.0.0.0b \ M0.0.0.0b \ s \ M0.0.0.0b$

$F \rightarrow as[NNNa]a$

}

**Iteration 0** (composition)

String: $Z$

The axiom $Z$ represents the highest structural level, the composition.

**Iteration 1** (periods)

String: $W4.0[ANNNNNNNA]B$

The composition develops into two identical musical units (periods), represented by the symbol $A$, separated by seven steps in the pitch dimension and then followed by the period $B$.

**Iteration 2** (phrases)

String: $W4.0[CCNNNNNNNCC]D$

The two types of periods $A$ and $B$ would develop into two simple sequences of phrases, $CC$ and $D$ respectively.

**Iteration 3** (ideas)

String: $W4.0[ENEnEENEnENNNNNNNENEnEENEnE]FFNF$

The phrases $C$ and D consist of a sequence of three ideas each.

**Iteration 4** (notes)

String: $W4.0[$

$< anaNNsa > M0.0.0.0b\ M0.0.0.0b\ s\ M0.0.0.0b\ N$

$< anaNNsa > M0.0.0.0b\ M0.0.0.0b\ s\ M0.0.0.0b\ n$

$< anaNNsa > M0.0.0.0b\ M0.0.0.0b\ s\ M0.0.0.0b$

$< anaNNsa > M0.0.0.0b\ M0.0.0.0b\ s\ M0.0.0.0b\ N$

$< anaNNsa > M0.0.0.0b\ M0.0.0.0b\ s\ M0.0.0.0b\ n$

$< anaNNsa > M0.0.0.0b\ M0.0.0.0b\ s\ M0.0.0.0b$

$NNNNNNN$

$< anaNNsa > M0.0.0.0b\ M0.0.0.0b\ s\ M0.0.0.0b\ N$

$< anaNNsa > M0.0.0.0b\ M0.0.0.0b\ s\ M0.0.0.0b\ n$

$< anaNNsa > M0.0.0.0b\ M0.0.0.0b\ s\ M0.0.0.0b$

$< anaNNsa > M0.0.0.0b\ M0.0.0.0b\ s\ M0.0.0.0b\ N$

$< anaNNsa > M0.0.0.0b\ M0.0.0.0b\ s\ M0.0.0.0b\ n$

$< anaNNsa > M0.0.0.0b\ M0.0.0.0b\ s\ M0.0.0.0b$

$]$

$as[NNNa]a\ as[NNNa]a\ as[NNNa]a$

The last idea, $F$, is interpreted three times only by the instrument linked to $a$, while the previous idea $E$ is performed by the instruments associated to $a$ and $b$, in polyphony; the latter always playing a harmony consisting of the root note of the current chord. To interpret the final string, we will use the following values:

Initial scale: *C major*
Tempo: 80 BPM
Default duration: *quarter note*
Default dynamic: *mezzo-piano*
Initial pitch: *middle C*
Initial chord: *major triad*
Initial root: *I*
Instruments: violin (symbol *a*), double bass (symbol *b*), musical rest (symbol *s*)

Resulting composition is shown in Figure 2.5.

**Figure 2.5.** Resulting score at Iteration 4.

- Sample 2.7. Tonal example. Resulting composition

# 2.3 Music notation

One of the main worries while designing the system was to count on adequate data structures to manage the musical content. Some of the main requirements for them were that they should be (a) standard, to store information coming from both the atonal and the tonal system; (b) easy and fast to operate with, since after the musical data is read from the string, there are still different operations to be performed, such as the instantiation of musical effects and adjusting the note durations and tessituras depending on the instruments and style specified (these can be understood as constraints from the physical context); and (c) easily readable and versatile, since we intended to produce automatically both music scores (and other symbolic formats) and synthesized audio content of high quality, emulating as close as possible a musician's interpretation of the scores.

## 2.3.1 Internal representation

The internal format is the way the symbolic information is stored in the system right after a genome has been developed and interpreted. Its structure consists of (1) a series of global fields and metadata:

- Composition name

- Composition ID

- Composition logo

- Name of the style which gave place to the composition

- Name of the author

- Date of creation

- Structure with the input values and ranges and their final instantiation

- Structure describing the genome development

- Structure describing the instrumental texture

- Tags and keywords associated to the composition (used for example when they are stored in the web server)

- Initial and default values of the musical variables (tempo, key, macro-dynamic…)

- Database of the virtual instruments used

- List of output formats

(2) a set of lists containing information associated to each track:

- Name of the instrument

- Role ID

- Instrument ID

- Instrument type ID

- Index of virtual instrument in the database

- Set of effects and configuration

and (3) a 3D structure where the first dimension corresponds to the tracks, the second dimension is associated to time and the third one contains the properties of each note, which includes:

- Onset

- Duration

- Sound or rest (binary field)

- Pitch

- Dynamic, velocity and volume (related parameters whose application depends on the targeted output)

- Note value and representation details for the score (keys, bars, note stems…)

- Instrumental effects and ornaments

- Subjective directions

## 2.3.2 Exchange format

The need to communicate certain information about the compositions to external systems eventually appeared. An example was the use of Melomics music in an application with dancing avatars. These actors followed the leading beat of the music, but also changed their choreography according to compositional features like structural or textural evolution. This information was provided separately with the composition itself. Storing compositions in the web server is another example that required a way to export additional information.

To satisfy these needs, we designed an exchange format containing meta-information about the compositions, based on the JSON format. To date, the fields that have been included are: the composition name, the style, applicable tags or keywords, the effects used on each track in the synthesis of the MP3, the hierarchy of structural entities, the global beats per minute (BPM), the textural evolution of instruments and the number of measures.

## 2.3.3 Output formats

According to the nature of the stored data, a music format can be of two types: (a) symbolic, representing abstract information, like pitches, tempos, durations, etc. and (b) audio, with the music being handled as sound waves. Examples of the first kind are: MIDI, LilyPond, MusicXML, Open Sound Control[10] or the proprietary formats used in notation software like Sibelius®, Finale® or Notion™. Examples of the second kind are WAV, FLAC, MP3 or OGG[11] and also platforms like Open Sound System[12] or Csound.[13]

In Melomics we have addressed the three main output layouts described next.

---

[10] https://en.wikipedia.org/wiki/Open_Sound_Control (accessed on May 16, 2021)
[11] https://xiph.org/ogg/ (accessed on May 16, 2021)
[12] http://www.opensound.com/ (accessed on May 16, 2021)
[13] http://csound.github.io/ (accessed on May 16, 2021)

### 2.3.3.1 Score

The musical score is the main target for the atonal system. We used the MusicXML format, despite being proprietary, because it is fully documented, can be used under a public license and it is compatible with most of the score writing software, digital audio workstation (DAW) and other music tools, including: Sibelius, Finale, Notion, MuseScore,[14] Cubase™, Logic Studio®, Mixcraft™, Reaper and LilyPond.

The functions that translate the information from our internal format to the score in MusicXML are fairly simple and mostly centered on formatting issues, since the required musical data is already computed and made explicit in the internal representation. In order to count on a music score ready to be performed, the system also generates the corresponding printed score in PDF format. The Lilypad[15] tool is used to convert the XML into a LY file, which is then printed in PDF with LilyPond.

### 2.3.3.2 MIDI

MIDI, commonly used with real time virtual synthesizers, is a kind of symbolic music format. It handles a more reduced set of parameters than MusicXML, but it is one of the main outputs because:

- Its use as an exchange format is wider than MusicXML's, in score writing and many other music tools, especially to import samples into DAWs.
- It is the most frequently used format to perform symbolic music information retrieval. It is used for example in jSymbolic, a component of the jMIR suite (see sections 3.1.1 Music Information Retrieval and 3.1.2 Measurement of similarity).
- Melomics automated music synthesis module is based on the MIDI format as an input.

In the primary versions of the system, MIDI files were obtained from XML through LilyPond, the same as with the PDF. Currently, we use the python library

---

[14] https://musescore.org (accessed on May 16, 2021)
[15] http://lilypond.org/download/gub-sources/lilypad/ (accessed on May 16, 2021)

MIDIUtil[16] to directly build a MIDI file from the internal representation, which makes the process simpler, faster and more controlled.

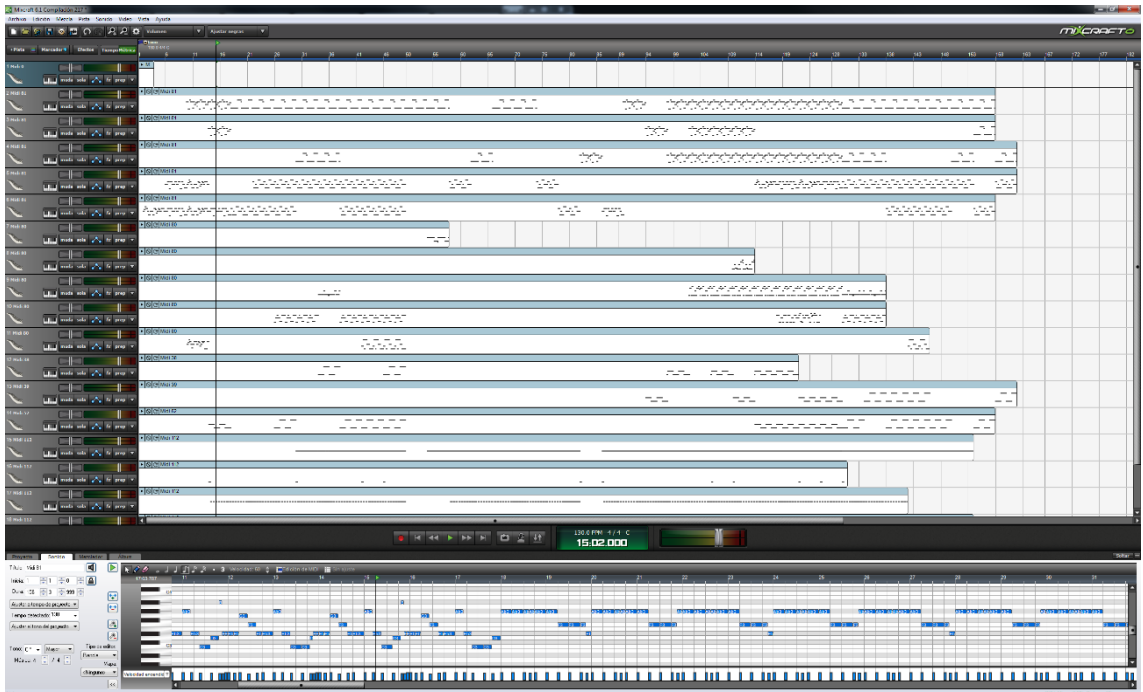

**Figure 2.6.** MIDI file loaded in the digital audio workstation Mixcraft.

### 2.3.3.3 Synthesis

In the compositional workflow, the audio is the most processed way of presenting the music, since it requires all the symbolic information to be produced in advance. The synthesized audio is an approximation of the sound waves related to symbolic content (pitch, volume, durations guidelines for instrumental performance...). At the beginning, the system was able to produce WAV files from the internal representation making use of synthesizers like FluidSynth[17] or TiMidity and virtual instruments in the SoundFont 2 standard (SF2) (De Vicente Milans, 2013). But later, this component was improved to get an approximation closer to what an expert musician can achieve using a DAW manually. In the newer version, the WAV file was obtained with a scriptable DAW called Reaper, which enabled a more complex editing and the ability to

---

handle VST, more advanced than SF2. From the WAV, the audio can be converted to any other audio format (FLAC, MP3, OGG...) with regular tools.

## 2.4 Composition process

Melomics uses a combination of formal grammars to represent music concepts of varying degrees of abstraction and evolutionary techniques to evolve the set of production rules. The system can be used to compose both atonal and tonal music and, although using slightly different encoding methods and a much stronger set of constraints in the second case, they both share a similar structure and execution workflow (see Figure 2.7). From a bio-inspired perspective, this can be thought of as an evolutionary process that operates over a developmental procedure, held by the formal grammar. The music "grows" from the initial seed, the axiom, through the production rules to an internal symbolic representation (similar to a MIDI file), where finally the compositions are subject to test by the set of constraints provided. The execution workflow can be described as follows:

- Filling the desired input parameters. These parameters represent musical specifications or directions at different levels of abstraction, such as instruments that can appear, amount of dissonance, duration of the composition, etc., with no need for creative input from the user.
- The initial gene pool is created randomly, using part of the input parameters as boundaries and filters.
- Each valid genome, based on deterministic grammar and stored as a plain text file, is read and developed into a resulting string of symbols.
- The string of symbols is rewritten after some adjustments of different forms: cleaning of the string, for example by removing sequences of idempotent operators; adjustments due to physical constraints of the instruments, like the maximum number of simultaneous notes playing; or the suppression (or emergence) of particular musical effects. This part in the evo-devo process can be seen as constraints in the developmental process due to epigenetic factors.

- Each symbol in the final string has a musical meaning with a low level of abstraction, which is interpreted by the system through a sequential reading from left to right and stored in an internal representation.

- The musical information is adjusted and stabilized. For example, shift the pitches to satisfy constraints in the instruments' tessituras or discretization of note durations.

- The input directions are used to assess the produced composition that might be discarded or pass the filter. In that case it is saved as a "valid" composition.

- A discarded theme's genome is replaced by a new random genome in the gene pool. On the other hand, a "valid" genome is taken back to the gene pool, after being subject to random mutations and possible crossover with another genome in the gene pool (see section 2.4.1 Mutation and crossover), until it passes the filters defined at genome level (the same as with a random genome), giving place to a sample of a new generation.

- If desired, any composition (usually the ones that pass the filter) can be translated, using the different implemented modules, to standard musical formats that can be symbolic (MIDI, MusicXML, PDF) or audio (WAV, MP3, etc.), after executing a synthesis procedure with virtual instruments, which is also led by the information encoded in the composition's genome.

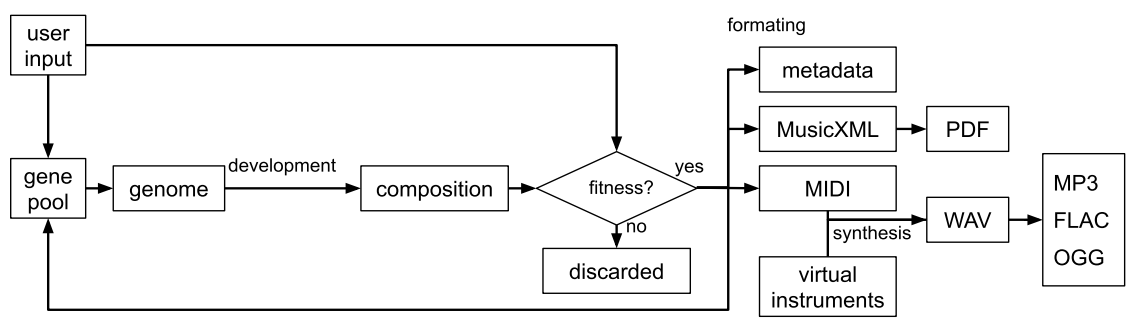

**Figure 2.7.** A graphical representation of the composition workflow.

The average time for the system to create a new valid genome varies very much depending on the style and constraints imposed. Using single core on an Intel Xeon E5645, for a very simple style it can be ready in less than one second, while for the most complex style, symphonic orchestra in the atonal system, the process takes 256 seconds on average. Given a built gene pool, the time to obtain a new valid composition, which includes executing the developmental process possibly repeatedly until passing the filters, varies a lot too. It can take 6 seconds for the simplest atonal style, while it takes 370 seconds on average for the most complex style in the tonal system. A study of execution times with different use-cases is given in Appendix C.

In a genetic algorithm, populations of candidate solutions evolve through generations. As many bioinspired methods, they can be inefficient to find the global optimum, but in general provide good solutions, a behavior that perfectly matches our needs. There are many operations and variations of the method, inspired by biology. Genetic representation of individuals (or solutions) is an important factor that can influence convergence of populations to the optimal. As discussed, we based it on formal grammars for our domain. Mutation of genomes and crossover of parent genomes to give place to individuals of a new generation are two relevant operations, which require to be balanced to manage both diversity and convergence. Population size is another parameter that influence those properties. Other genetic operators, such as inversion of chromosomes have not been implemented in our system. Finally, selection of fit individuals is one of the main operations in genetic algorithms. In our particular implementation of the metaheuristic, the Boolean fitness that simply discard the unfit samples, but accepts the ones that comply with the minimum requirements, acts as a mechanism that values originality.

## 2.4.1 Mutation and crossover

In Melomics, the genomes can be altered in any way possible in the valid set of symbols defined: by removing, adding or altering symbols in any position; the developmental procedure always gives place to a valid musical composition. However, the disruption provided by genetic operators (mutation and crossover) should be balanced in order for the system to converge properly to the given directions. We have observed that a small amount of disruption combined with the hierarchical structure granted by the encoding makes the system produce reasonable results since early iterations, typically less than 10.

For atonal music we allow stronger alterations, since the styles addressed in general are less constraining than the popular styles pursued with the tonal system. The implemented mutations are: (1) changing a global parameter (e.g., default tempo or dynamics); (2) changing the instrument associated to a symbol in the grammar; (3) adding, removing or changing a symbol on the right hand side of a production rule; (4) removing a production rule; and (5) adding a new production rule, copying an existing one or generated randomly, and introducing the new symbol randomly in some of the existing production rules. The implemented crossover mechanism consists of building a new genome taking elements from other two. The values for the global parameters and the list of instruments are taken randomly from any of the source genomes. The set of production rules is taken from one of the parent genomes and then the right-hand side can be replaced with material from the other parent.

The disruption introduced with these mechanisms is still too high, even for atonal music, hence the mutation operations (4) and (5) are used with less probability and the rule with the symbol associated to the musical rest is always kept unaltered. The experiment Genetic engineering: Nokia tune is an example of the use of these mechanisms.

For tonal music the mutation and crossover operations are similar, but more restricted and executing less of them at each iteration. Mutations allowed are: (1) changing a global parameter; (2) changing an instrument for another valid for the same role; (3) adding, removing or changing a non-terminal symbol or a terminal reserved symbol (operators, not instruments associated) on the right hand side of a production rule, provided that it does not alter the five-level hierarchical development; and (4) removing or adding a new terminal non-reserved symbol (Instrument), in the second case duplicating appearances and forcing polyphony with an existing one (enclosing the new one with the symbols <, > and placing it to the left) and assigning an instrument of the same role. The crossover is also similar, taking the genome of one of the parents and only replacing a few rules with material of the same structural level from the other. The global parameters are taken randomly from any of them and the same for the instruments, respecting role constraints. Sample 2.23 from the experiment Genetic engineering: SBS Logo illustrates an example of crossover between Melomics piece generated in the style *DiscoWow2* and a tune of Korean television that was reversed engineer into the tonal system.

## 2.4.2 Fitness

Both the atonal and the tonal systems count on a set of parameterized rules to guide the composition process, by allowing the development of those compositions that comply with the rules, while filtering the rest of them out. There are (a) global rules that basically constitute physical constraints of the musical instruments, such as the impossibility for a single instrument to play more than a certain number of notes simultaneously or to play a note too short or too long; and (b) style based constraints that encode expert knowledge and are used to assure the emergence of a particular kind of music, resembling the way a human musician is requested to create music in a certain style. This latter kind of rules, in general looser in the atonal system, can be grouped as follows:

- **Duration**. Lower and upper boundaries for the duration of a composition. In the tonal system there are also duration boundaries for the different compositional units. These constraints are assessed at the end of the developmental process, on the phenotype, when the musical information is explicitly written in the internal symbolic format.

- **Structure**. Constraints to the number of compositional units. They are applied while building the genome, through boundaries to the number of each non operator symbol. In the tonal system there are four additional types of rules to control the distribution: (1) *symbol X must never appear right before symbol Y*, (2) *X must never appear right after Y*, (3) *X cannot appear in a determined position* (e.g., first, third or last) and (4) *X must appear in a determined position*.

- **Texture**. It is referred to the density and type of instruments playing at each point of the composition. In the atonal system it is checked at the end of the developmental process through a function given as input that specifies the desired instrumental density at each point in time of the composition normalized. The function can be specified globally and by group of instruments. There is a tolerance parameter that should be specified. In the tonal system it is done at genome level. Similar to the case of structure, to manage instruments there are different types of rules: (1) *instrument X is mandatory in unit Y*, (2) *instrument X is forbidden in unit Y*, (3) *unit Y should have instrumental density D*. This can be specified by groups of instruments and can include a degree of tolerance. These rules in the tonal system are managed in the genome through the gene suppressor *x* (see A.2 Genome in the tonal model).

- **Dynamics**. It is the intensity or volume of the different instruments playing at each moment. The management of dynamics is done in a similar way as for texture in both the atonal and the tonal system.

- **Harmony**. It is referred to the relationship between pitches or frequency of the notes being played at a given time by one or multiple instruments and their progression in time during the composition. In the atonal system we implemented a mechanism to assess the developed composition based on a time window, which inspects all the instruments and counts the number of occurrences for each of the different pitch steps in the chromatic scale. These are then weighed and aggregated to provide a measure of the dissonance for that window, which has an upper and a lower boundary. For the tonal system there is a more rigid mechanism implemented at genome level. A particular style is defined with (a) a set of valid modes and tones, (b) a set of valid chords, (c) a set of valid harmonic transitions and (d) a set of instruments that need to copy the harmony from another specified instrument. These directions are used to build the production rules of the lowest level in the grammar (the ideas). However, that alone cannot assure the harmony rules and it has to be checked numerically on the developed composition as well.

- **Rhythm**. It is understood as how note durations and rests evolve during the composition and the relationship between the different instruments. For the atonal system the assessment is done by instrument at phenotype level, similar to harmony: with a time window that keeps a count of duration steps executed, aggregating these values and filtering through an upper and a lower boundary. For the tonal system, some instruments are specified with a set of valid patterns of notes and rests durations and some others are specified to copy the rhythm from another instrument. These directions are used to build the production rules of the lowest level, but rhythm needs to be assured by checking the developed phenotype as well.

- **Instrumental roles**. This section includes relationships between participant roles, such as melody, accompaniment, percussions, etc. It only exists in the tonal system and it comprises four types of rules: (a) *an instrument of role A cannot play at the same time as an instrument of role B*, (b) *an instrument of role A can only play when there is an instrument of role B playing*, (c) *instruments of role A cannot start playing in the composition at the same time as an instrument of role B* and (d) *at least one instrument of role B is required to be playing before an instrument of role A starts playing*. These rules are used to build genomes, however, due to the number of changes that can occur during the developmental process, the directions cannot be guaranteed at this level, hence a later check on the phenotype is necessary.

## 2.5 Relevant experiments

### 2.5.1 Generating atonal music

**Iamus Opus #1**

After building the core of the system we were obtaining some promising preliminary results. In this context we started collaborating with Gustavo Díaz-Jerez, a Spanish pianist, composer and professor in Centro Superior de Música del País Vasco Musikene and Centro Superior de Música Katarina Gurska, who would later describe some of his work within the project (Diaz-Jerez, 2011).

Iamus *Opus #1*, written for flute, clarinet, horn, violin and violoncello, was created on October 15, 2010 and it can be considered the first piece of professional contemporary classical music composed by a computer in its own style. The main contribution of this piece to the system was the introduction of the constraints that make the produced music able to be performed with real instruments. We also introduced other important features as a consequence of this experiment: assessment of the textural evolution, the amount of dissonance and the repetition of note-values and note-pitches. The method to evaluate the wellness of a composition was fine-tuned in an iterative fashion with the help of the expert. The main pursued goal for this case was to obtain pieces of music with a limited amount of perceived dissonance and whose fundamental musical materials (specific melodic constructions, structure...) were evoked along the time and among the different participant instruments.

Figure 2.8. Excerpt of Iamus Opus #1.

- ▪ Sample 2.8. Iamus Opus #1 full score
- ▪ Sample 2.9. Iamus Opus #1 audio emulation

## Hello World!

One year after *Opus #1* was created, we worked on four main aspects of the system: (1) refinement and development of the fitness function; (2) increase of the type of real instruments and performing techniques able to handle; (3) enhancement of MusicXML writing, being able to produce richer scores that contained most of the elements and directions used in standard music notation; and (4) improved process of genome building; divided in two phases, one controlling the structure and another to build the low level material (phrases and motives).

The system was tested with a particular configuration of parameters and ensemble of instruments. The computer cluster Iamus[18] (see Appendix B.2 Computer clusters) was programmed to run for about ten hours, creating independent compositions for clarinet, violin and piano. At the end, one composition was arbitrary picked among the bundle that was produced and it was called *Hello World!*, making reference to the computer program Hello World!.[19] This piece of contemporary classical music was composed on September 2011 and premiered on October 15, 2011 at the Keroxen music festival in Santa Cruz de Tenerife, Spain (Keroxen aúna coreografía y música sobre el escenario de El Tanque, 2011) (Redacción Creativa, 2011),[20] being arguably the first full-scale work entirely composed by a computer and using conventional music notation.

---

[18] https://en.wikipedia.org/wiki/Iamus_(computer) (accessed on May 16, 2021)

**Figure 2.9.** Excerpt of Hello World!.

- Sample 2.10. Hello World! full score
- Sample 2.11. Audio of Hello World! premiered in Tenerife

---

[19] https://en.wikipedia.org/wiki/%22Hello,_World!%22_program (accessed on May 16, 2021)
[20] https://www.youtube.com/watch?v=bD7l4Kg1Rt8 (accessed on May 16, 2021)

### Genetic engineering: Nokia tune

One of the main advantages of compositions having a genetic representation, encoding not only the final notes of each instrument, but its inner structure as well, is the ability to produce coherent variations that affect the whole piece, by applying small changes in the genome. We choose the Nokia tune[21] for this first experiment for being a simple and very identifiable theme.

Since any musical composition can be written in an infinite number of ways in our genomic model, we tried to represent the tune's simple structure as well, when we manually reverse engineered it into the system (see Figure 2.10). Then we run a series of restricted mutations on the genome: change of instruments or scale; addition, deletion or replacement of effects and performance techniques symbols; addition, deletion or replacement of non-reserved symbols (structure or notes); and addition, deletion or replacement of tempo and dynamics symbols. This resulted in a coherent collection of musical pieces, relatives of the Nokia tune; some of them being genetically closer to the original than others. Sample 2.13 and followings show some resulting examples.

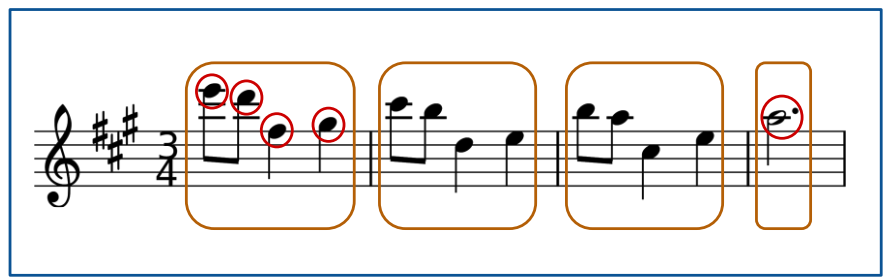

**Figure 2.10.** Reverse engineering experiment. The tune was reengineered using three compositional levels: Phrase, idea and notes. For clarity, we have omitted the global parameters, the track parameters and some of the music operators in the representation of the genome.

---

- Sample 2.12. Audio of Nokia tune reengineered
- Sample 2.13. Audio of Nokia tune relative 1
- Sample 2.14. Audio of Nokia tune relative 2
- Sample 2.15. Audio of Nokia tune relative 3
- Sample 2.16. Audio of Nokia tune relative 4

## Themes for relaxing applications

Besides the targeted contemporary classical style, Melomics software was used to provide with music within a research project of the same name. This project's goal was to create a method based in music therapy with biofeedback to reduce anxiety. The system that had been designed to generate full compositions in contemporary classical music, was set up for this new purpose. The approach was to allow to produce simpler samples, in general, by restricting the search space.

- Sample 2.17. Audio of theme for relaxing applications. Example 1
- Sample 2.18. Audio of theme for relaxing applications. Example 2

It was also a first test of producing a genetic family of themes that could be used together as different versions of one central theme. Each version would be selected during the real-time therapy, according to certain rules. This was the seed of the concept of hypersong introduced above and the applications using adaptive music described in Chapter 4.

- Sample 2.19. Evolving audio for relaxing applications

## Web repository and Iamus album

The input mechanisms used to create *Hello World!* were refined and a higher level user interface was implemented, including a graphic tool that allows to describe the expected evolution of some parameters along the duration of the composition. These improvements made easier to describe compositional styles, establishing the basis for the future tool used in the tonal system, described in section 2.2.3.1 Musical styles.

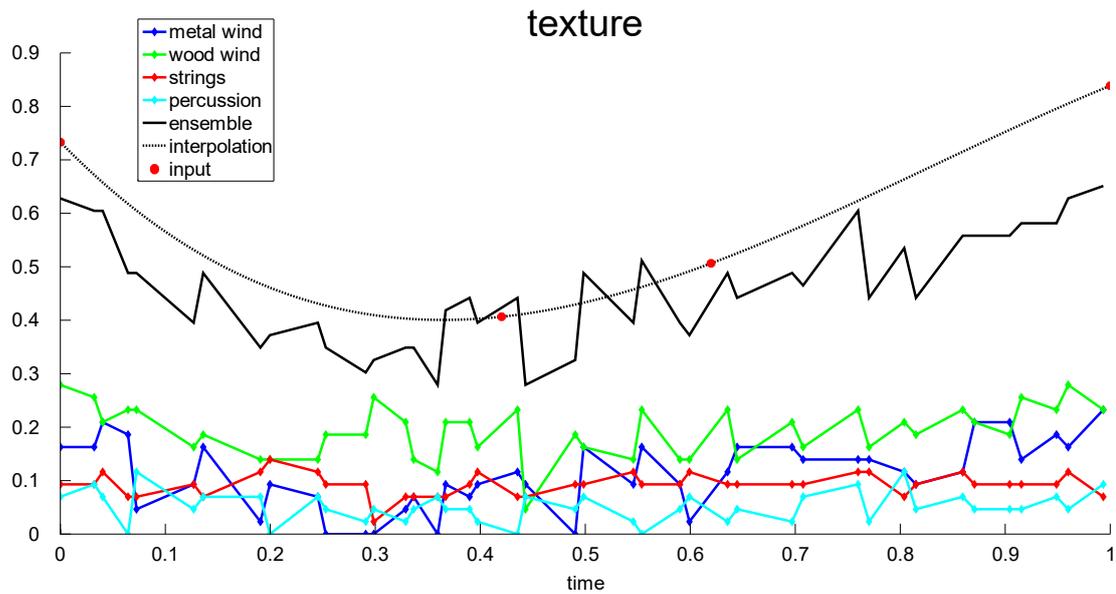

**Figure 2.11.** Interface to specify one of the compositional parameters. After the user places the desired points, the system interpolates them to obtain the whole curve.

A proposed goal in contemporary classical music was to produce a full album with different ensembles. With the help of the expert musician, we prepared seven specification settings and the result was the Iamus album, released in 2012 and containing the composition *Hello World!*, as well as a piece for orchestra that was recorded by the London Symphony Orchestra;[22] and some chamber pieces with the following ensembles: viola d'amore and harpsichord; clarinet; violin and piano; piano; violin; and voice and piano. These recorded by recognized musicians in Sala María Cristina, Málaga.[23]

---

[22] http://lso.co.uk/ (accessed on May 16, 2021)
[23] https://es.wikipedia.org/wiki/Sala_Mar%C3%ADa_Cristina (accessed on May 16, 2021)

**Figure 2.12.** First page of the score of Transitos.

**Figure 2.13.** First page of the score of Colossus.

**Figure 2.14.** First page of the score of Ugadi.

**Figure 2.15.** First page of the score of Mutability.

At the same time, we used these musical settings to produce music in the computer cluster Iamus. During several days, it produced tens of thousands of compositions in the contemporary classical style. They were stored in a web server, labeled with the generated metadata and available to download in four different formats: MIDI, XML, PDF and MP3.

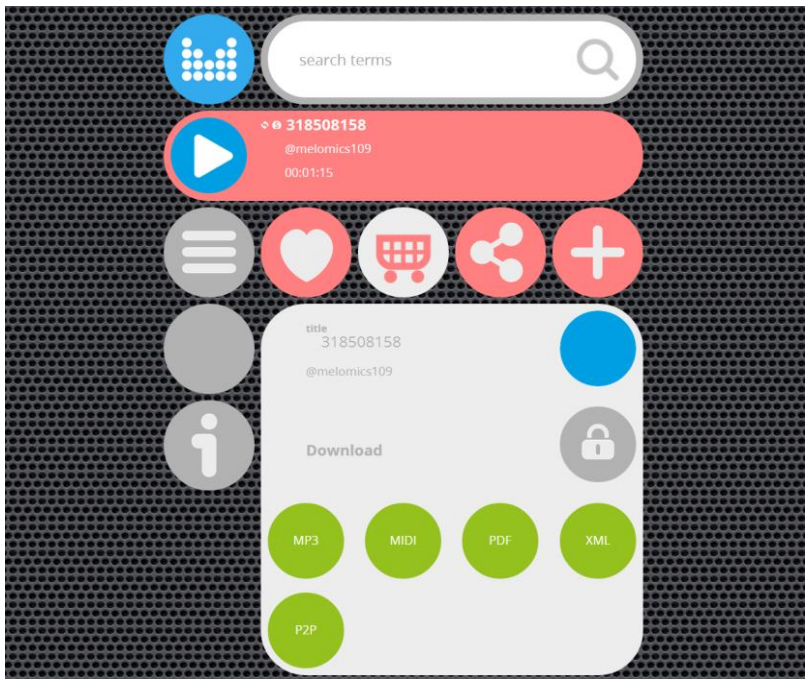

**Figure 2.16.** Melomics website. Page of a random composition.

## 2.5.2 Generating tonal music

### Genetic engineering: SBS Logo

We did a similar experiment to the one with the Nokia tune in the atonal model. Melomics was going to be presented at the event Seoul Digital Forum 2015 (SDF2015) [24] in Seoul and it was going to be broadcasted by the Korean SBS TV channel, so we decided to make use of their musical logo[25] in the experiment.

The first stage was to reverse engineer the musical piece. This time we counted on the help of an expert musician to analyze the theme and find the best way to reverse engineer it within the tonal system, managing different structural levels,

---

[24] http://www.sdf.or.kr/2015/visionaries/visionaries.jsp?lang=en (accessed on May 16, 2021)
[25] https://www.youtube.com/results?search_query=SNSD+SBS+Logo+Song (accessed on May 16, 2021)

global harmony, chord progressions, rhythms, melody and intended evolution of these parameters. We wrote an artificial genome that would be the base SBS Logo, formed by a melody, a bass and a pad.

The second stage was to generate a family of themes. We wanted to illustrate different uses:

1. Building a possible evolutionary path in our system, from a very elementary piece of music, to result into the SBS logo. We obtained this path reversely, by manually simplifying the engineered genotype in an iterative fashion. First, we deleted the melody, then, we simplified the structure and finally we removed the bass role, with one single instrument playing a simplified version of the original harmony remaining.

2. Generating a variety of themes by mutating the SBS Logo, continuing the phylogenetic evolution that had been started previously. Many random but restricted mutations were executed, including duplication of structures, altering pitches, altering progressions (root, chords), changing global parameters (tempo, dynamics, scales...), commuting roles, changing instruments, adding new kind of roles (percussions and harmonizations), adding effects, and changing the textural evolution (Figure 3.2 illustrates some of these evolutionary paths).

3. Contaminating the genome of an independent musical theme with genetic material from the SBS Logo. The idea was to preserve the identity of the original theme, while being possible to identify the logo as part of it. We produced some compositions in the defined style *Disco02* and then tried several ways of introducing low level genetic material from the SBS Logo, replacing parts of the low-level structure of the pre-generated compositions. One example is shown in the phylogenetic tree of Figure 3.2, marked as "d" and can be listened to in Sample 2.23. In this case, one of the ideas of the original theme was replaced by what can be recognized as a shortened and altered version of the SBS tune (because of having been developed into the host theme's style).

- Sample 2.20. Audio of SBS Logo reengineered
- Sample 2.21. Audio of SBS Logo relative 1

- Sample 2.22. Audio of SBS Logo relative 2
- Sample 2.23. Audio of SBS and Disco02 crossover (human synthesis)

## Chronic pain app

The chronic pain app eMTCP initiated the creation of the standardized tool to create music arrangements or hypersongs for adaptive applications. Continuing the work started with the atonal system for the application to reduce anxiety, apart from the capability to produce multiple versions of the same song, we introduced into the style-tool the functionality to generate each composition in a fragmented way, with multiple genomes each one giving place to a small part of the global composition. Further explanation and examples are given in Chapter 4.

## Web repository and 0Music album

We wanted to populate the web repository with music from the tonal system, as it was done before with the contemporary classical style. We used some of the styles that we had set up so far, including disco, minimalist, relax, modern band, world music, synth music, pop, fusion and symphonic music. About fifty thousand themes were generated and made available on the website.

- Sample 2.24. Tonal music from the web repository. Example 1
- Sample 2.25. Tonal music from the web repository. Example 2
- Sample 2.26. Tonal music from the web repository. Example 3

Generating this collection of themes served to test the ability to control harmony in compositions with many interrelated roles and to produce enough diversity of themes in these styles, despite the amount of constraints imposed. To show-case the current state of the system, a small sample considered representative of the different styles was selected. They were released as the 0music album during a one-day symposium at MIMMA, Málaga, on July 21, 2014.[26]

---

[26] http://geb.uma.es/0music (accessed on May 16, 2021)

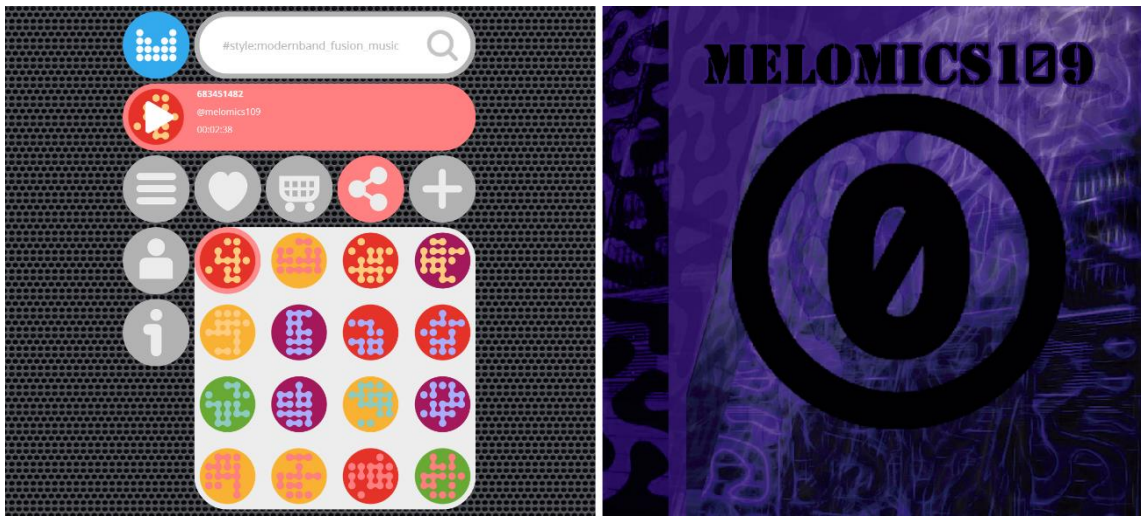

**Figure 2.17.** Search tool of the website and 0music album cover.

- Sample 2.27. 0music album

Styles DocumentaryPop and DocumentarySymphonic

In 2014 we were contacted by a company that wanted us to produce music for media productions, such as documentaries. To satisfy the specifications, we designed two styles: one to generate classical music and another to produce a more popular kind of music. This highlights the fact that the system is robust enough that there is no need to implement any new functionality in order to satisfy a random demand for a new kind of music, only to fill in the correct values within the style-tool. We generated 200 musical themes, 100 of each style, in both high-quality MP3 and uncompressed WAV, having been synthesized, mixed and mastered automatically.

- Sample 2.28. Theme from the style DocumentaryPop
- Sample 2.29. Theme of the style DocumentarySymphonic

# 2.6 Main technologies used

Among the software used during the development of this thesis, these are the most relevant:

- **MATLAB® R2010b**[27] is the language and development environment used to code most of the algorithms, including the core system. We

---

[27] www.mathworks.com/products/matlab/ (accessed on May 16, 2021)

mostly use the standard libraries, but also take advantage of the parallel and computer vision toolboxes.

- **Python 2.7**[28] was used to support the backend of the tonal system, from interpreting the resulting string of symbols (after the genome is unfolded) to produce the different output files.

- **Lilypond**[29] is a suite for score writing. We made use of its format (.ly) and software, mainly as a middle format between XML and MIDI or PDF.

- **TiMidity++**[30] was used in the atonal system to synthesize MIDI into a WAV file.

- **SoundFont® 2**[31] is the virtual instrument technology used to synthesize with TiMidity.

- **Reaper**[32] is a graphical DAW with the possibility of programing scripts to automate functions editing (in Python language). We use this tool within the latest version of the synthesis module.

- **VST™ instruments**[33] (VSTi) is a virtual instrument technology, more advanced than SF2, which we used with Reaper in the latest version of the synthesis.

- To convert WAV to other synthesized audio formats we have made use of the tools **ffmpeg**,[34] **lame**,[35] **sox**[36] and **flac**.

## 2.7 Summary and discussion

In this chapter we have exposed the foundations of Melomics technology, a tool that may be considered a noteworthy contribution to the field of algorithmic composition. We achieved a number of milestones, some of them summarized next: in 2010, a start-up proposal based on adaptive music therapy through

---

[28] https://www.python.org/ (accessed on May 16, 2021)
[29] http://lilypond.org/ (accessed on May 16, 2021)
[30] http://timidity.sourceforge.net/ (accessed on May 16, 2021)
[31] http://www.synthfont.com/sfspec24.pdf (accessed on May 16, 2021)
[32] http://www.reaper.fm/ (accessed on May 16, 2021)
[33] http://ygrabit.steinberg.de/~ygrabit/public_html/index.html (accessed on May 16, 2021)
[34] https://www.ffmpeg.org/ (accessed on May 16, 2021)
[35] http://lame.sourceforge.net/ (accessed on May 16, 2021)
[36] http://sox.sourceforge.net/ (accessed on May 16, 2021)

Melomics was awarded by the University of Málaga;[37] later, the first piece on its own style (*Opus #1*) was created; MELOMICS research project was funded by the Spanish ministry of science; in 2011, the full work *Hello World!* was premiered; an article about composing with Melomics by Professor Gustavo Diaz-Jerez was published in Leonardo Music Journal (Diaz-Jerez, 2011); in 2012, Iamus album pieces were created and after recorded by the LSO and other top-shelf musicians; a live concert premiering works from Melomics was celebrated at the ETSII of the University of Malaga, commemorating Alan Turin's birth centenary, and a great media attention was shown during these years; Melomics technology made #70 in the 100 top stories of Discover Magazine (Berger, 2013); Melomics Hackathon was celebrated at UC Berkeley and Melomics' technology was presented at Googleplex during the SciFoo[38] session *music-as-it-could-be*; Melomics music was used in therapeutic applications and experiments, and the eMTCP app was recommended by the American Chronic Pain Association;[39] in summer 2014 the second album 0music was released after a one day symposium; Melomics music was tested against human composed music at MIMMA and the music conservatory of Malaga; in the middle of 2015 Melomics was presented in the event SDF2015; and, in march 2016, four new pieces were premiered in MaerzMusik, a festival hosted in Berlin.

The designed computational methods have allowed us to produce thousands of original compositions. Any of these can be altered into consistent mutations or modified in a directed fashion to get a specific version of it. We can reuse the genetic materials to produce new one or to create contaminated versions of preexisting pieces. The proposed genetic models are at least capable of representing any kind of music that has been required, in the same way it can be done with standard music notation. At the present, the system also includes a set of tools and the musical knowledge to produce music in a wide range of styles. Apart from the music produced with specific purposes or the one made available to the general public, Melomics music is used in other contexts, such as in therapeutic systems and mobile applications. Additionally, it is intended to map the defined style-tool into a sort of online questionnaire or an API, so the community of user can interact directly with Melomics and be able to design their own new styles.

---

[37] http://www.uma.es/emprendedores/cms/menu/spin-OFF/spin-2010-2011/ (accessed on May 16, 2021)

[38] http://www.digital-science.com/events/science-foo-camp-2013/ (accessed on May 16, 2021)

[39] http://theacpa.org/ (accessed on May 16, 2021)

# Chapter 3

# Assessment of the generated music

This chapter is focused on the study of the music created by Melomics. We first show a series of experiments performing a statistical analysis of the musical features over both symbolic and audio formats. Secondly, we present a study from a more subjective perspective, facing Melomics music against human made compositions.

The chapter is organized as follows: In section 3.1 we introduce the concept of Music Information Retrieval (MIR) and describe some MIR tools and methods; then we present some studies performed over the music coming from the system. In section 3.2 we describe an experiment with Melomics music in the real world, comparing with music made in the same style by a human composer. Then, in section 3.3 we comment some additional assessments and finally we expose some conclusions about all these studies in section 3.4.

## 3.1 Numerical study

We have run two types of tests: On the one hand, we wanted to know how the system really managed the relationship between codified and resulting music, studying how changes in the genotypes would have an impact into their corresponding developed forms. Especially we wanted to check whether small mutations in the genome would produce small changes in the resulting theme, as expected; and also, whether genotypes from a certain input configuration (expected style) actually produced music that could be classified into the same musical style. On the other hand, by using the same analytical tools, we studied where Melomics music can be placed in a space of musical parameters, in relation to the preexisting human made music and to other kinds of sounds.

## 3.1.1 Music Information Retrieval

MIR is an interdisciplinary field of knowledge dedicated to obtaining information from different sources of musical data. The most obvious source is the audio recording, from which the information (musical style or genre, participating instruments or voices, arousal and valence, rhythms performed, etc.) is gathered after analyzing the wave signal. Nevertheless, the data sources subject to MIR analysis comprises the whole variety of formats, including symbolic standards and even cultural information related to musical pieces that can be provided by people in websites or social networks (Lamere, 2008) (Schedl, Gómez, & Urbano, 2014). Some of the most common applications of MIR are: recommendation systems; music transcription systems, to reverse audio recordings into scores; theme recognition tools, to identify songs or related versions, like the Shazam system (Wang, 2003); and categorization tools, that most music providers in the present use to arrange songs according to their musical properties. And there is a high interest in the matter in research, counting on the fairly well-known ISMIR,[40] an international organization coordinating conferences and committed to MIR.

Some relevant utilities to study music found in literature are Marsyas (Tzanetakis & Cook, 2000), an open-source software written in C++, containing a large set of functions to analyze audio signals and designed to be highly accessible and interoperable; MIDI Toolbox (Eerola & Toiviainen, 2004), published under GNUGPL and implemented in Matlab, with different functionalities to analyze and visualize music in symbolic format or Humdrum Toolkit (Huron, 2002), which comprises a large collection of command-line tools based on music information retrieval and provided under a free license. Below we describe two MIR frameworks that we have used to study Melomics music:

---

[40] http://www.ismir.net/ (accessed on May 16, 2021)

- jMIR (McKay C. , 2010). It is an open-source application for MIR implemented in Java. Its architecture is highly compartmented and its modules communicate with each other through the ARFF format (Witten & Frank, 2005) used in WEKA[41] or through the built-in ACE XML. The system's core is divided into two blocks: (1) components to obtain features, including jAudio, to analyze audio waves (compactness, peak detection, moments, strongest frequency...); jSymbolic, to study the symbolic MIDI format (instrumentation, texture, rhythm, dynamics, pitches, chords...); jWebMiner, to extract cultural information from the web (author, genres or styles, similarities between songs or artists...) and jLyrics, for mining information from lyrics (number and frequencies of words, syllables and letters, number of lines, punctuation, misspelling, vocabulary richness...). (2) components to process the obtained information, in particular jMIRUtilities, to combine the extracted features, labelling and performing other miscellaneous tasks, and ACE, to execute a variety of classifications depending on each particular problem.

- The Echo Nest. It was a company acquired by Spotify in 2014, owning one of the world's greatest platforms to provide developers and media companies with music information. They used MIR to help discover new music, provide updated and extensive information about musical content and give support for fingerprinting of musical content among others. To developers, they gave access to external sandboxes, making possible to build applications by exploiting these services and they also provided an API and libraries in different programming languages, to interact with the information stored in their database. The web-based API contained a collection of methods responding in JSON or XML format. These methods are grouped into (1) artist methods, providing biographies, related images, genres, terms, songs, reviews, hotness and others; (2) genre methods, with artists, profiles, similarities, etc.; (3) song methods, to access the gathered information and the analyzed musical features, such as loudness, hotness, danceability, energy, mood and countless others; (4) track methods, for analyzing or getting information about the tracks of a piece of music; and (5) playlisting methods, to allow managing playlists and user profiles.

---

[41] http://www.cs.waikato.ac.nz/ml/weka/ (accessed on May 16, 2021)

## 3.1.2 Measurement of similarity

To deal with music in a symbolic way, we used jMIR[42], in particular jSymbolic (version 1.2), which is able to extract information from MIDI files. We set up the tool to get features classified into the categories of instrumentation, texture, rhythm, dynamics, pitch statistics, melody and chords adding up to 111 in total; including both single-value and multi-dimensional features. Most of MIR tools work on audio waves (like jAudio, from the same toolbox), partly because it provides with additional information about the actual piece under analysis (performance attributes, etc.), but mostly because music is usually found in this type of formats. However, analysis using symbolic notation is more adequate to get information of high level of abstraction, more meaningful in compositional terms.

Among the compositional characteristics that jSymbolic can extract, we have selected all the single-value features and some of the multi-dimensional: "Fifths Pitch Histogram" and "Pitch Class Histogram", which provide some information about tonality and harmony; and "Pitched Instruments Present", indicating which of the MIDI instruments are used. Since many of the features are not normalized, we developed a simple method to assess distances, based on first obtaining the centroid of a set of themes and then giving the measures relative to that point.

### 3.1.2.1 Compositional analysis to assess variations produced by mutation and crossover operations

A first study was made using Melomics subsystem for atonal music. We took the bundle of compositions that resulted from the experiment Genetic engineering: Nokia tune, with the aim of checking whether a small set of changes in the genome, yields a similar composition with respect to the original one and how the compositional distance increases as the set of mutations and crossovers becomes greater too. To assess how the resulting phenotypes differ from the original theme, we compared each theme with the bundle's centroid and with the original tune. In particular, for a given feature, if the current theme being evaluated is further from the Nokia tune than it is from the centroid, that would add one unit to the measure of distance for this theme (see the equation below). This was done in order to equalize the importance of all the considered features, without having to appeal to a more complex method or dig further into the meaning of the musical features.

---

[42] http://jmir.sourceforge.net/ (accessed on May 16, 2021)

$$d_{current} = \sum_{i=1}^{n} abs(Reference_i - Current_i) > abs(Centroid_i - Current_i)$$

where

$d_{current}$ is the computed distance from the current theme to the theme of reference (the Nokia tune),

$n$ is the number of considered features,

$Reference_i$ is the value for feature $i$ of the theme of reference,

$Current_i$ is the value for feature $i$ of the theme being evaluated,

$Centroid_i$ is the value for feature $i$ of the considered bundle's centroid,

$abs$ is the function absolute value.

The analysis was performed over a bundle of about 1000 pieces and the results (see Figure 3.1) showed that themes produced by slights variations of the Nokia tune, for example, mutating the gene responsible for choosing the instrument or establishing the tempo, were measured very near to the original piece. In the middle of this aggregated dimension, we find themes that apart from changing the instrument, they have a few appearances of the genetic operator that split notes or altered pitches, for example theme55 or theme209 respectively. A little farther we find themes with many operators to alter the duration, the pitch and even introducing an additional instrument in polyphony, like theme625. At the opposite side of the space there are themes whose genome have suffered so many mutations, that the resulting pieces have almost nothing in common with the Nokia tune, like the theme249, as the jSymbolic analysis prove and as we can clearly perceive.

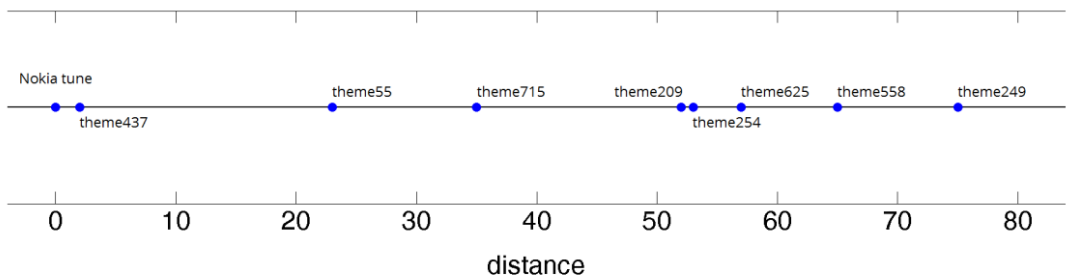

**Figure 3.1.** Nokia tune experiment. Illustration of the musical distance.

▪ Sample 3.1. Examples of the Nokia tune experiment (plain synthesis)

With the tonal system we made a similar experiment, this time over the phylogenetic tree described in the section Genetic engineering: SBS Logo and using the same method to assess the distances. The original SBS Logo (node 4 in Figure 3.2) was taken as the reference; then we wanted to check whether variations score higher, the farthest they were from this center. Figure 3.2 shows that the themes close to node 4 got low values and, on every branch, the more steps further you go, the higher the scores are. There is only one exception, in the branch representing the crossover of a SBS variation with a theme produced by Melomics. Here the scores are distorted by the fact that these three themes (nodes d, 22 and 18) last a few minutes, instead of a few seconds, like the rest of the pieces.

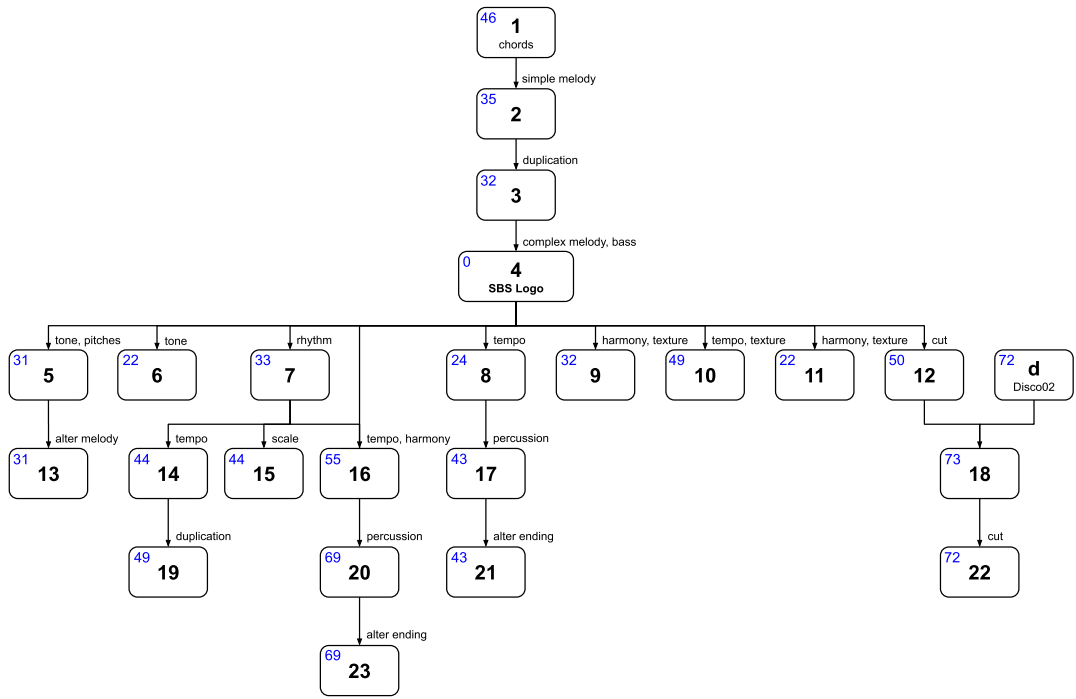

**Figure 3.2.** Phylogenetic tree of the SBS Logo. The main changes of a theme with respect to their parents are indicated to the right of the incoming arrow. The scores given by the analysis are placed inside each node, in blue.

- ▪ Sample 3.2. Audio of the SBS Logo experiment (plain synthesis)

## 3.1.2.2 Compositional analysis of the produced musical styles

Another test was aimed to check how themes produced with the same input configuration (targeted style) were similar to each other and in relation to compositions made with other configurations and to themes created by humans. We computed the centroid of each style, as before, and then the distances for all the considered themes, applying the formula:

$$d_{current} = \sum_{i=1}^{n} DIndex_i$$

where

$d_{current}$ is the computed distance from the current theme to the centroid of the considered bundle,

$n$ is the number of considered features,

$DIndex_i$ is, in a list sorted in ascending with the distances for feature $i$ (measured from the centroid to all the considered themes), the index of the value for the current theme. As in the previous case, this was done as a simple method to equalize the importance of all the considered features.

For the atonal system, we picked a collection of 656 pieces of classical contemporary music with different ensembles, then computed the average value for each feature, obtaining the centroid. After that, we added to the bundle 10 more pieces from the same contemporary classical substyle; 10 from the style *Disco02*, produced by the tonal system; and 25 pieces from the Bodhidharma dataset (McKay & Fujinaga, The Bodhidharma system and the results of the MIREX 2005 symbolic genre classification contest, 2005) tagged as "Modern Classical" and created by human composers; representing 701 pieces in total.

Figure 3.3 shows a representation of the distances of each theme to the computed centroid. The closest composition is located at 13788 units and the farthest at 26902. There are a few themes of the initial set close to the centroid, some others notably far from it and most of them are located at a medium distance, between 16000u and 21000u. The extra generated pieces of contemporary music appear scattered around the centroid of the initial group, with a maximum distance of 20285 (mean and standard deviation for the initial and the new group: $\mu = 1842.54$, $\sigma = 1838.3$ and $\mu = 1819.7$, $\sigma = 1603.2$ respectively); the human made compositions in similar styles appear close but shifted to farther distances ($\mu = 2101.88$, $\sigma = 2158.76$); while music created with different directions appear at the farthest distance to the centroid ($\mu = 2363.76$, $\sigma = 1755.02$).

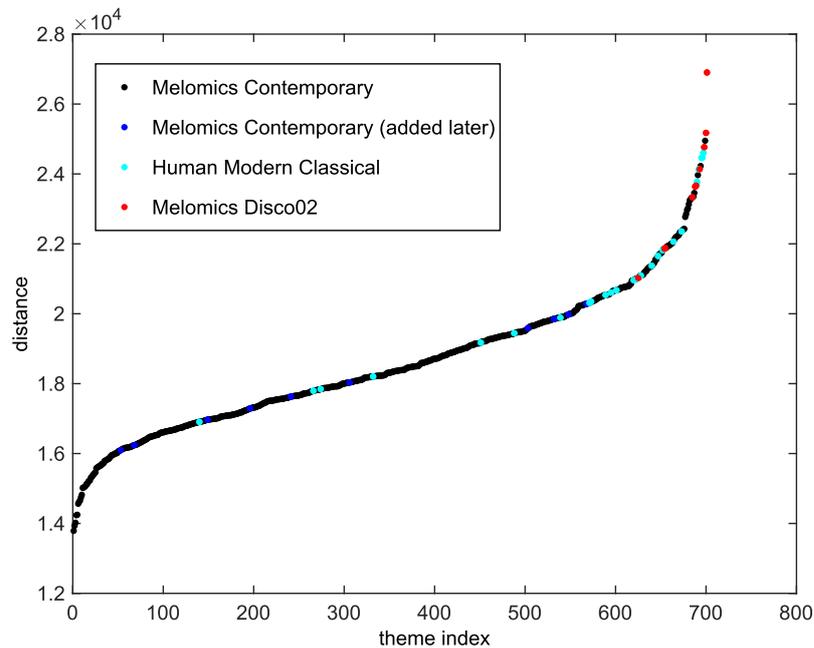

**Figure 3.3.** Chart of distances to the centroid of Melomics Contemporary style. The set studied contains pieces of different styles configured with Melomics and a collection of themes composed by human artists from the Bodhidharma dataset, tagged as "Modern Classical".

For the tonal system we performed a similar test. We chose 220 themes from Melomics' *DancePop*, excluding 14 themes to compute the centroid. Then we added: those 14 pieces; 10 more from the style Melomics' *Disco02*, which we consider the predecessor of *DancePop;* and 10 more from Melomics atonal's *DocumentarySymphonic*.

Figure 3.4 shows a representation of the distances for this study. The pieces of the same style that were added later are close to the centroid of the group of reference (mean and standard deviation for the initial and the new group respectively: $\mu = 8273.57$ , $\sigma = 1071.6$ and $\mu = 8091.64$ , $\sigma = 977.7$ ); the compositions of a similar style appear close but shifted farther ($\mu = 8879.0$, $\sigma = 1065.41$) and the music from a more different style are located at the farthest distance ($\mu = 10{,}141.4$, $\sigma = 575.53$).

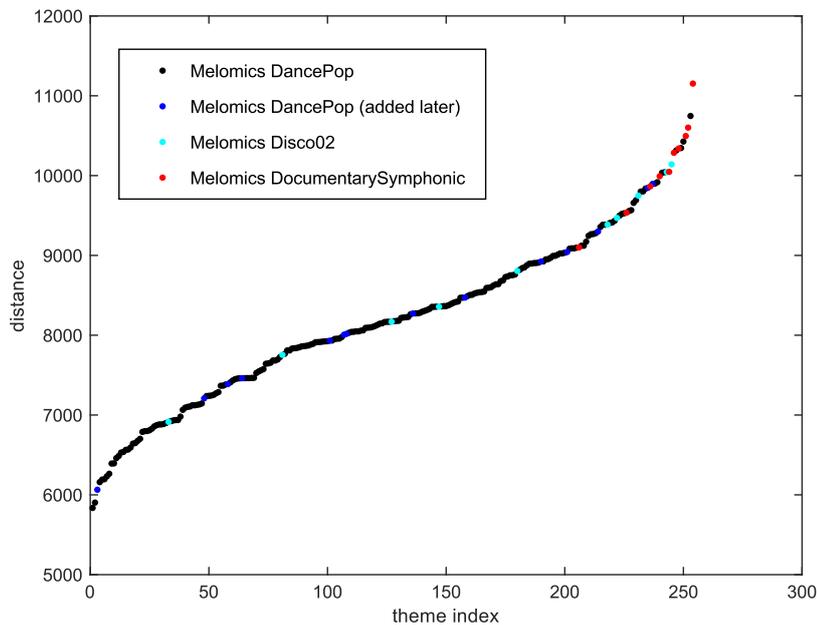

**Figure 3.4.** Chart of distances to the centroid of Melomics DancePop style. The set studied contains pieces of different styles configured with Melomics atonal and tonal systems.

## 3.1.3 Melomics music among other kind of sounds

In order to put in context different musical outputs coming from Melomics, we made use of The Echo Nest platform. We installed Pyechonest,[43] one of their official libraries to access the API, and we used some of their Track API Methods to upload and analyze some audio samples. Besides the expected low-level data, such as tempo, loudness or key; the analyzer returns other values with a more abstract meaning, what they called *acoustic attributes*. These aggregated values, each one computed using the previously mentioned low-level parameters, range between 0.0 and 1.0, and these are: *danceability*, to describe how the audio is suitable for dancing; *energy*, a perceptual measure of how fast, loud or noisy is the track; *speechiness*, indicating the presence of spoken words; *liveness*, meaning the identification of an audience in the recording; *acousticness*, indicating if the audio was created with acoustic instruments as opposed to more artificial means; and *valence*, describing the degree of positivity or negativity conveyed by a piece. With the goal in mind of establishing the basis for a further study, a first exercise was to compare against these variables different types of music made by humans, music by Melomics and other kind of sounds (natural sounds and noises). The samples analyzed and the abbreviations used in the figures are listed in the table below.

---

[43] https://github.com/echonest/pyechonest (accessed on May 16, 2021)

| Abbreviation | Sample name | Composer type |
| --- | --- | --- |
| WN | White Noise 30 seconds | Artificially generated |
| Tur | 09 Turkey Talk | Natural sound |
| Rap | 17 Small Rapid | Natural sound |
| Atl | Joan Albert Amargós - Atlantic Trio I mov | Human |
| Exe | Gustavo Diaz Jerez – Exedrae | Human |
| Op1 | Iamus - Opus #1 | Melomics |
| HW! | Iamus - Hello World! | Melomics |
| Req | Mozart - Requiem | Human |
| Smo | Michael Jackson - Smooth Criminal | Human |
| Sat | Rolling Stones - Satisfaction | Human |
| 0m6 | Melomics - 0music 06 | Melomics |
| 0m10 | Melomics - 0music 10 | Melomics |

**Table 3.1.** Samples analyzed and abbreviations used in the figures.

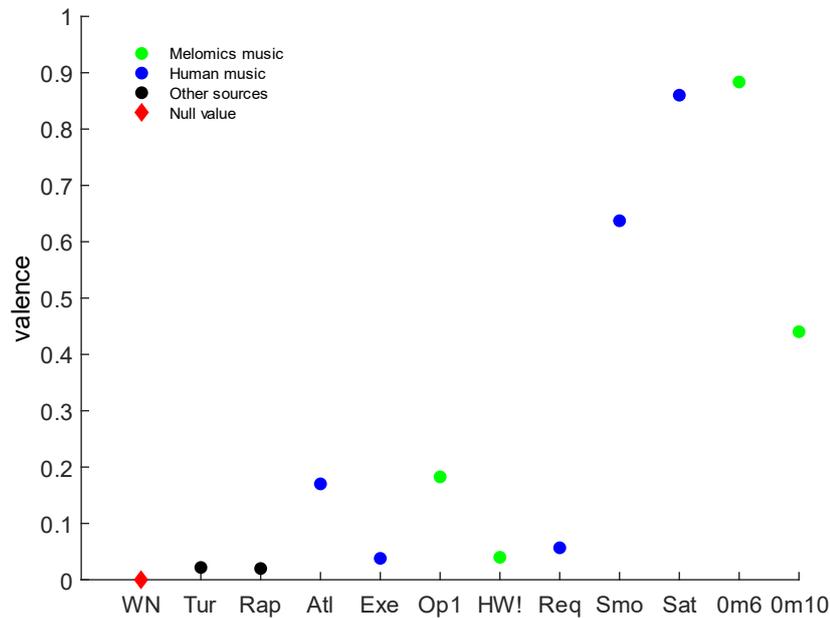

**Figure 3.5.** Attribute *valence* assessed in the 15 considered samples.

The aggregate value *valence* seems to provide some useful information to our purpose. It probably makes use of the detected harmony to assess this emotional feature. Figure 3.5 shows that the system was not even able to compute this value for the sample with white noise. Nature sounds ("Tur" and "Rap") have a notably low score, all below 0.022. Above these values, starting from 0.037, there are some of the contemporary classical compositions and then, at a higher range, the rest of the pieces.

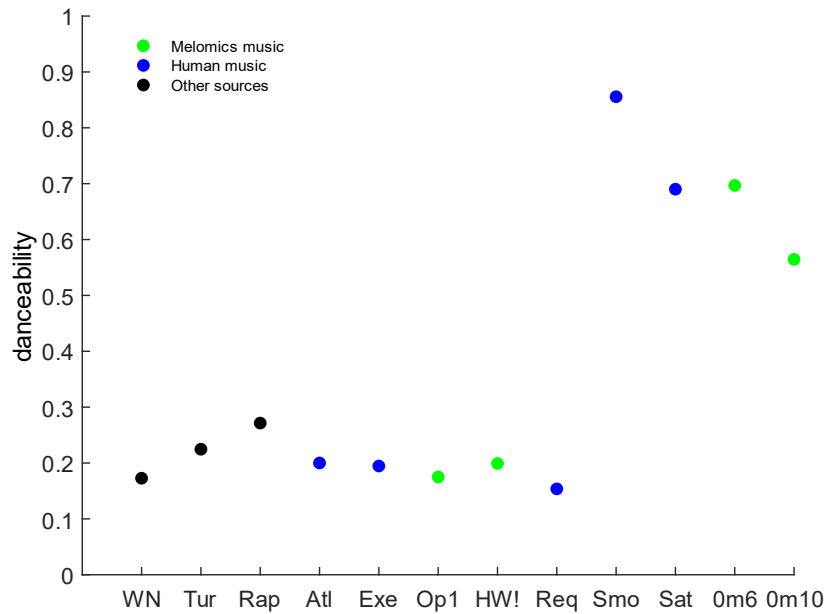

**Figure 3.6.** Attribute *danceability* assessed in the 15 considered samples.

*Danceability* combines tempo, rhythm, beat strength and data about regularity. Figure 3.6 shows how this attribute puts classical music, especially the contemporary or atonal style, below the 0.3 bound, both for human and Melomics music. However, this parameter is not very useful to distinguish between this kind of music and non-musical samples, with the white noise being equally "danceable" as Mozart's Requiem, for example, and Small Rapid natural sound more danceable than most of the classical compositions that were analyzed. Above the 0.3 threshold there is the more "danceable" music.

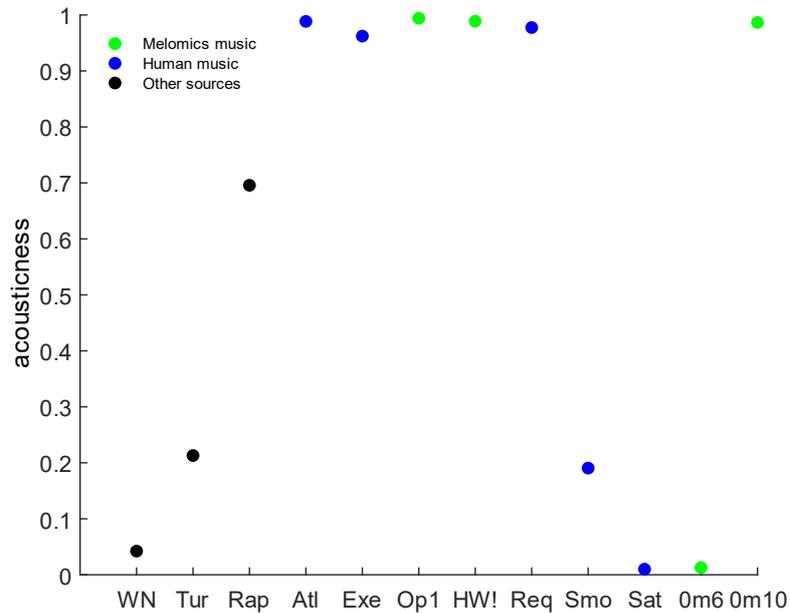

**Figure 3.7.** Attribute *acousticness* assessed in the 15 considered samples.

Another interesting indicator is *acousticness*. It separates very accurately the music played by orchestral instruments from those themes made with more artificial instruments. Above 0.8 we can see the first group, no matter if they were created by Melomics or a human composer, or even if the piece was synthesized by Melomics with acoustic virtual instruments or recorded in live concert, they are all in the same group. Then, below 0.1 the genres using electric and electronic instruments can be found (rock, hard rock or electronic music). Between the 0.1 and 0.8 thresholds there is the music whose assessment of *acousticness* is unclear. It can be because they include a combination of different types of instruments, because they do not include artificial instruments, but they do not use orchestral ones either or because they include lyrics with much prominence. Natural sounds appear in the middle category and white noise is placed into the category of artificial sounds.

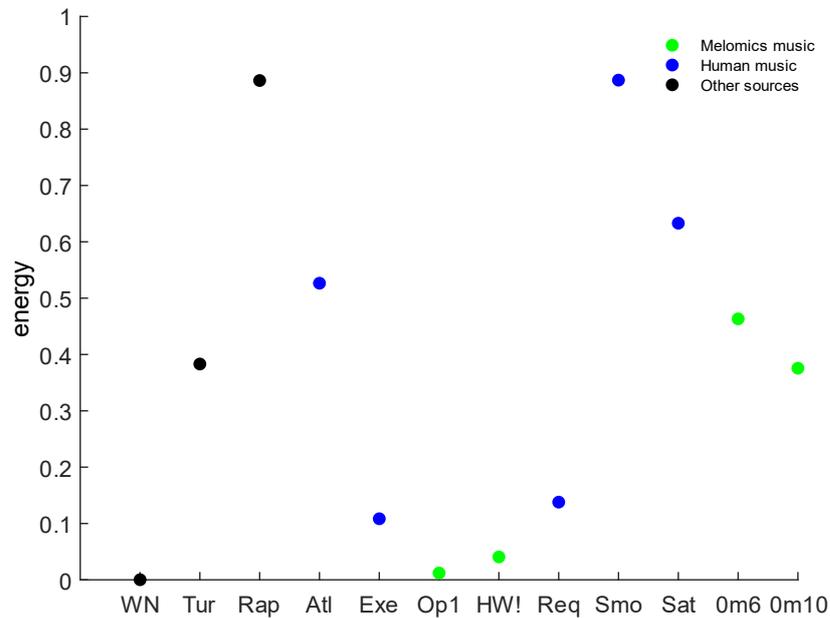

**Figure 3.8.** Attribute *energy* assessed in the 15 considered samples.

The attribute *energy* is computed taking into account the detected rhythm, timbres, loudness and general entropy, which is why the white noise sample is scored with almost 0, the music perceived as calmed has a low value (mostly classical music in this analysis) and the popular genres are located above (see Figure 3.8).

The study of the other two attributes given by The Echo Nest do not provide much information. *Speechiness* resulted to be low in all compositions from Melomics, as expected, since they had no lyrics, as well as the instrumental pieces composed by human musicians. *Liveness* was also low for all Melomics music, even for *Hello World!*, which, although being an audio from an actual concert, was obtained by using professional means, with noise cancellation, etc., the same way it is done in a recording studio.

## 3.2 Perceptive study

This work is very concerned with how "human" the production can be considered. For this reason, we proposed and developed an additional study consisting of the assessment of Melomics music by human subjects in a blind test.

While measuring the performance of an assembly robot in a production line is straightforward, assessing the quality of other AI systems, particularly those involving simulated cognition or creativity, is always controversial. To date, the Turing Test (TT) (Turing, 1950) is still considered the best way to determine if an artificial system involves some sort of thought. The validity of this approach stands, even if the TT was epistemic rather than ontic or more heuristic than demonstrative (Harnad, 2000). That is the reason why a wide range of variations have been proposed to measure the success of AI systems in related areas. Some of them consider the term "creative" in a very restrictive sense, like the Lovelace Test (LT) (Bringsjord, Bello, & Ferrucci, 2001) that demands full creativeness on the machine under study, which is hard to meet (and verify) for an artificial system. Some interesting and more practical variants are the Musical Directive Toy Test (MDtT) and the Musical Output Toy Test (MOtT) (Ariza, The interrogator as critic: The turing test and the evaluation of generative music systems, 2009), with the second one focusing more on the results of the composition and paying less attention to how the composers were commissioned. Both the MDtT and MOtT remove the essential component of natural language discourse, while keeping the blind indistinguishability of the TT. As such, they must be considered surveys of musical judgments, not a measure of thought or intelligence.

Another important aspect of generative music systems is their lack of experimental methodology (Papadopoulos & Wiggins, 1999) and the fact that typically there is no evaluation of the output by real experts. Delgado (Delgado, Fajardo, & Molina-Solana, 2009) considered a compositional system based on expert systems with information about emotions as input and described an interesting study where the extent to which the listeners match the original emotions is evaluated, but the test was only completed by 20 participants. Monteith (Monteith, Martinez, & Ventura, 2010) presented another system and tested it in an experiment where the participants had to identify the emotions evoked for each sample, as well as assess how real and unique they sounded as musical pieces. There were two major limitations in this case: the sample contained only 13 individuals and they knew in advance that there were computer-generated pieces. A more recent example was described by Roig (Roig, Tardón, Barbancho, & Barbancho, 2014), with a questionnaire including three sets of compositions, each with two compositions made by human and the other generated by their machine learning system, which the participants had to identify. In that experiment, there was a bigger sample (88 subjects), however it suffered from lack of control, since the test was performed through the web, and simplicity, because the participants only had to answer which of the three samples was computer generated.

Affective algorithmic composition (Williams, et al., 2015) has promoted new strategies of evaluating the results of automatic composition systems, as compared to human produced music. Here, emotions are understood as relatively short and intense responses to potentially important events that change quickly in the external and internal ambience (generally of social nature). This implies a series of effects (cognitive changes, subjective feelings, expressive behavior or action tendencies) occurring in a more or less synchronized way (Juslin & Sloboda, 2011). This theory of musical emotion (Meyer L. B., 1956) asserts that music can elicit emotions from two different involuntary mechanisms of perception and interpretation of the human being: (a) the (conscious or unconscious) assessment of musical elements; and (b) the activation of mental representations, related to memories with significant emotional component (Koelsch, Brain correlates of music-evoked emotions, 2014) (Koelsch, Music-evoked emotions: Principles, brain correlates, and implications for therapy, 2015). Even though self-report has been associated with some obstacles, like a quantitative demand of lexicon or the presence of biases, it is considered one of the most effective ways of identifying and describing the elicited emotions (Juslin & Sloboda, 2011). This work has opted for an open-answer modality of the cited method, since the fact that establishing explicit emotional dimensions might itself condition their appearance due to the human empathy (Juslin & Sloboda, 2011).

The aim, through an MDtT design, is quantifying to what extent are Melomics and human music perceived as the same entity and how similarly do they affect the emotions of a human listener, considering professional musicians and non-musicians. The hypothesis is that a computer-made piece is indeed equivalent to conventional music in terms of the reactions elicited in a suitable audience.

As compared to previous works, the main contributions here are the rigor of the study and the significance of the results.

In an effort to assess the hypothesis, an experiment registered the mental representations and emotional states elicited in an audience while listening to samples of computer-generated music, human-generated music and environmental sounds. It involved 251 participants (half of them professional musicians) who reported on the mental representations and emotions that were evoked while listening to the various samples. The subjects were also asked to differentiate the piece composed by computer from the one created by a human composer.

## 3.2.1 Methodology

Two musical pieces were commissioned in a specific style to both a human composer and the Melomics tonal system. In the first stage a comparison between both pieces was made using an approach that focused on human perception: listeners were asked what mental images and emotions were evoked on hearing each sample. In this phase of the experiment the listeners were not aware that some of the samples had been composed by a computer. In an effort to gauge whether or not both works were perceived in a similar way, the nature of the composer was not revealed until the second stage.

Data analysis has been performed with a fourfold contingency table (two by two) and the differences between groups were evaluated by the chi-squared test with continuity correction. The Fisher test has been used only in those cases where the expected frequency was lower than 5. The significance level has been established at $p < 0.05$. In the second phase of the study, the sensitivity (i.e., the capacity to correctly identify the pieces composed by the computer) and the specificity (i.e., correct classification of pieces composed by humans) were also evaluated for both musicians and non-musicians, at a confidence level of 95%.

## 3.2.2 The audio samples

The specifications for the musical pieces were kept simple and the duration short, in order to ease their assessment, especially by non-musician participants:

style: guitar ballad
instruments: piano, electric piano, bass guitar and electric guitar
bar: 4/4
BPM: 90
duration: 120s
structure: A B A'
scale: major

Guitar ballad was chosen also because its compositional rules were already coded within the system. The final audio files were obtained in MP3 format with a constant bitrate of 128 kbps and shortened to a duration of 1 m 42 s, meaning the beginnings and endings of the pieces (around 9 s each) were removed since they typically render constant composition patterns in popular music.

Following composition, both the human- and the computer-composed ballads were doubly instantiated by means of, on the one hand, performance by a professional player and, on the other hand, through computerized reproduction. For the purposes of the performance the human player interpreted the scores of both works while the computer automatically synthesized and mixed a MIDI representation of both pieces using a similar configuration of virtual instruments and effects, the four combinations resulting in corresponding music samples. Table 3.2 shows how the pieces were labelled: HH stands for human-composed and human performance, CC for computer-composed and synthesized, HC for human-composed and computer-synthesized, CH for computer-composed and human performance, and NS for natural sounds. In contrast to the musical samples, this final sample has been introduced in an effort to gauge the listeners' response to non-musical sounds and consists of a two-minute excerpt from natural sounds (Jungle River, Jungle birdsong and Showers from The Sounds of Nature Collection (The sounds of Nature Collection (Jungle River, Jungle birdsong and Showers)) combined with animal sounds (Animals/Insects, n.d.). Sample 3.3 contains the five audio samples used in the study.

|  |  | composed | | |
|---|---|---|---|---|
|  |  | **human** | **computer** | **nature** |
|  | **human** | HH | CH | - |
| **interpreted** | **computer** | HC | CC | - |
|  | **nature** | - | - | NS |

**Table 3.2.** Classification of audio samples according to the composer and the interpreter.

- Sample 3.3. Bundle of themes used in the perceptive study

## 3.2.3 Participants

The experiment was carried out in two facilities in Malaga (Spain): the Museum for Interactive Music and the Music Conservatory. Subjects were recruited via posters in the museum facilities and internal calls among students and educators at the conservatory. Selected participants ranged in age from 20 to 60 and the answers were processed differently according to music expertise: subjects with five or more years training were labelled as "musicians" while subjects with less or no training were classified as "non-musicians". Musicians are assumed to process music in a far more elaborate manner and possess a wider knowledge of musical structure, so they would be expected to outperform non-musicians in a musical classification task.

The final sample consisted of 251 subjects, the mean age being 30.24 ($SD = 10.7$) years, with more musicians ($n = 149$) than non-musicians $n = 102$). By gender the sample featured marginally more women ($n = 127$) than men ($n = 124$) and, in terms of nationality, the majority of the participants were Spanish $n = 204$), the remainder being from other, mainly European countries ($n = 47$).

## 3.2.4 Materials

In both facilities the equipment used to perform the test was assembled in a quiet, isolated room in order to prevent the participants from being disturbed. Each of the three stations used for the experiment consisted of a chair, a table and a tablet (iPad 2). The tablet contained both the web application that provided the audio samples and the questionnaire that was to be completed by means of the touch screen. The text content (presentation, informed consent, questions and possible answers) was presented primarily in Spanish with English translations located below each paragraph. The device was configured with a WIFI connection to store the data in a cloud document and featured headphones (Sennheiser EH-150) both for listening to the recordings and to avoid external interferences. The iPad device was configured to perform only this task and users were redirected to the home screen when all the answers had been saved. The possibility of exiting the app was also disabled. The test is publicly available at http://www.geb.uma.es/mimma/ (accessed on April 30, 2021).

### 3.2.5 Procedure

Supported by neuroscientific studies (Steinbeis & Koelsch, 2009), critiques to computer-made compositions are suitable to be affected by anti-computer prejudice, if knowing in advance the non-human nature of the author. Hence, during the test, each subject was informed that they were undergoing an experiment in music psychology, but the fact that it involved computer-composed music was not mentioned at the beginning so as not to bias the results (Moffat & Kelly, 2006). Participants were also randomly assigned to one of two different groups and the compositions were distributed between these two groups in such a way that each subject listened to the musical pieces as rendered by the same interpreter (Table 3.3). In this way, the responses were independent of the quality of the execution since each subject would listen to both the human and computer compositions interpreted either by the artist or by the computer, meaning that potential differences in composition and performance became irrelevant.

|  | Group A | Group B |
|---|---|---|
| Phase I | HH / CH / NS | CC / HC / NS |
| Phase II | HH / CH | CC / HC |

**Table 3.3.** Distribution of musical pieces into groups according to the interpreter.

The workflow of the test detailed in Figure 3.9 shows that the subject is first introduced to the presentation and the informed consent screens and then prompted for personal data relevant to the experiment (see specific questions in Table 3.4). The subject is then assigned to either group A or B, listens to five sequential audio recordings and answers a number of questions. The listening and answering process is divided into two phases:

- During Phase I each subject in group A listens to the three pieces in random order. HH and CH have been composed by a human and by our computer system, respectively, and both are performed by a human musician. Subjects assigned to group B proceed similarly, listening to the same compositions, but in this case synthesized by computer. Both groups also listened to the same natural sounds recording (NS). Having listened to each piece the subject is then asked whether the excerpt could be considered music and what mental representation or emotional states it has evoked in them. This final question requires an open answer as the subject was not given a list of specific responses.

- In Phase II, the subject listens to the same musical pieces (but not the natural sounds), following which they are asked whether they think the piece was composed by a human or by a computer. As previously stated, it is important that identification of the composer is withheld until the second phase so that subjects can provide their assessment of the music in Phase I without the potential bias of Phase II, in which the subject becomes aware that the music might have been composed by a computer.

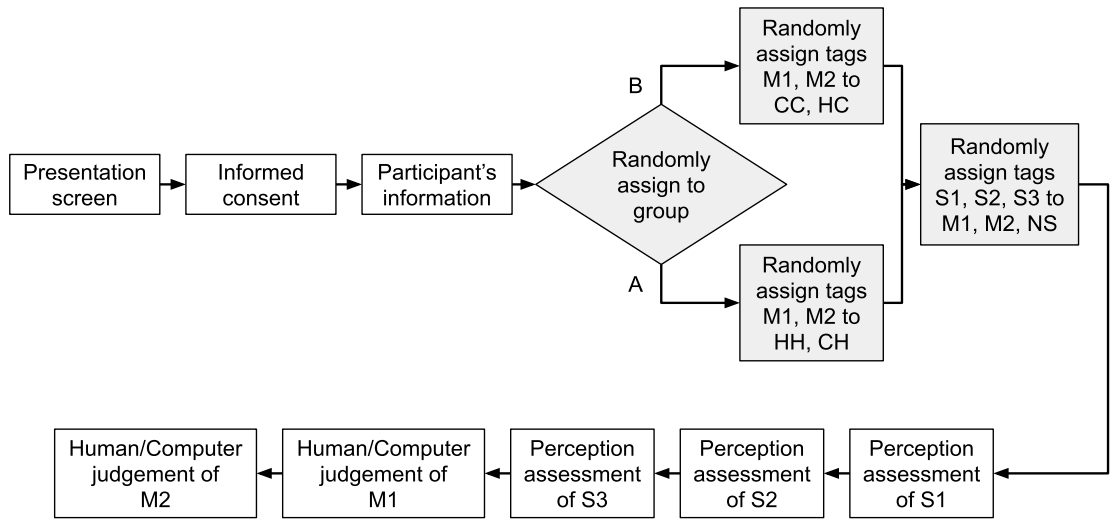

**Figure 3.9.** Test workflow. After the initial pages, the participants are randomly assigned to one group (A or B). Then, the samples are played and evaluated in a random order. Finally, the human or artificial nature of the musical pieces is assessed.

| Stage | Question | Input Type |
|---|---|---|
| Participant's information | Age | Numeric |
| | Sex | Option button |
| | Country | Text box |
| | Email address (optional) | Text box |
| | Do you have at least 5 years of academic musical training? | Option button |
| Perception assessments | Q1. Would you say that what you have listened to is music? | Option button |
| | Q2. Does it generate mental representations when you listen to it? | Option button |
| | Q3. If so, could you write three mental representations that have come to mind listening to it? | Text box |
| | Q4. Does it generate any feelings? | Option button |
| | Q5. If so, could you write three feelings that you have felt while listening to it? | Text box |
| Human/Computer judgements | Q6. Do you think that the piece you have listened to has been composed by a composer or by a computer? | Option button |

**Table 3.4.** Information requested to participants and questions to interrogate the personal opinion and feelings elicited by the recordings.

In summarizing the process, a subject assigned to group A might proceed as follows:

1. Presentation screen
2. Informed consent
3. Participant's information
   **Perception assessment of S1**
4. Listen to HH (labelled as M1, and then S1)
5. Q1-Q5
   **Perception assessment of S2**
6. Listen to NS (labelled as S2)
7. Q1-Q5
   **Perception assessment of S3**
8. Listen to CH (labelled as M2, then S3)
9. Q1-Q5
   **Human/computer judgement of M1**
10. Listen to HH
11. Q6
    **Human/computer judgement of M2**
12. Listen to CH
13. Q6

## 3.2.6 Experiment results

Table 3.5 shows the results in percentage of the *yes/no* questions provided in the first (blind) phase of the experiment. All the questions asked to the subject referred to the piece that had just been listened to: in random order, HH, CH and NS for subjects of group A, and CC, HC and NS for group B. None of the contrasts performed, for the total sample or for each type of listener, was found significant.

| | Sample | | Musicians | | Non-musicians | |
|---|---|---|---|---|---|---|
| | Group A ($n$ = 134) | Group B ($n$ = 117) | Group A ($n$ = 89) | Group B ($n$ = 60) | Group A ($n$ = 45) | Group B ($n$ = 57) |
| | HH | HC | HH | HC | HH | HC |
| Q1 | 98.5 | 96.6 | 98.9 | 100.0 | 97.8 | 93.0 |
| Q2 | 70.1 | 62.4 | 68.5 | 60.0 | 73.3 | 64.9 |
| Q4 | 82.1 | 82.1 | 84.3 | 81.7 | 77.8 | 82.5 |
| | CH | CC | CH | CC | CH | CC |
| Q1 | 97.8 | 99.1 | 97.8 | 100.0 | 97.8 | 98.2 |
| Q2 | 63.4 | 55.6 | 66.3 | 51.7 | 57.8 | 59.6 |
| Q4 | 87.3 | 80.3 | 89.9 | 76.7 | 82.2 | 84.5 |
| | NS | NS | NS | NS | NS | NS |
| Q1 | 36.6 | 38.5 | 41.6 | 41.7 | 26.7 | 35.1 |
| Q2 | 91.8 | 95.7 | 94.4 | 95.0 | 86.7 | 96.5 |
| Q4 | 83.6 | 84.6 | 84.3 | 81.7 | 82.2 | 87.7 |

**Table 3.5.** Results of the *yes/no* questions proposed in Phase I of the experiment. The percentages of `"yes"' are shown. Q1: Would you say that what you have listened to is music? Q2: Does it generate mental representations when you listen to it? Q3: Does it generate any feelings?

An affirmative answer was almost unanimously given to the first question (Q1) by the subjects after listening to a sample of a musical piece (as opposed to NS), independently of who composed or interpreted it. By contrast, the NS sample was classified as non-music, although the results were narrower. Regarding the second question (Q2), all the musical pieces elicited mental images, with global percentages ranging from 55.6% to 70.1% and without significant differences observed in terms of who composed or interpreted the pieces ($p > 0.05$). Meanwhile, natural sounds elicited mental images in more than 90% of the cases, more than with any of the musical samples. Regarding the responses elicited by the recordings (Q4), subjects answered affirmatively (up to 80%) to both musical pieces -regardless of who was the composer or the interpreter- and the natural sounds. Only in the case of the music sample composed by computer and played by human musicians did the global percentage approach 90%. Most of the answers (179) were given in Spanish. Among the responses provided in English (23), three were given by US citizens, three by UK citizens, one by a Canadian and the remaining sixteen by subjects from non-English speaking countries. One subject provided the answers in German and the remainder (48) left the text boxes empty.

Regarding mental representations, three different categories were established: "nature/naturaleza", "self/sí mismo" and "others/otros". This taxonomy aims to clearly distinguish whether the representational object is associated with the subjective realm or with the world (Castellaro, 2011). Regarding emotions, the model proposed by Diaz & Flores (Díaz & Enrique, 2001) contains 28 polarized categories and an associated thesaurus for the task of grouping and was adopted for the purposes of classification. The aim was, first of all, to provide sufficient classes in which to include most of the participants' subjective characterizations while avoiding oversimplification and, secondly, to reduce the number of classes in the analysis where possible. The particular clustering model can be described as a combination of Plutchik's affective wheel (Plutchik, 2001), with the terms arranged together based on similarity, and the valence-arousal plane (Russell J. A., 1980) that settles polarities for each of the present terms.

In order to analyze the distribution of frequencies in the assembled descriptive data, the different groups were compared using a chi-squared test with continuity correction in two by two tables, establishing the significance level at $p < 0.05$. Among the 753 descriptive entries for mental representations (251 subjects $\times$ 3 samples), 501 were classified into the categories defined above: nature ($n = 300$), others ($n = 112$) and self ($n = 89$); 23 were not classified and subjects did not offer a response (the box was left empty) in 229 cases. With respect to the emotional states, among the 753 entries, 600 were classified, 9 were not classified and 144 were left empty. Table 3.6 shows the contingency table for the five audio samples and the 501 classified mental representations, those descriptions relating to "nature" (59.9%) being the most abundant. In the study of distributions there exist significant differences, since the mental representations elicited by the NS sample are mostly described using terms relating to "nature" (91.7%). In the rest of the samples (the musical ones) the percentages are distributed more evenly among the categories, with a slight predominance of the type "others".

| | | | Type of mental representation | | | Total |
|---|---|---|---|---|---|---|
| | | | Nature | Others | Self | |
| sample | CC | $n$ | 19 | 21 | 12 | 52 |
| | | % | 36.5 | 40.4 | 23.1 | 100.0 |
| | CH | $n$ | 30 | 24 | 24 | 78 |
| | | % | 38.5 | 30.8 | 30.8 | 100.0 |
| | HC | $n$ | 15 | 29 | 17 | 61 |
| | | % | 24.6 | 47.5 | 27.9 | 100.0 |
| | HH | $n$ | 26 | 36 | 19 | 81 |
| | | % | 32.1 | 44.4 | 23.5 | 100.0 |
| | NS | $n$ | 210 | 2 | 17 | 229 |
| | | % | 91.7 | 0.9 | 7.4 | 100.0 |
| total | | $n$ | 300 | 112 | 89 | 501 |
| | | % | 59.9 | 22.4 | 17.8 | 100.0 |

**Table 3.6.** Contingency table audio sample, type of mental representation.

In the study of the described emotions, the 600 valid descriptions of emotional stimuli were classified into 20 of the 28 established categories. The most frequent emotional state was "calm" (57.5%), followed by "sadness" (13.3%) and the rest of them appearing with a notably smaller percentage. Since there are several categories and groups of study, they have been arranged into three classes, the two most frequent emotional states, "calm" and "sadness", and "other emotions". Table 3.7 shows the results. There is a significant difference in the group NS ($p < 0.001$), since most of the emotional descriptions are associated with "calm" (89.7%), while the rest of groups show similar percentages.

| | | | Affective System (Grouped) | | | Total |
|---|---|---|---|---|---|---|
| | | | Calm | Sadness | Other Emotions | |
| sample | CC | $n$ | 33 | 16 | 40 | 89 |
| | | % | 37.1 | 18.0 | 44.9 | 100.0 |
| | CH | $n$ | 53 | 27 | 32 | 112 |
| | | % | 47.3 | 24.1 | 28.6 | 100.0 |
| | HC | $n$ | 34 | 11 | 47 | 92 |
| | | % | 37.0 | 12.0 | 51.1 | 100.0 |
| | HH | $n$ | 42 | 24 | 37 | 103 |
| | | % | 40.8 | 23.3 | 35.9 | 100.0 |
| | NS | $n$ | 183 | 2 | 19 | 204 |
| | | % | 89.7 | 1.0 | 9.3 | 100.0 |
| total | | $n$ | 345 | 80 | 175 | 600 |
| | | % | 57.5 | 13.3 | 29.2 | 100.0 |

**Table 3.7.** Contingency table audio sample, affective system (grouped).

In addition, the emotional states were analyzed according to the polarity (or valence) in the 14 defined axes. Table 3.8 shows the results of this arrangement. The majority of emotional descriptions fit into the "pleasant" category or positive valence (75.7%). Nevertheless, there is once again a significant difference with the group NS ($p < 0.001$), since 96.1% of the descriptions come under the category of "pleasant", while in the music samples this figure ranges from 60.7% to 71.7%.

|  |  |  | Unpleasant | Pleasant | Total |
|---|---|---|---|---|---|
|  |  |  | **Valence** |  | **Total** |
| sample | CC | *n* | 28 | 61 | 89 |
|  |  | % | 31.5 | 68.5 | 100.0 |
|  | CH | *n* | 44 | 68 | 112 |
|  |  | % | 39.3 | 60.7 | 100.0 |
|  | HC | *n* | 26 | 66 | 92 |
|  |  | % | 28.3 | 71.7 | 100.0 |
|  | HH | *n* | 40 | 63 | 103 |
|  |  | % | 38.8 | 61.2 | 100.0 |
|  | NS | *n* | 8 | 196 | 204 |
|  |  | % | 3.9 | 96.1 | 100.0 |
| total |  | *n* | 146 | 454 | 600 |
|  |  | % | 24.3 | 75.7 | 100.0 |

**Table 3.8.** Contingency table audio sample, valence.

Table 3.9 provides a summary of the final stage of the experiment in which the participants responded to the question regarding the identity of the composer of each of the two musical pieces they had previously listened to. These questions were again presented in random order (HH, CH for group A subjects and HC, CC for group B).

| Q6 | Sample (*n* = 251) | | | | | Musicians (*n* = 149) | | | | | Non-musicians (*n* = 102) | | | | |
|---|---|---|---|---|---|---|---|---|---|---|---|---|---|---|---|
|  | Group A | | Group B | | | Group A | | Group B | | | Group A | | Group B | | |
|  | *n* | % | *n* | % | *p* | *n* | % | *n* | % | *p* | *n* | % | *n* | % | *p* |
|  | HH | | HC | | | HH | | HC | | | HH | | HC | | |
| computer | 65 | 48.5 | 58 | 49.5 | 0.967 | 44 | 49.4 | 31 | 51.7 | 0.921 | 21 | 46.7 | 27 | 47.4 | 0.897 |
| human | 69 | 51.5 | 59 | 50.4 |  | 45 | 50.6 | 29 | 48.3 |  | 24 | 53.3 | 30 | 52.6 |  |
|  | CH | | CC | | | CH | | CC | | | CH | | CC | | |
| computer | 61 | 45.5 | 66 | 56.4 | 0.111 | 46 | 51.7 | 39 | 65.0 | 0.149 | 15 | 33.3 | 27 | 47.4 | 0.220 |
| human | 73 | 54.5 | 51 | 43.6 |  | 43 | 48.3 | 21 | 35.0 |  | 30 | 66.7 | 30 | 52.6 |  |

**Table 3.9.** Results of the explicit inquiry about the nature of the composer in Phase II of the experiment.

In the analysis, no differences were observed regarding the composer or interpreter of the music (the human or the computer) and this lack of bias was observed for both professional musicians and non-musicians. The results are shown in Table 3.10. Musicians showed marginally higher values and a slight tendency to classify the piece CC as computer made. In general, all subjects failed to correctly identify the compositional source of the two musical samples.

| | Musicians | | | Non-Musicians | | |
|---|---|---|---|---|---|---|
| | Value | CI (95%) | | Value | CI (95%) | |
| Sensitivity (%) | 57.05 | 48.76 | 65.33 | 41.18 | 31.14 | 51.22 |
| Specificity (%) | 49.66 | 41.3 | 58.03 | 52.94 | 42.76 | 63.12 |

**Table 3.10.** Sensitivity and specificity in the human-computer judgement.

## 3.3 Further assessments

Apart from the described experiment, Melomics music has been tested in several other less controlled situations:

On May 2015, a professor of the Conservatory of Malaga performed *Just out of curiosity*, a newly generated composition for clarinet by Melomics, in front of his students and, additionally, he played back another Melomics composition to the same audience. Following this, and not having mentioned the source of the music, the professor asked them to highlight any relevant aspect of these compositions, for instance whether they could identify the historical context, name a possible author or just give their opinion. After a number of answers, the students were told the nature of the actual composer. The general conclusion was that they did not imagine that these compositions came from a computer, with no human intervention or having been fed with previous pieces.

- Sample 3.4. Compositions played at the Conservatory of Malaga

Another test with Melomics music was carried out during the SDF2015 event. It was a similar version of the experiment described in section 3.2. Subjects also ranged in age 20 to 60 years, with a population of 150 musicians and 350 non-musicians and they were asked the same questions about the same music and natural sounds. This experiment was performed with the help of the platform

OpenSurvey,[44] through a web-based questionnaire and results showed in this case too that Melomics music elicited the same kind of feelings and mental representations than human music, different than what is evoked by natural sounds. Also, success rates when trying to distinguish human-composed from computer-composed music was around 50%. Only in the case of pieces synthesized by the computer, professional musicians had a slight tendency that it was computer generated.

---

[44] https://web.opensurvey.com/ (accessed on May 16, 2021)

## 3.4 Chapter summary

In this chapter we have analyzed the music created by Melomics. We have done this by reviewing a wide range of its compositional styles and following different approaches. By analytical methods, we have concluded that the similarity between genomes is tightly connected to the similarity measured in the corresponding musical phenotypes; and we have seen that Melomics compositions, produced in a variety of styles, are numerically close to their human counterparts and distant to other kind of sounds. The second half of the chapter has been focused on a more empirical analysis, based on human subjective assessments. The results seem to point into a similar conclusion: Melomics music can be perceived in the same way as human-composed music, and clearly different from other types of sounds, at least when the critic is not biased by the fact of knowing the computer nature of the composer (see section 5.2 for an extended discussion).

# Chapter 4

# Adaptive music composition

In this chapter we introduce a way of creating and providing music, capable of responding dynamically to the circumstances of a considered environment or the current listener's state. In section 4.1 we show ways to take advantage of automated composition and present our proposal of adaptive music; in section 4.2 we summarize some previous attempts in the area; in section 4.3 we define a specific implementation of the adaptive music system; and in section 4.4, we show and comment the different implemented applications based on it, designed for different scenarios and capable of producing sentient responses through music.

## 4.1 Motivation

Counting on a computational system, capable of creating music of the same kind than an expert human composer, can enable a set of applications and lines of work that may have not even been considered or developed until now. An example of so is the possibility to create highly tailored music for the tastes of a particular person. A human composer could do it for a limited amount of people, but there are no composers enough in the world that could satisfy a global demand. Besides, it would probably not be a task that they would like to spend their time on. In this case, it would not be as much a human activity being replaced by a computer, but the computer filling in a niche that is not meant for people. Music tasks that involve just satisfying a bunch of given specification would no longer be an obligation of composers, in the same way it occurred with the invention of photography, in regard to the tasks of creating portraits or capturing scenes; quite the opposite, automated composition should bring new ways of creating art.

Some identified advantages of a computer composer are: (a) fast production of compositions; (b) production of themes without deviations from the established expectations; and (c) music versioning, by modifying a few elements in a precise way or through major transformations (affecting the structural basis, harmony, texture, etc.).

As commented in previous sections, we have experimented with all these possibilities, but on the later stage of this work we have focused our efforts on *adaptive music*. This consists of making music in real time, by manipulating musical parameters responding to certain values or states captured from the environment. A simple example to illustrate this idea is to consider the stage of a disco and the average speed of movement detected as the input from the environment. The music to provide could be adjusted, by bending the tempo, volume or textural density, for example, to be proportional to the detected intensity of the people's movement, being the played music a reflection of the "energy" perceived on the stage.

## 4.2 Related work

Audio being adjusted in real time to answer to certain input parameters can be found in different fields of application. The most widespread case might be its use in computer games. Practically since the beginning, playing music according to the situation presented to the player was seen as an important part of the dynamic. Late in the 90s, when most video games were made with 3D engines and wanted to provide an immersive experience, music also acquired a greater importance, with titles like Silent Hill or Resident Evil making use of specific music to create the desired fearful atmosphere or stealth games like Metal Gear Solid, where music indicated a delicate moment or told the player that something was going to happen and needed to prepare for a fight. Experimenting with adaptive audio, not pre-synthesized, but created in real time according to certain rules and inputs, in electronic entertainment and other interactive settings, has been a tendency in recent times. This matter has also been studied in science, usually referred to as non-linear music (Buttram, 2003) or procedural audio (Garner, Grimshaw, & Nabi, 2010); an example technique consists of using Markov chains to generate musical transitions between two different pieces (Wooller & Brown, 2005).

Regarding inventions that exploit this idea, we have found a patent consisting of a device for emitting acoustic signals (sounds, music and/or speech) in order to promote sleep, with the signals controlled by a physiological parameter registered in an individual (Patent No. DE3338649 A1, 1985). Another patent describes a similar system but with a more complex analysis of the physiological parameter, characterizing the sleep phases in the subject, to act in consequence (Patent No. US20090149699 A1, 2009). A last patent found is about a system that produces acoustic signals and is addressed to treat stress pathologies (Patent No. WO2003018097 A1, 2003).

A more recent application of adaptive music and nearer to the idea that we present includes an algorithm that is designed to produce MIDI output in an interactive context (Morreale, Masu, & De Angeli, 2013). The inputs considered are based on the emotions detected on a certain audience, for example extracting information from the movements of people while dancing (Morreale, De Angeli, Masu, Rota, & Conci, 2014), so the algorithm can adjust some global parameters like volume or tempo and some specific features like amount of dissonance and density of notes.

As a summary of the cases explored, we observed that there are two types of systems: either (a) based on pieces pre-composed by human artists in the traditional way, which are played sequentially, selected with certain criteria and possibly using transition effects (typically crossfade); or (b) systems for live generation of audio signals, but these later being far from real music. Our contribution to this area is a method that can provide the audience with genuine music, responding in real time and with the possibility to deal with any musical parameter of any level of abstraction. Rhythmic evolution, harmonic curves, instrumental density, relationships between instrumental roles or compositional structure are some of the parameters that can be controlled.

## 4.3 Feedback-based music: The hypersong

In this section we propose an implementation of an adaptive musical system, taking advantage of being able to produce actual genuine music. To illustrate the concept, we consider a case example consisting of a mechanism that can characterize the relaxation level of an individual, through monitoring the heart rate, and play music in response. The aim is to reduce the perceived level of anxiety. The system can be structured in the next elements:

- A component to measure and characterize the environment. In the example, the heart rate sensor and the function that summarizes the provided data in different levels, for instance: anxious, active and relaxed.
- A mechanism that correlates the state detected in the environment to an appropriate type of music to be played back.
- A system capable of producing the convenient flow of music.

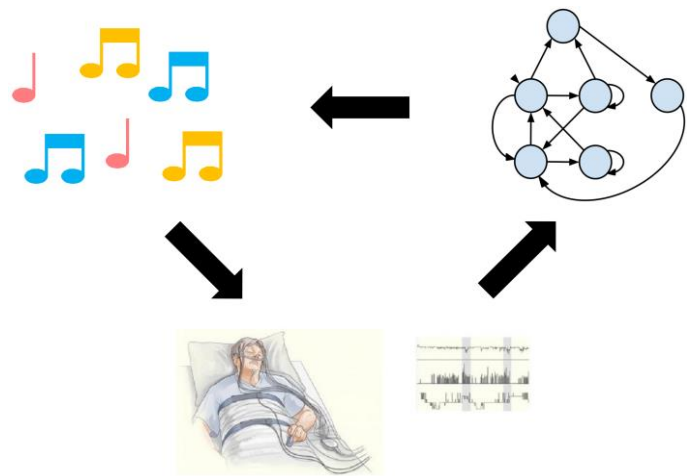

**Figure 4.1.** Feedback-based music system. It comprises characterization of the user or the environment, selection of music and production of music.

While a first obvious way of implementing this behavior could be by making Melomics manipulate genes and generate audio in real time according to the present evaluation of the environment, in practice this approach suffers from technical limitations. The system is intended to work on a small laptop or preferably on a mobile device. It can be optimized to run on these devices, but the composition operations are still very computationally demanding and hence would heat up and drain their battery quick. In addition, synthesizing music with the desired level of quality requires a professional library of virtual instruments. This supposes the need for a prohibitive amount of storage for these devices. Even if we ignored all those problems and supposed that we counted on a high-end computer, the composition process is not immediate and would introduce certain latency, which can be critical depending on the scenario.

To solve these limitations, we propose a different approach: For a given application, the type of music and the different versions required are settled in advance. These compositions with the different versions are pre-generated and synthesized. Each version is different internally, but the general structure is preserved in all of them. They are divided (in the time dimension), cutting in between different structural entities (see section 2.2.3.1 Musical styles). The same temporal fragments of different versions may have different durations. However, each version has the same amount of fragments since they all share the same structure. As a result, for each composition we produce a two-dimensional musical entity that we call an *hypersong*. It also contains some meta-information, as the usual music produced with Melomics.

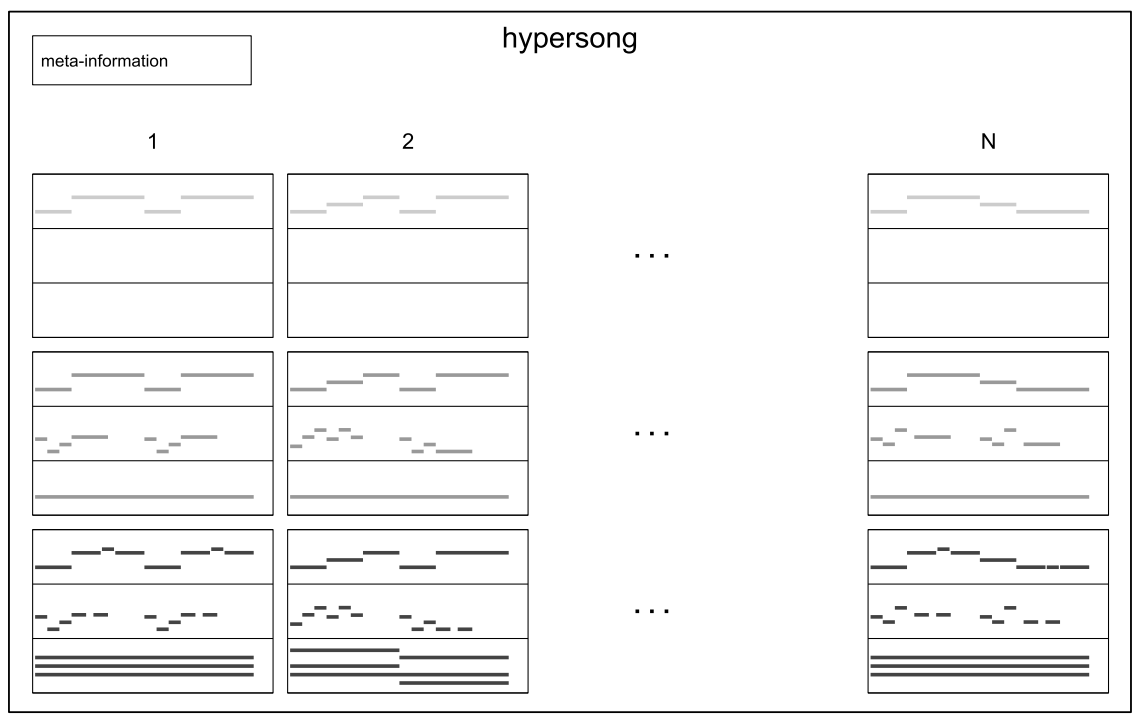

**Figure 4.2.** Illustration of the hypersong. Each of the three versions in the example (represented vertically), is divided in N fragments in time (horizontally). The versions comprise from one to three instruments, indicated by the divisions inside the boxes. The different fragments are made of different musical content,

although each instrument is supposed to behave in a similar way along the time. The difference between the versions are: (a) from the first to the second one, the addition of two instruments, all playing at a higher volume (represented by the brightness of the lines); and (b) from the second to the third version, the increase of complexity and volume of the three performing instruments.

At first, this method may seem computationally more expensive since all the versions must be produced in advance and stored. Nevertheless, in practice it results to be more efficient because most fragments are played eventually and, actually, more than once. Hence, we avoid having to compute each fragment of the hypersong every time that is needed, as it would be needed with the first approach. To mitigate the inconvenience of having to store a lot of music in the devices, the hypersongs can be stored in a remote server, where they can be downloaded on demand. The mobile device can cache them as well, to save on data cost in case there is no WIFI available. For us, there are two main advantages with this solution: (a) Very little computer power is required from the client devices, which was one of the main worries, because of battery consumption and because high-end devices are expensive. These only have to run the environment characterization, send the processed information and get the corresponding music through streaming; deciding what fragment to play next and generating or synthesizing the music is the server's job. (b) Because both the software in charge of generating music and the software that manages how the music evolves with the environment's state run in a remote server, this approach helps to preserve the intellectual property.

We adapted the style-tool to make it capable of producing a valid composition, then different controlled versions of it and finally split everything in fragments that would be treated as separate pieces by the system. Cutting in fragments could not be based on time marks, since different versions might have different durations, because of different tempos or rhythmic patterns. Cutting should be based on compositional information, making sure that transitions between different versions are smooth and do not break any musical idea (see Musical structure). An extra silence was needed at the end of each fragment, to assure the proper decay of the notes being played.

- Sample 4.1. Same temporal fragment of a hypersong in its three levels
- Sample 4.2. Evolution of a hypersong through its three levels

The resulting adapting system has three main elements: (1) a software to identify states in real time, whose general description is outlined in the next section and some specific designs for different applications are given in section 4.4; (2) a mechanism to map the detected states to a version of the hypersong, which are explained in section 4.3.2 Control of compositional paths by DFA; and (3) a system to produce the flow of music; in this case comprising the hypersongs pre-generated with Melomics and played by the client application according to the directions given by the second component.

## 4.3.1 Characterization of states

The component in charge of characterizing the state of the environment has a global architecture shared by all the applications, divided in four stages: (1) one or more sensors take measures from the environment; (2) each measured raw signal is processed to obtain more stable and meaningful data; (3) the data is analyzed through functions designed for each application, to obtain information of a higher level of abstraction; (4) this information is translated jointly into one of the allowed input states of the system in charge of modulating the music, based on finite automata, as described in the next section. In section 4.4 some examples are described.

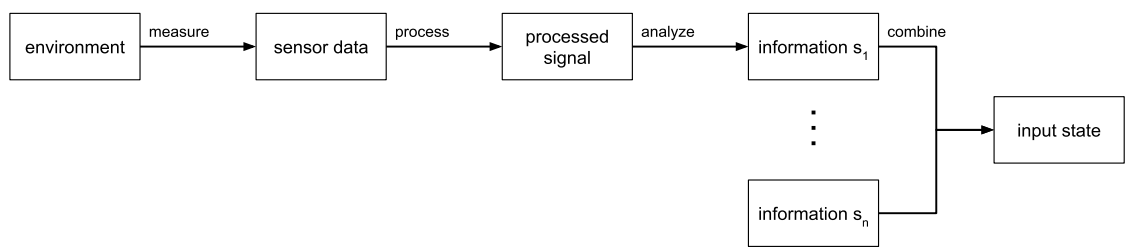

**Figure 4.3.** Process of characterization.

## 4.3.2 Control of compositional paths by DFA

### 4.3.2.1 A model based on Moore Machines

A mechanism is needed to decide what version of the music to play, depending on the intended purpose of the application and the current characterized state. In order to implement this functionality in a standardized way for all the applications based on adaptive music, we proposed a system based on the Moore Machines, a deterministic finite automaton (DFA). To the standard Moore model, we added (a) a *time counter*, that resets after every transition; (b) an *insensitivity time* for each state that forces to stay in a state for a certain period of time, before letting any new transition to occur; and (c) a *null-input timeout* parameter for each state that forces a default transition from the current state to another, if no input is received by the machine during this period. The model can be defined as an 8-tuple $(Q, \Sigma, \Delta, q_0, F, T, G, M)$ where:

$Q$ is the set of states

$\Sigma$ is the input alphabet, including the null value

$\Delta$ is the transition function, $\Delta: Q \times \Sigma \times T \to Q \times T$, such that:

$\Delta(q_i, \Sigma, t) = (q_i, t + 1)$, if $t < I_i$, being $I_i$ the *insensitivity time* given for the state $q_i$

$\Delta(q_i, \emptyset, t) = (q_i, t + 1)$, if $t < \lambda_i$, being $\lambda_i$ the given *null-input timeout* for the state $q_i$

$\Delta(q_i, \emptyset, t) = (q_{\lambda i}, 0)$, if $t \geq \lambda_i$, being $\lambda_i$ the *null-input timeout* for the state $q_i$ and $q_{\lambda i}$ the *null-input timeout* transition state defined for $q_i$

$\Delta(q_i, \Sigma_k, t) = (\delta(q_i), 0)$, otherwise, with $\delta(q_i)$ the state defined by a regular Moore Machine transition function $\delta: Q \times \Sigma \to Q$

with $I_i < \lambda_i$ $\forall i$ and $t, I_i, \lambda_i \in T$

$q_0$ is the initial state, $q_0 \in Q$

$F$ is the set of final states, $F \subseteq Q$

$T$ is the set of discrete instants of time, $T \subseteq \mathbb{N}$

$G$ is the output function $G: Q \to M$

$M$ is the output alphabet

The system to control the adaptive playback is composed by this mechanism based on finite state machines; the logic to manage the playlist of hypersongs, including the theme and fragment currently being played; and a custom media player for hypersongs, formed by two normal media players that work in an alternate fashion: one plays the current fragment, while the other is preparing the next fragment and starts playing right when the first one is playing the decay section of the previous fragment. When the first player is finished, it prepares the fragment that comes after the one being played by the second player and the process starts over, swapping roles every time.

We used JFLAP,[45] an open source free software developed at Duke University, to ease the process of designing our DFAs. The tool was modified to allow us to create and edit our particular kind of automatons. These designs are stored in XML files, which can be easily converted to JSON and interpreted by the software in charge of controlling the adaptive system.

### 4.3.2.2 Implementation through web services

The implementation has evolved from the first prototype in MATLAB, where the transition functions for each application had to be hardcoded every time in a separate ".m" file. Now it is mobile oriented and the design of new state machines is standardized and simple, using a graphic tool that produces a configuration file that can be saved and exported.

```
{
    "_id" : <DFA_UUID>,
    "model" : {
        "outputs" : <List_with_the_output_for_each_state>,
        "numMetaThemes" : <List_of_hypersong_indexes_of_the_metaThemePrefix_class>,
        "metaThemePrefix" : <Hypersong_class_to_be_used>,
        "initialState" : <Initial_state>,
        "delay" : <List_with_the_insensitivity_time_for_each_state>,
        "finalState" : <Final_state>,
        "table" : <Transition_function_based_on_a_list_of_2-tuples_for_each_state>,
        "introTheme" : <Possible_initial_theme_to_be_played_before_the_DFA_starts>,
        "lambdas" : <NullInput_timeout_based_on_a_list_of_a_2-tupla_for_each_state>
    },
    "name" : <DFA_NAME>
}
```

**Figure 4.4.** Melomics DFA template for adaptive music applications.

---

[45] http://www.jflap.org/ (accessed on May 16, 2021)

```
{
    "_id" : "88e5f38b-123c-447d-99af-c217d0c15213",
    "model" : {
        "outputs" : [ 0, 1, 2, null ],
        "numMetaThemes" : [ 0, 1, 2, 3, 4, 5, 6, 7, 8, 9, 10, 11, 12, 13, 14 ],
        "metaThemePrefix" : "painnightalter",
        "initialState" : 0,
        "delay" : [ 60, 0, 0, 0 ],
        "finalState" : 3,
        "table" : [ [ [ 0, 0 ] ], [ [ 1, 0 ] ], [ [ 2, 1 ] ], [ [ 3, 3 ] ] ],
        "introTheme" : "",
        "lambdas" : [ [ 60, 1 ], [ 600, 2 ], [ 600, 3 ], null ]
    },
    "name" : "sleep"
}
```

**Figure 4.5.** Example of instantiation of DFA.

The DFAs are stored in a remote server (using mongoDB[46]) and the mobile application interacts with it through a web RESTful API. For better scalability, no information about the state of the client app is stored in the server. The app logs into the system, tells about its context, receives the order about the music that must be played next and closes the connection. Then, if necessary, the app downloads the specific fragments of audio by opening a new connection. In summary, the mobile application is responsible for the direct interaction with the user, through sensing, playing music and a simple graphic interface, while the server acts as the decision maker and content provider.

---

[46] https://www.mongodb.org/ (accessed on May 16, 2021)

# 4.4 Applications

In this section we describe different instances of the adaptive music system. Most of them work in real time, responding to the user's or the environment's characterization, following the exact mechanisms based on DFA described before.

## 4.4.1 Easement of sleep initiation

eMTSD

The name comes from Empathetic Music Therapy for Sleep Disorder and the app uses the adaptive model for self-management of sleep related disorders like insomnia, somniphobia, sleep apnea, RLS and KLS. These diseases affect more than 20% of the population in developed countries (Stranges, Tigbe, Gómez-Olivé, Thorogood, & Kandala, 2012) and have been linked to the increase of depression and anxiety. The connection of poor sleep quality with hypertension, diabetes and obesity has also been reported in literature (Peppard, Young, Palta, & Skatrud, 2000) (Ioja & Rennert, 2012) (Barceló, et al., 2005). For this reason, and since music can help to relax (Lai & Good, 2005) (Huang, Good, & Zauszniewski, 2010), we designed eMTSD for mobile devices, implementing a 14-day trial[47] studying the benefits of empathetic music for sleep initiation and sustaining.

The app works by placing the smartphone over the mattress, making possible to identify the user's sleepiness state by characterizing their movements. This is done by combining the 3-axis data from the accelerometer that is then processed.

The sleepiness level characterized is used to feed an automaton whose purpose is just to provide music with a level of activation proportional to the detected state, namely, the more relaxed and nearer to falling asleep the user is, the softer and quieter the music should be.

---

[47] http://geb.uma.es/emtsd (accessed on May 16, 2021)

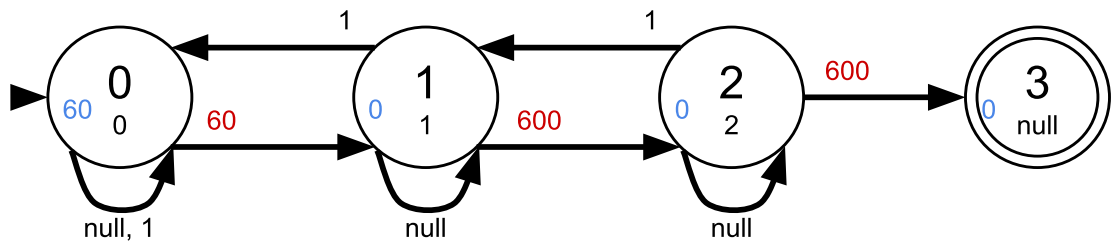

**Figure 4.6.** Visual representation of the automaton for eMTSD. Each state is represented by a circle, labeled with an ID. The associated output is located below and the insensitivity time to the left, in blue. The arrows represent the transitions, marked with the input or the null-input timeout (red) that triggers them. The initial state is preceded by an arrow with no origin and the final state is illustrated with two concentric circles.

All versions of the hypersongs for this application were specified to have a low arousal and a positive valence. After having been generated, the resulting themes were reviewed and approved by the experts that gave the specifications. The main musical features configured comprised: a small set of instruments playing the roles of melody, harmonic accompaniments, percussion and FX; medium and low tessituras; long rhythmic patterns; instruments with a soft timbre; harmony simple and familiar, using only major or Dorian modes and only major triad, minor triad and maj9 chords; and a very simple structure and textural evolution, favoring the repetition of ideas. The hypersongs are divided in three levels, corresponding to the states of the automaton, excluding the final one, which will not play music at all. The highest level includes the whole instrument set (their actual appearance is conditioned to the inner theme's textural evolution) and a higher tempo and volume; the middle level does not contain the percussion roles, the bass role is simpler, the tempo and general volume is lower and the set of VSTi includes softer instruments; finally, the third level only includes the melody role and two accompaniments and the tempo, volume and strength of instruments is even lower.

- Sample 4.3. Fast evolution of a hypersong for eMTSD

The app eMTSD includes a simple questionnaire for trial purposes and a binary selector to indicate the type of use: *nap* or *sleep*. The user has also access to a small set of buttons: *play/pause*, *forward* and *settings*. The *nap* mode has different configuration of times in the DFA and different music than *sleep*.

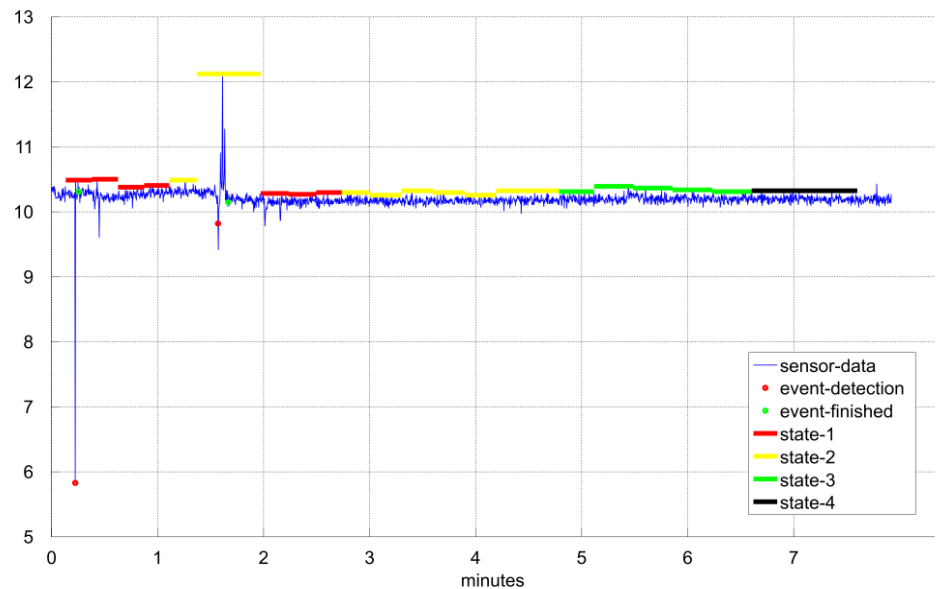

**Figure 4.7.** Recorded session with eMTSD. Events of motion, detected with the accelerometer, triggered the transition to a higher state. When no events were detected, the system automatically moved progressively to lower states, ending in state-4 (asleep).

eMTSD evolved from one initial simpler app called @sleep that was latter embedded as a scenario of a more general app, @life. The analysis of the sleepiness was improved as well as the music specifications and this was finally separated as a standalone app called eMTSD, to be used in the experiments. Later it was versioned into duermeteya, to facilitate sleep in babies, with hypersongs with a simpler structure and harmony and different instruments, trying to resemble lullabies.

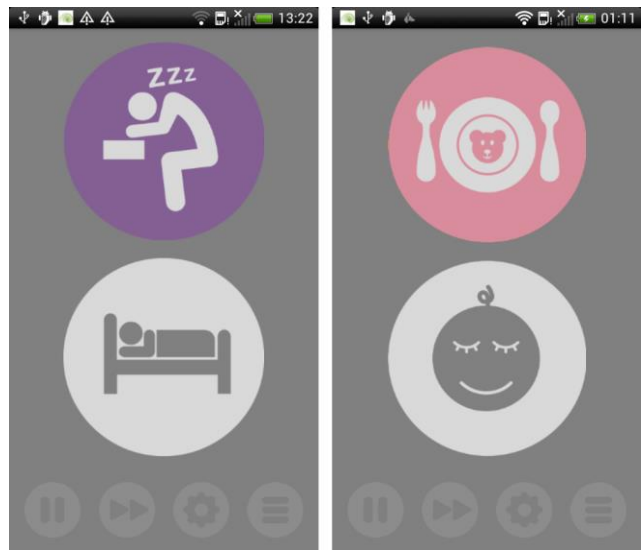

**Figure 4.8.** User interface of eMTSD and duermeteya.

## 4.4.2 Improvement of concentration and coordination

### PsMelody

PsMelody purpose is to promote performance at work and, at the same time, try to decrease the level of stress by providing adaptive relaxing music. It is a prototype implemented for Windows. It takes keyboard and mouse usage, key pressing frequency and movement, clicks and scrolls respectively, to interpret the user's activity level. The general intended behavior implemented through the DFA is to play more energetic music when the user is too relaxed and softer music when the user seems to go too fast. The goal is to bring the user to an optimal state of performance and wellbeing.

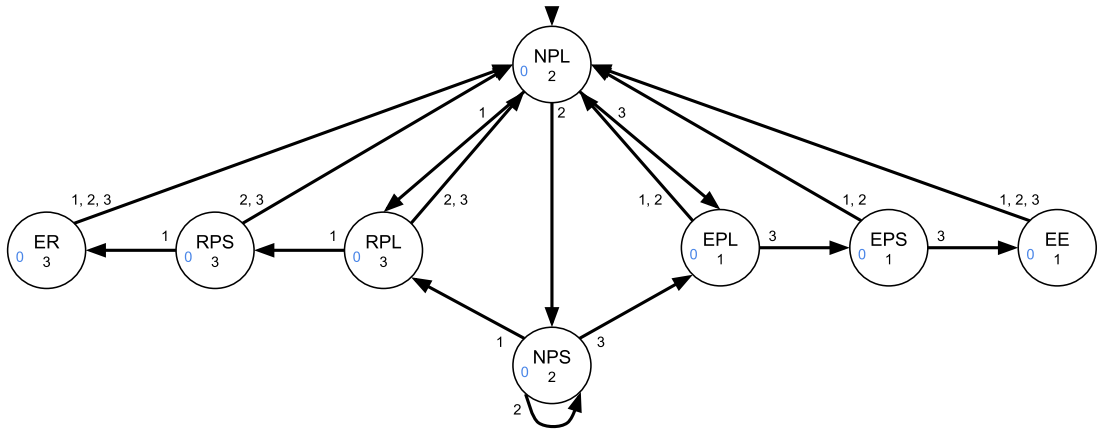

**Figure 4.9.** Visual representation of the automaton for PsMelody. Each state is represented by a circle, labeled with an ID, the associated output is located below and the insensitivity time to the left, in blue. The arrows represent the transitions, marked with the input that triggers them. The initial state is indicated by an arrow with no origin. The ID are acronyms for: Emergency Relaxed (meaning too relaxed), Relaxed Period Short, Relaxed Period Long, Normal Period Short, Normal Period Long, Excited Period Long, Excited Period Short and Emergency Excited.

Figure 4.9 shows the default behavior of PsMelody. The activity detected by analyzing the keyboard and mouse usage is mapped to three levels ("1", "2" and "3") and the hypersong is formed by three versions (also labeled as "1", "2" and "3").

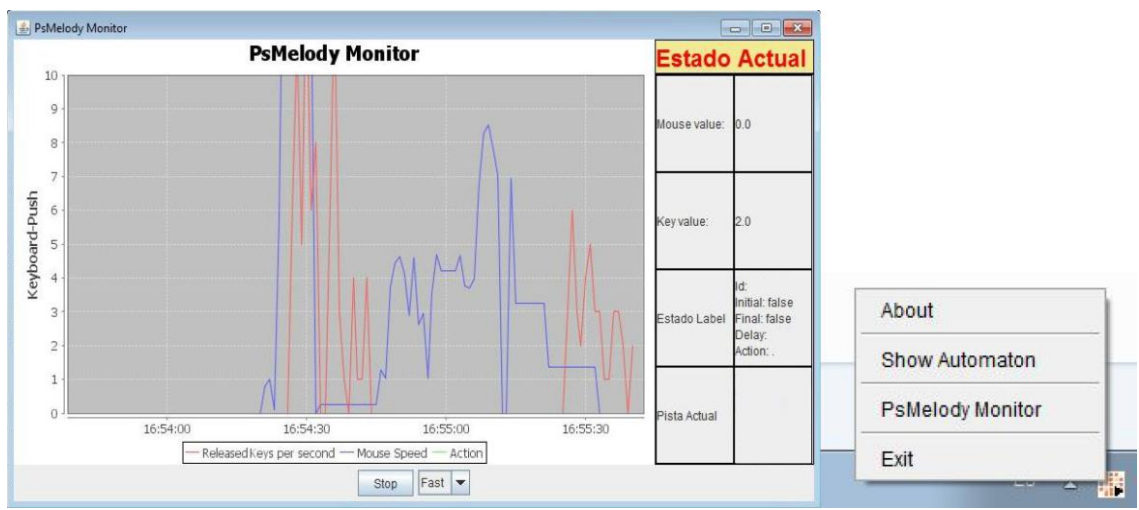

**Figure 4.10.** User interface of PsMelody.

This prototype works locally and comes with the tool JFLAP integrated to easily try new DFA. The user interface is simplified in the form of a tray icon on the taskbar, which allows four actions: go to "About"; manage the automaton through JFLAP; show the PsMelody Monitor, with information and graphics with the registered data in real time; and "Exit".

The music used is simple and with a small set of instruments and it puts emphasis on the rhythm, suggesting a proper tempo to the user, but trying not to be annoying or distracting. To manage different preferences and to prevent from boredom, we included different styles (pop, rock, jazz, blues, soul, classic, ambient, indie styles...). The tempo changes depending on the piece and on the hypersong level, globally from 40 to 130 BPM. The harmony progressions, chords, rhythmic patterns and global structure are simple to not attract much the user's attention.

@car

@car (Gutiérrez-Gutiérrez, 2013) is a mobile app in Android whose goal, through playing adaptive music, is to favor concentration and reduce fatigue while driving, especially under irregular conditions, like extreme weather or saturated traffic.

The state characterization considers data from multiple sensors present in most smartphones: GPS, accelerometer, photometer and magnetometer. The raw data allows to infer information classified in four groups: (1) type of road (e.g., highway, city street, etc.), (2) present state or events on the road, (3) state and events on the driver and (4) weather conditions. All the states and events score with a positive or negative sign and different magnitudes, depending on how they affect to the driving activity (for example bad weather or rough driving score negative and regular speed on a highway score positive). The automaton takes the scores and plays music to try to relax the driver under bad conditions and more energetic music when conditions are good.

For the application to work, the mobile device must be secured in a fixed position inside the car, in any desired orientation. @car is able to initialize and regularly update the car's movement axes. In the figure below, we show the speed, the accelerometer data and the level of the music played during a session recorded while driving first through a residential area, then a highway, a main avenue in the city and some narrow streets at the end of the route.

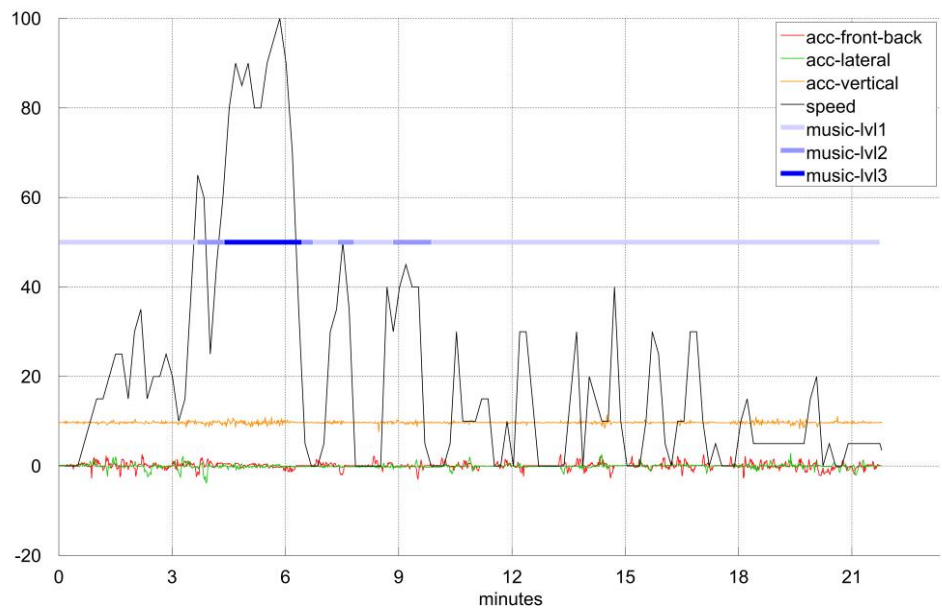

**Figure 4.11.** Recorded session with @car. The 3-axis accelerometer data is shown in orange, red and green (in m/s$^2$). The speed measured by the GPS (km/h) is shown in black. The levels of the music played by the system are represented in different tones of blue.

The virtual instruments are softer in the more relaxed states and more percussive in the more activating ones. Volume, rhythm, instrumental density and composition complexity are proportional to the score that characterizes the driving conditions. Different musical genres were used to satisfy different preferences and to provide a varied playlist, preventing monotony while driving.

@car helped to design and implement much of the functionality that was the base for the rest of apps, including working and processing data from sensors, the overall architecture and the adaptive player itself. This media player can acquire musical fragments from different sources (internal resources, expansion files and remote servers), implementing different policies (normal streaming, file cache, use WIFI connections or cellular networks...). Then, in coordination with the app's core, it is able to provide with an uninterrupted adaptive musical experience, by properly overlapping one fragment after the other. The app's user interface was designed intentionally simple, only with a big play/pause button on the center, with no need for any additional action, in order to keep it comfortable and safe for the driver.

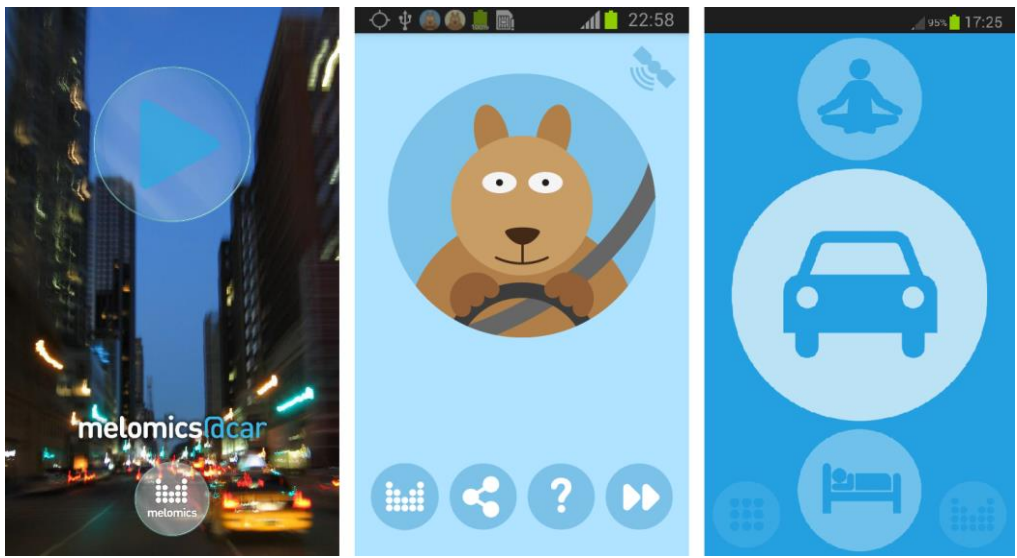

**Figure 4.12.** User interface of three versions of @car. The original, the most recent, *commuting*, and the car scenario of @life.

## @memory

This app uses Melomics music to exercise the associative memory. Using a method to generate simple forms, it proposes a game based on matching melodies to figures, through a sequence of different stages with increasing difficulty.

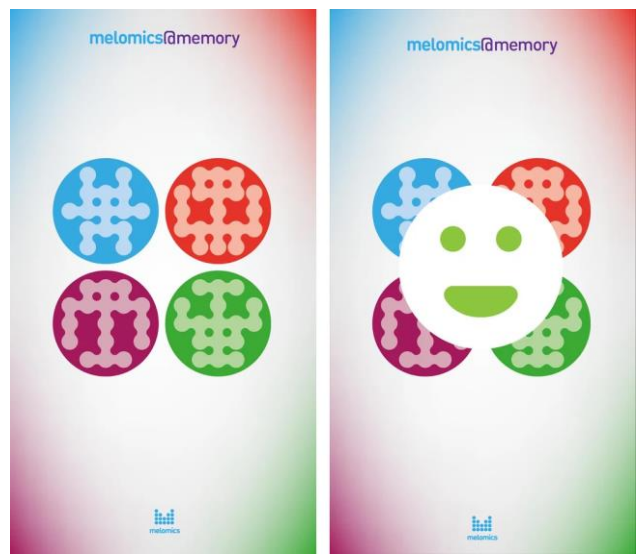

**Figure 4.13.** User interface of @memory.

The music for this app is bundle of short pieces (a few seconds), configured with a relatively high BPM and predictable rhythms, simple and familiar harmonies, medium to high pitch, clear melody and a small set of very recognizable instruments.

Some targets of @memory are managing emotional states of anxiety and agitation, reactivating decayed cognitive processes, stimulating the maintenance of decayed physical processes and improving concentration, attention and recognition of people and objects. Alzheimer or Multiple Sclerosis are two examples that could be tackled.

## @rhythm

The app is designed to improve concentration and coordination. It presents short pieces of music and the users have to identify the leading beat, tapping on a specific point on the screen surface. It has a simple interface, providing a series of stages with an increasing difficulty. On each stage, the users must start tapping on the screen as soon as they identify the beat. After a short while, they are given a score, based on their speed and precision.

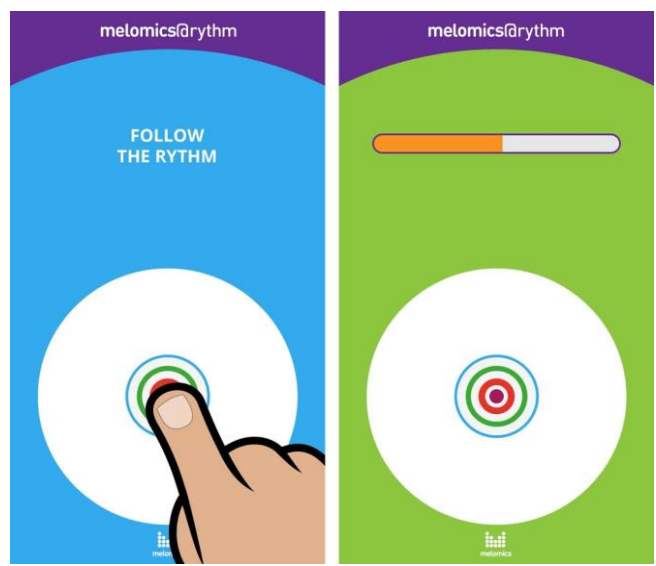

**Figure 4.14.** User interface of @rhythm.

Music consists of a bundle of music pieces, with a duration from a few seconds to near a minute, depending on the level of difficulty. For the same reason the tempo may vary between 40 to 120 BPM and the ensemble may be formed by 2, 3 or 4 instruments. Tempo will not change in a musical piece, compositional structure and harmony remain simple, tessituras are medium to high and harmony is built using major modes and simple chords.

### 4.4.3 Management of pain and anxiety

eMTCP

Because music is a powerful distractor, it can be used to reduce pain perception. In order to learn how music can help in self-managing of chronic pain, eMTCP (for Empathetic Music Therapy for Chronic Pain) is a mobile app that was originally designed to perform a 14 day ambulatory clinical trial,[48] without the supervision of specialized personnel or control in a clinical sense. The hypotheses to be tested were: (a) empathetic music therapy improves sleep quality of subjects with chronic pain and this correlates with reduced pain perception; and (b) it reduces the perception of pain when applied interactively. eMTCP applied to sleep promotion, behaves in a similar way as eMTSD. In eMTCP used during the day, subjects report through the screen the perceived level of pain, using a numeric scale or visual analog scale (VAS) (Ritter, González, Laurent, & Lorig, 2006), and the music adjusts accordingly in real time. The application can be used anywhere and anytime the user feels pain.

eMTCP was promoted by the American Chronic Pain Association in order to test the system on its large community of users, involving thousands of people. Chronic pain affects 100 million people in the US alone and it is associated with a large number of conditions (Institute of Medicine (US), 2011). An estimated 20% of American adults report that pain or physical discomfort disrupts their sleep a few nights a week or more (Sleep in America Polls, n.d.), so the goal was to create an inexpensive, but effective method to manage pain.

The state machine for eMTCP during the day has 25 states and 57 transitions, it makes extensive use of the parameters null-input timeout and insensitivity time and it considers a wider range of inputs and outputs, with 11 states of pain and 8 levels of music.

---

[48] http://geb.uma.es/emtcp (accessed on May 16, 2021)

Parameters studied and values set up by musical therapists to produce compositions for eMTCP comprise: soft virtual instruments for the synthesis; a high number of roles, up to 17; simple harmony, with Dorian and major modes and major, minor, sus2 and maj9 chords; simple internal accentuations and rhythmic patterns; long duration, 10 minutes on average and only three fixed models of textural evolution: (1) starting from the minimum number of instruments playing simultaneously, gradually incorporating roles to the maximum allowed number and then gradually again to the minimum; (2) starting from the maximum, gradually to the minimum and then back to the maximum; and (3) starting from the minimum, gradually to the maximum, instantly to the minimum and then gradually to the maximum again. The hypersongs have eight levels. Lower levels user lower volume and tempo and vice versa. Textural density and complexity vary from using only one simple melody and one accompaniment in the softest level to two melodies and some accompaniments in the medium levels and to using most of the accompaniments in the higher levels.

- Sample 4.4. Evolution of a hypersong for eMTCP through all its levels

eMTCP includes a login system designed to manage the ambulatory trial, including sending an email with an identifying code, and a system to collect the usage and operational data to be analyzed.

In 2013 the application was used as the basis for the system Melomics Music Medicine ($M^3$), to perform a clinical trial in Hospital Materno-Infantil, Málaga, in the pediatric allergy unit (Requena, et al., 2014). The experiment was aimed to reduce pain perception during the pediatric skin prick test (SPT), ending with a high success compared to the control group, as summarized in Table 4.1.

| Pain (VAS) | Experimental | Control | $p$ |
|---|---|---|---|
| No pain (0-2) | 22 (71.0%) | 7 (18.9%) | <0.001 |
| Moderate acute (3-10) | 9 (29.0%) | 30 (81.1%) | |
| Mean/SD | 2.1/1.1 | 4.9/2.7 | <0.001 |

**Table 4.1.** Pain reported by the two groups during the skin prick test.

After the clinical trial period, the app was released under open access terms.[49] Later, a simplified version, named *chronic pain*, was designed for a more general public. The login requirements and the survey to introduce personal and medical data were suppressed and social interaction mechanisms were implemented. The automaton was simplified, operating with three stages of pain and hypersongs of five levels. The application was released during the event SDF2015 celebrated in Seoul, South Korea, in May 2015.

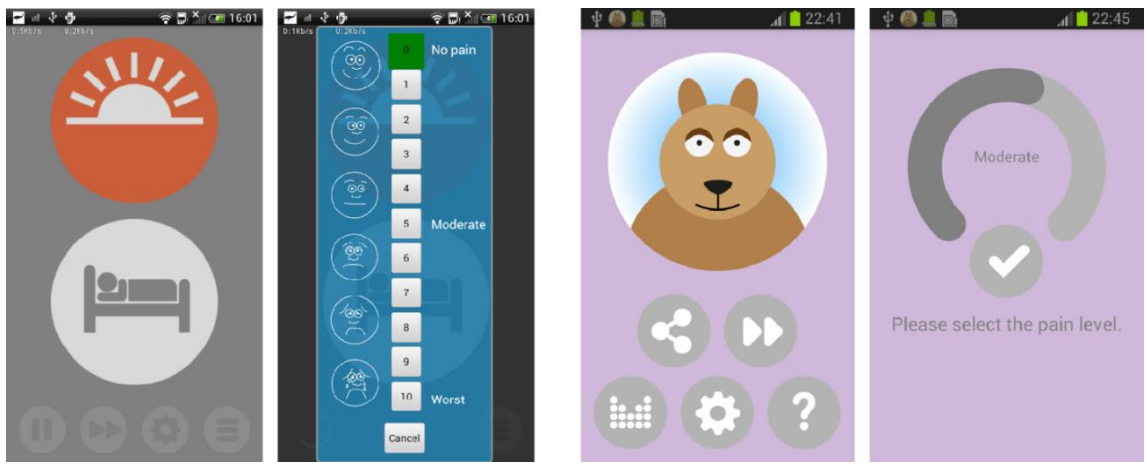

**Figure 4.15.** User interface of the two versions of eMTCP.

## STM

The very first use of adaptive music with Melomics consisted of a therapeutic application for hospitals and clinics that worked on locally Windows machines (a netbook in the trials). This primary system could manage different types of biosensors, analyze the corresponding biosignals to characterize the patient's state, manage the automaton and play the music adapted in real time.

STM, Spanish for Sistema de Terapia Musical and later transcribed into English as Empathetic Music Therapy (EMT), was developed under the research project "MELOMICS Optimización de la respuesta terapéutica a la modulación de estímulos auditivos"[50] with the purpose of developing and testing the adaptive music system in hospitals, as described by Vico (Vico, Sánchez-Quintana, & Albarracín, 2011).

---

[49] https://en.wikipedia.org/wiki/Open_access (accessed on May 16, 2021)
[50] http://geb.uma.es/research/melomics.html (accessed on May 16, 2021)

The system was built to carry out experiments in different units at the Hospital viamed Montecanal, a partner in the research project. The starting trial was performed in the neonatal unit and consisted of providing mothers of newborns with adaptive music during breastfeeding. The heart rate measured on moms was taken as the input parameter to characterize their state of anxiety. The aim was to reduce the stress level, which was hypothesized to lead to a low level of stress on the newborns, increasing the breastmilk sucking rate, hence improving their feeding. Later other experiments were proposed and performed over patients undergoing unpleasant or painful procedures, for example in the hemodialysis unit of Hospital Costa del Sol in Marbella.

STM was prototyped using the MATLAB. It comprised a small database; a set of classes to acquire the data coming from the heart rate monitor, with the possibility of adding extra sensors just by implementing the proper methods; a component in charge of processing the signal and characterizing the patient's current state; the code implementing the state machine model; an early version of the adaptive player; and a user interface with an initial survey to be filled in before the first session and a window showing the ongoing therapy, with the current heart rate, the oximetry status and other information.

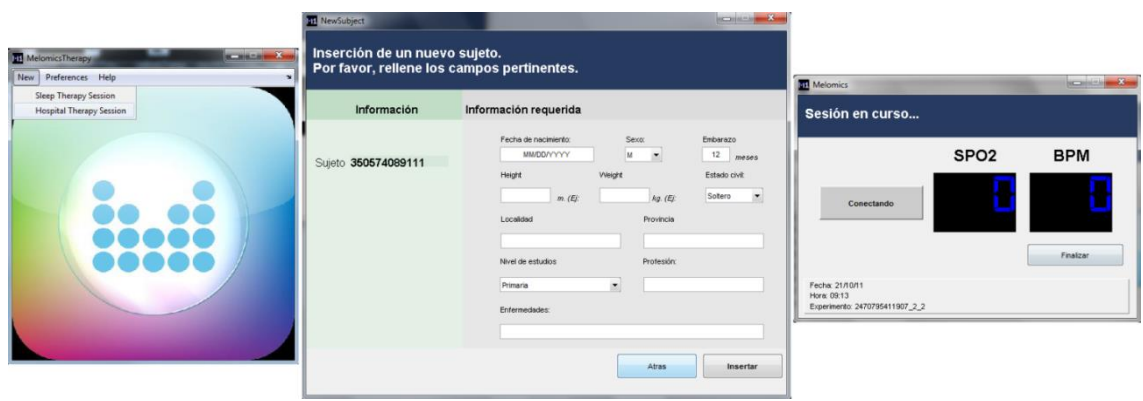

**Figure 4.16.** User interface of the prototype STM.

A double-blind experiment was performed including three groups of patients: (1) a group with headset but no music being played at all; (2) a group supplied with static music, unresponsive to the patient's state (a single level of the hypersongs was used for this group); and (3) the group supplied with the adaptive therapy, characterizing the state in real time and deciding the audio content to be played next. The MATLAB program was compiled and stored in an Asus EeePC netbook connected to the internet. The database and logs from each session were stored in a folder synchronized via Dropbox.[51]

To characterize the state of anxiety, at the beginning of each session, the basal heart rate is computed, then the states are established as follow:

$f(t)$ is the heart rate over the time $t$, in seconds.

The basal heart rate is estimated in a time $T_B$ (usually 2 to 5 minutes, $T_B = 120$ or $T_B = 300$).

$f_b = \frac{\sum_{t=1}^{T_b} f(t)}{b}$, is the basal heart rate.

$d = \sqrt{\frac{\sum_{t=1}^{T_b}(f(t)-f_b)^2}{b-1}}$, is the data deviation while measuring the basal

$C_e = 0.4$

$C_r = 0.2$

$T_e = f_b + C_e \cdot d$

$T_r = f_b - C_r \cdot d$

The patient is *anxious* if $f(t) > T_e$

The patient is *normal* if $f(t) \leq T_e \wedge f(t) \geq T_r$

The patient is *relaxed* if $f(t) < T_r$

The coefficients $C_e$ and $C_r$ were adjusted empirically, after 30 tests in total involving 10 people.

---

[51] https://www.dropbox.com (accessed on May 16, 2021)

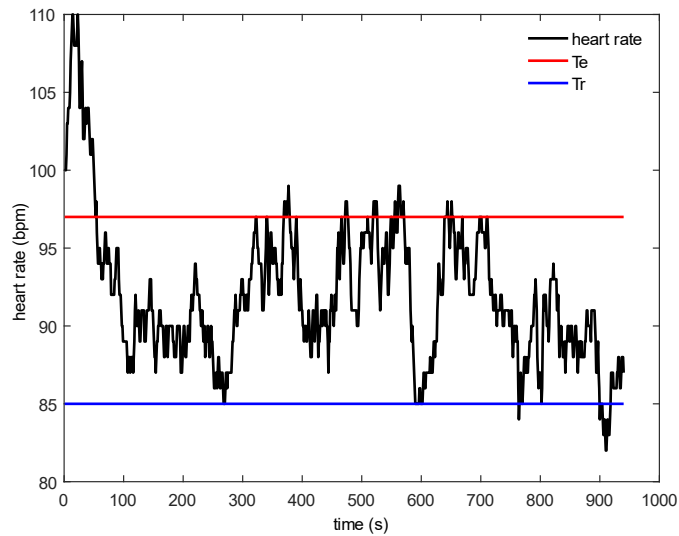

**Figure 4.17.** Data registered during a session with the STM system.

The automaton is divided in $M$ stages (in our experiment $M = 5$), each one representing a different attempt to catch the attention from the user, with the goal of achieving a level of complete relaxation ($f(t) < T_r$). The general behavior is described as follows: If the anxiety level is high, the system tries to attract the user's attention by executing any of the defined acoustics actions. This is done in a state of type E, which can be maintained only for a very short period of time, then the DFA moves to an N-state. These are associated to normal input (the present heart rate is near the basal). If the detected state after a while is normal, the automaton remains in the N-state.; if the state is relaxed, the automaton moves to an R-state, reproducing a softer version of the hypersong; but if the state is still anxious, the system moves back to the E-state of the following stage, executing the next acoustic action established to catch the user's attention. The process continues like this until the session is over. The acoustic actions defined for the different stages are (1) an abrupt pause of 10 seconds inserted in the music flow; (2) play the fourth level of the hypersong (not relaxing); (3) play the third level of the hypersong (relaxing, but at high volume); (4) play an independent short melody interrupting the current music flow; (5) immediate change of hypersong. The automaton moves along the stages every time that it passes through an E-state.

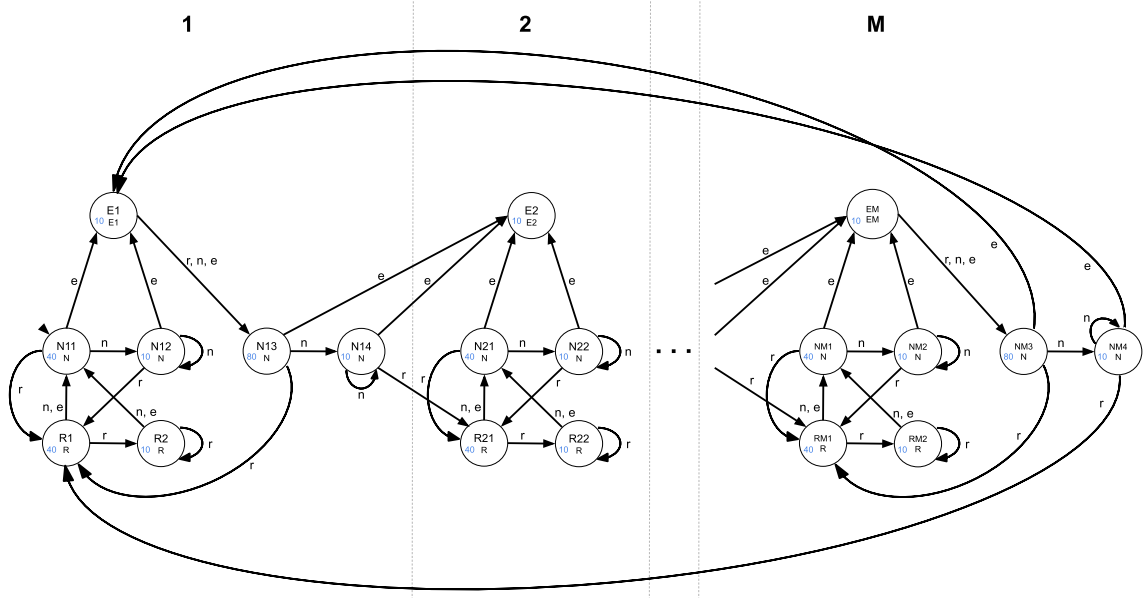

**Figure 4.18.** Visual representation of a template of the automaton for STM. Each state is represented by a circle labeled with an ID. The associated output is located below and the insensitivity time to the left, in blue. The arrows represent the transitions, marked with the input that triggers them (r, n, e). The initial state is indicated by an arrow with no origin. Each of the *M* stages is composed by 2 states of type R (Relaxed), 4 states of type N (Normal) and 1 state of type E (Excited).

The music for STM has been designed with a positive valence and a low arousal, according to the directions given by the music therapists. The hypersongs contain four levels from very relaxing to not relaxing. The participant roles can include two types of melodies, a pad, a role of chords, an arpeggiator and a bass. They use major modes and sometimes Dorian. The allowed chords comprise major triad, minor triad, sus2 and maj9. Low tessituras are favored over high pitches and the internal accentuation and rhythmic patterns are simple, as well as the textural evolution and the compositional structure, trying to favor repetitiveness.

### 4.4.4 Sport

@jog

@jog is a mobile application that performs podometry to detect the runner's current leading pace and plays music accordingly in real time. The user interface is simple, the screen for the current session shows the time spent and a big play/pause button in the middle. The setting screen allows you to establish the session time and choose between three difficulty levels. The easy mode just plays music fitting the user's real-time running rhythm. In the regular mode the application computes the average pace and, if the current pace falls under a certain percentage of the average, it plays music slightly more intense than it would correspond, to encourage the user to maintain at least the initial rhythm. The hard mode includes an extra behavior: eventually the system increases the music intensity, in order to encourage a higher performance. This boost state lasts for a short period of time, after which the automaton returns to the normal state. If the boost took effect and the user is running at a higher pace, the normal state actually plays music that is more intense than before. A visual representation of the state machine for the hard mode is shown below. The regular mode is identical, just removing the states B, N2 and the corresponding transitions; and the easy mode consists of only the state N1.

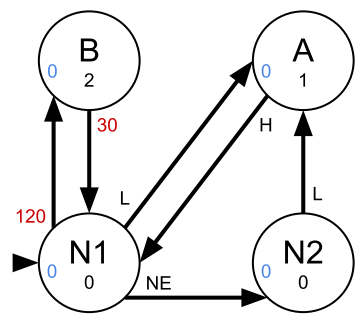

**Figure 4.19.** Visual representation of the automaton for @jog (*hard* session). Each state is represented by a circle, labeled with an ID. The associated output is located below and the insensitivity time to the left, in blue. The arrows represent the transitions, marked with the input or the null-input timeout that triggers them (red). The initial state is indicated by an arrow with no origin. The input NE is associated with an internal counter of times to encourage a higher pace to the user. The inputs L and H are associated to the present pace compared to the average pace during the session.

Detecting the leading pace requires the device to be placed attached to the body, for example inside a tight pocket or a bracelet for smartphones. The 3-axis accelerometer signal is filtered, combined and analyzed, using the Short-time Fourier transform, to detect the right current rhythm. A few parameters are adjusted to achieve a good balance between precision and battery consumption.

The current setting includes hypersongs of 25 levels. Apart from musical texture, BPM is the other main changing parameter. It starts from 80 (corresponding to a slow walk) to 200 (an extremely fast sprint). The step between BPM, after a large series of tests, was established in 5.

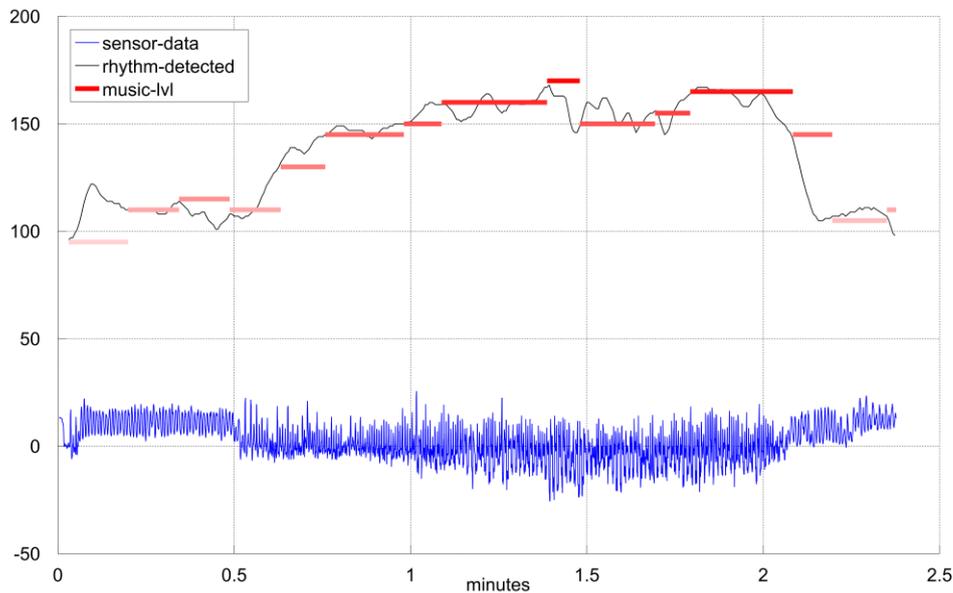

**Figure 4.20.** Recorded session with @jog on easy mode. The accelerometer combined data is shown in blue. The real-time detected leading rhythm (beats per minute) in black; and the level of the music played by the system is presented in different tones of red.

The music style for the @jog has the following specifications: (a) a medium-sized set of roles, including two melodies interchanged depending on the range of BPM, harmonic accompaniments, a bass rhythmically coordinated with the percussion, a percussion set and an FX track; (b) major, Mixolydian and Dorian modes, with major and minor triads and sus2, sus4, maj7 and maj9 chords and different harmonic progressions; (c) very restricted compositional structures and textural variations; (d) small number of alterations between compositional units, pitch shift, tune modulation and modal progression are the only valid operators; (e) appearance of one type of musical fill, with a duration of 1 or 2 measures, in certain moments of the composition to produce a break in the normal evolution, creating expectation.

@jog was versioned as a scenario in the app @life and it was also included as part of the later app *sport*. The idea of detecting the user's pace while doing a particular activity and use this information to play music that acts as a guide or lead for the user's rhythm, was exported to other uses. @bike is a variation targeting indoor and outdoor cycling, where the smartphone is placed in contact with one of the legs and the hypersongs designed required more levels than @jog. Additionally, several other applications and research experiments have been proposed to help to treat different pathologies, such as hypertension, Parkinson, disabling stroke, Alzheimer, multiple sclerosis, asthma, depression, neuromuscular damages or helping to recover from surgery involving the locomotor system.

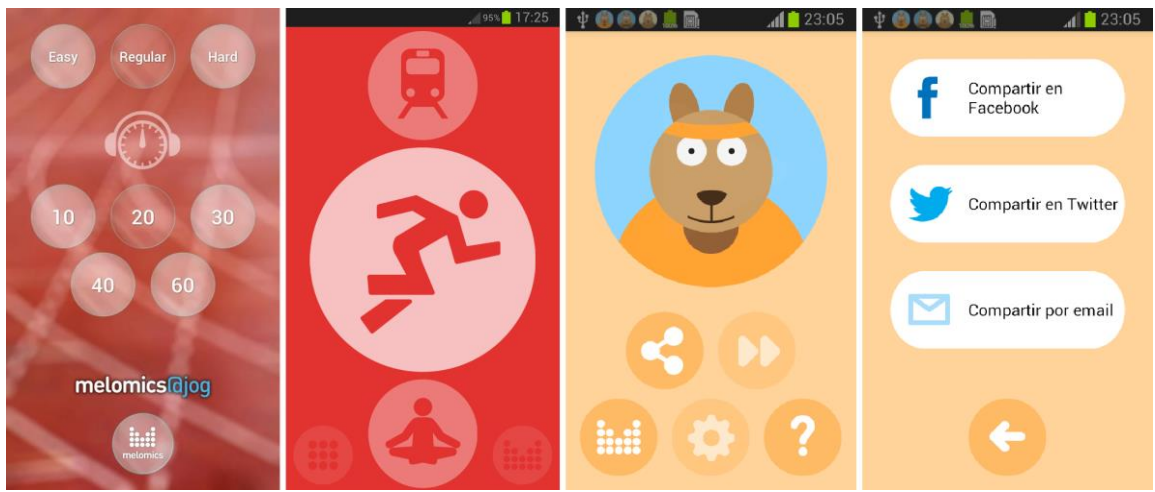

**Figure 4.21.** User interface of the three version of @jog. The original, its inclusion as a scenario of @life and two captures of the app *sports*.

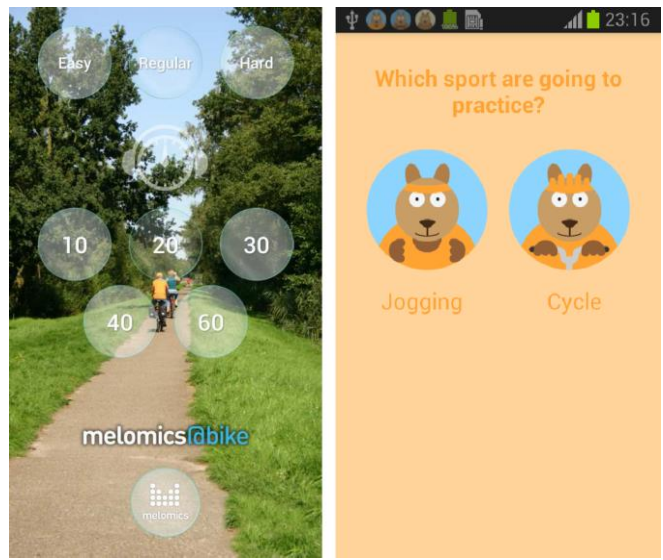

**Figure 4.22.** User interface of @bike. The original version and the menu to choose between the scenarios *Jogging* and *Cycle* in the app *sport*.

### 4.4.5 Other uses

Apart from the applications presented before, we implemented other prototypes with different targets, for example the apps @trip, *school* or @life, the last one including many different scenarios (*Waking-up*, *Starting-out*, *Walking*, *Break*, *Relaxing*...).

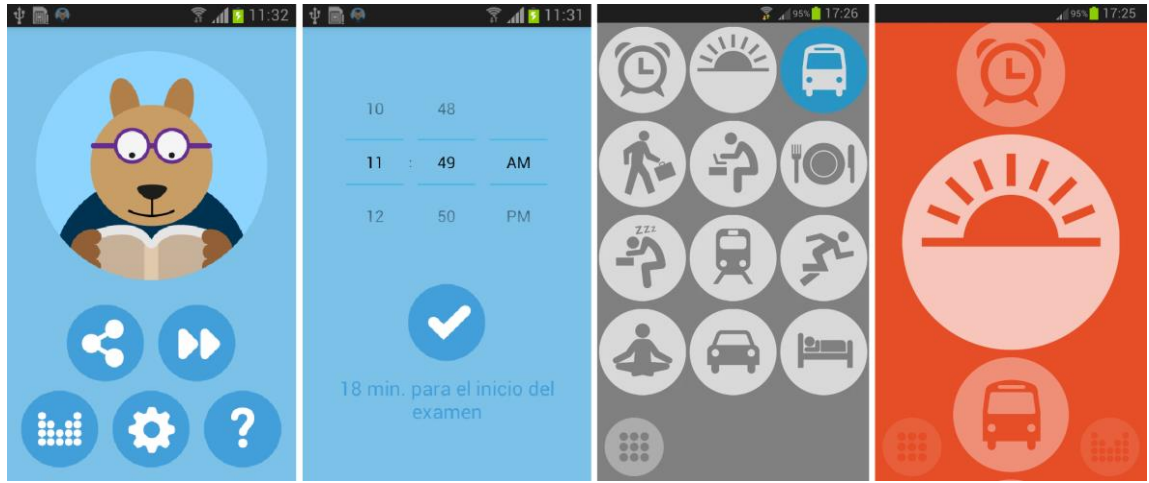

**Figure 4.23.** Screen captures of different apps. Main screen of the app *school*, setting menu of *school*, the menu of all the scenarios in @life and the scenario *Starting-out* selected.

The main focus developing applications that make use of Melomics music was put on the feedback-based adaptive music system. In particular we developed the therapeutic applications, since it was part of the target of the research project MELOMICS. However, there are several other uses of adaptive music that have been considered and might be developed in the future, for example its use in video games, where adapted music can help to create a more immersive experience, considering thematic, concrete scenarios and real time events. Another related example is its use to create movie soundtracks. The method would include an automatic or semi-automatic pre-analysis of frames, to establish how the music should evolve along the scenes of the movie. One last proposal, called *IDMelody*, would consist of producing short pieces of music that could be acquired by people or brands, for example, and become part of their identity. These pieces could actually be hypersongs, with different versions that the owners could use in social apps, for example, to express their current mood or anything else.

# Chapter 5

# Conclusions

## 5.1 Summary and contributions

In this dissertation we have presented the compositional system named Melomics, a tool that can produce objectively novel and valuable musical pieces without using any pre-existing material, but trying to model a creative process instead. In particular, we have described the design of the two versions of the compositional algorithm, based on a bioinspired method. We have detailed the main mechanisms that are involved during the execution and the way the information is stored along the process. We have shown how music in several styles can be produced. We have evaluated those results from different approaches and compared them with traditional music made by human composers. Finally, we have exposed a way to take advantage of this actual music being produced by a computer, consisting of a completely new way of distributing and playing music.

The way Melomics has been designed, combining formal grammars and evolutionary algorithms, and the lack of needing any existing dataset of compositions bring together a set of advantages:

- The product is innovative, since the method is not based on imitation and has complete freedom to explore the search space defined by the –either restrictive or looser– input rules. These act merely as checks that the produced samples comply with the commission. Apart from the fitness and the evolutive mechanisms, the other essential part of the system that enables this free but efficient search is the implicit encoding based on formal grammars. They impose a hierarchical structure and favor behaviors such as repetition, satisfying some of the basic requirements of music composition.

- The syntax to write the genomes –and hence the music– is (a) highly expressive, since it allows the representation of any music piece that can be expressed in regular music notation; (b) flexible, any musical sequence can be written in infinite forms; (c) compact, meaning that in spite of including all the compositional and performing information, it consumes between a half and a third less storage than the equivalent in MIDI format, between a third and a quarter of the corresponding MusicXML file and definitely less than any audio based format; and (d) robust, meaning that if a genome is altered in any way, not only it still produces a valid piece, but it also shares many elements in common with the original, as being a mutation of it.

- The system is affordable: (a) to set up, since there is no need to search and obtain samples from any external source; (b) regarding memory space, since there is no need to store or move large amounts of data to train; (c) while it is true that it requires the intervention of an expert to set up the rules and achieve convergence to a particular style, once it is done, the execution is not very computationally demanding. Using a single CPU thread on a current computer, both the atonal and the tonal systems produce roughly one genuine composition in their most complex style in 6.5 min; and it is possible to run in parallel as many of these tasks as desired.

Melomics has produced *Opus #1*, which can be considered the first piece of professional contemporary classical music composed by a computer in its own style; it has also created the first full-scale work entirely composed by a computer using conventional music notation, *Hello World!*; it has produced a large repository with music of many styles, accessible in the usual symbolic and audio formats, under a CC0 license; and two albums, Iamus album, with contemporary music, and 0Music, with music in more popular styles.

We developed a tool that allows specifying a musical style by filling in a questionnaire of specifications. It includes global values, structural directions and more specific parameters, which Melomics can use to produce genotypes and drive the whole compositional process. The way the form has been designed makes it possible for people with just a little bit of musical knowledge to provide directions to the system to create music. We confirmed that when different people were asked to specify the same musical target the instructions given through the questionnaire were very similar and hence the music obtained. The tool was actually exported to a web form where the answers are directly transcribed to an input file that can be interpreted by the system. This is the first step of a future utility that can be made publicly available for the community of users.

Finally, we have presented the hypersong, a musical structure comprising a musical theme and a bundle of versions of it, all of them fragmented in the time, giving place to a two-dimensional array of music. This concept let us develop a number of applications based on creating an adaptive flow of music, which is only practical if it can be created by automated means. This technology was also made publicly available in the form of an API, fed by a repository of standard music and a collection of hypersongs, making possible for the community to develop new applications.

Automated music composition being fully developed and mainstream will have many implications. Maybe an obvious one is that the general cost of music production will be reduced, especially when the intervention of a human being is not required per se, but just to provide a few directions at the beginning, which are most of the cases. We might witness a change in the paradigm of how we understand music, disengaged in two categories: one viewed as a form of expression of the feelings, emotions, ideas or thoughts of the author and another one more focused on the audience. To compare it with something, it could be similar as it happens with photography, understood as a form of art, as opposed to viewed as a practical tool to remember or register things, people or places. At the same time, automated composition is, and will be more in the future, a useful tool for human artists, to provide them with samples and variations that they can incorporate in their creations or use as brainstorming material. It will have an impact in the industry, replacing human musicians in many niches. The first will be tasks that can be automated more easily, such as composing background music, music for advertising, videogames, etc. Essentially when composing music is done almost exclusively with the audience in mind and the musical specifications are already very narrow. Ultimately, automated systems will cover niches that are not fit for human musicians. For example, a new kind of personalized advertising, through music adapted to each audience, in the same way that Netflix or YouTube use different covers for the same content according to the user profile; also, the concept of adaptive music, introduced in this thesis, and probably many other applications that we have not imagined yet.

## 5.2 An experiment in the real world

Novelty and, especially, value of Melomics pieces as true musical creations would ultimately need to be determined by the –unbiased– judgement of people and experts, as it is always done, in any form of art. The Turing Test was designed to identify traces of thought in computer processing, but its interactive nature makes the adaptation to a musical version difficult. Nevertheless, the underlying principles remain a valid inspiration for new tests that measure how close artificial music from human music is.

Melomics is set up, if desired, by using a collection of abstract parameters, such as amount of dissonance, repetitiveness, instruments, duration of the composition, etc., which are then translated to genomes that undergo the evolving process. It neither requires any existing creative input from the user nor does it require a repeated interaction from them (i.e., interactive fitness function). The initial input acts merely as musical directions, nothing more than a musician would be given to compose.

However, normally the composition of music is regarded as a creative process where humans can express and evoke sensations. Then, could we ever consider this artificially-created music, actual music? The definition of music is a tricky question, even if only instrumental music is being considered. For example, thinking of it as an organized set of sounds can be too broad an understanding, as it includes a collection of sounds that, while organized, are not music and, at the same time, it takes the risk of excluding the compositions made by Melomics, since until they are synthesized –or performed– they are just scores. Two properties that are usually required in the definition of music are either tonality –or the presence of certain musical features– and an appeal to esthetic properties. The first can be guaranteed by Melomics through the encoding and the fitness function. The second property is certainly more complex to be respected since we are unable to ascribe any esthetic intention to the software. These properties are not the only way to define music and, indeed, they have problems to capture all and only what humans usually consider music. There are multiple possible definitions, each one with strengths and weaknesses, and their discussion is more of the domain of the philosophy of music (Kania, 2007). We needed to take a more practical approach, instead of looking at Melomics from a rather philosophical perspective. If we moved the focus from the composer to the listener, the fact that the composer is a machine, is not as relevant. Among the different comments on Melomics music, Peter Russell's positive judgement (Ball, Artificial music: The computers that create melodies, 2014) was interesting to us as he, who had no knowledge of the composition's origin, did not express any doubt on the musical nature of the piece. This encouraged us to follow this approach to establish whether the system produces actual music, looking at the opinion of critics and the general public. If they considered the end result as music, then it could be considered music.

In contrast with previous works, the experiment presented illustrates a controlled and rigorous methodology for a trial performed over a large sample of participants. The first question of the questionnaire ("Would you say that what you have listened to is music?") was motivated to measure potential differences perceived from the original music sample composed by a musician and the other samples. In this sense, it is worth noting that the natural sounds sample was classified as music by 41.7% of the professional musicians. In contrast, both music samples were classified as music by most of the subjects (over 90%). With respect to the capability of eliciting mental images, it also endorsed the hypothesis of the experiment, as there was no significance when evaluating the musical samples, eliciting images in around 50% of the subjects. Regarding the natural sounds sample, this measure raised to 90%, both in the specific question and in the presence of terms in the descriptions. This is not surprising: natural sounds are associated with concrete and recognizable physical sources, while music is produced by musical instruments and images arise in the form of memories not directly related to the perceived sounds, following a more abstract channel. The study of qualitative data confirmed this fact: most of the terms used in these descriptions fit the category "nature", differing from those used to describe the musical samples, which did not point to any of the defined categories. These results highlighted the difference between natural sounds and the presented music, which appears to generate a more complex and wider set of mental images, with independence of the musical training of the listener or who composed or interpreted the pieces. With respect to the evoked emotional states, one of the most revealing results was that natural sounds had a significant rate of 89.7% of descriptions assigned to the state "calm", in contrast to the music recordings, with a maximum rate of 47.3% in this category, even though the music style presented was arguably calm. As in the case of mental images, all the musical pieces seemed to elicit a wider range of emotions, with independence of the listener, the composer or the interpreter. The second interesting result came from the study of valence in the descriptions. They turned out to be significantly positive ($p < 0.001$) when describing sounds of nature, while in the case of music, with no relevant differences among the groups of study, they also elicited unpleasant feelings.

The final part of the test confirmed that the subjects were unable to distinguish the source of the composition. Even if it was done only with two different musical pieces, the sample was wide and the fact that about a half of it was made of professional musicians is an indicator of the robustness of the conclusions, which rejected the null hypothesis, suggesting that computer compositions might be considered as "true music". This was, of course, a first result in this line of quality assessment and the model can be extended to different musical styles.

## 5.3 Future work

Because of the multidisciplinary nature of this thesis and the wide scope defined, which included building a music composition system with two variants, a formal assessment of its outcome compared to human productions and the definition of a new kind of applications based on automated composition; there are many additional areas that we want to tackle and could be developed as a continuation of this work. They include both research topics and business cases; below we introduce some of them.

## 5.3.1 Extend Melomics capabilities

In spite of the possibility of representing any kind of regular musical content with Melomics, in practice it can only produce music in the styles that have been specified. There are some musical genres that would require further work to incorporate them into the system. Examples of these are jazz, country, rap, rhythm & blues, rock or folk music. Some would only require an extended analysis and find the best way to specify them, through the style-tool, into the system. However, some other styles and genres come with a bigger problem: they include vocal tracks. We have little experience with introducing voice to compositions. We did a few experiments in choral music (Sample 5.1) where the lyrics consisted either of a sequence of "Aahs" and "Oohs" or random fragments chosen from some available texts (like the Bible). We would need to tackle the problem of building lyrics, which is an ongoing research (Pudaruth, Amourdon, & Anseline, 2014) (Malmi, Takala, Toivonen, Raiko, & Gionis, 2016) and the major problem, and far from being resolved, of vocal synthesis, much more complex than normal speech synthesis.

- Sample 5.1. Experiment in choral music

Also related to the management of styles, even though the templates can be reused as a reference to build different known styles or even to try and explore new things, this potentially distributed but manual procedure seems limited to search the whole music space. It would be interesting to provide the system with the capability of extracting global specifications, in the terms used in our style templates, from a set of input samples. The system could then generate new pieces in that style. This would enable a new way to interact with the system. Users could provide a set of keywords that Melomics would use to search for music samples, initializing the composition process. A second more ambition step would be to build a new system to encode the style definitions, in a similar way that we did for the music itself. Then, we could develop an additional layer of abstraction, for example based on another genetic algorithm or a neural network, that would deal with these styles. This could enhance, in many ways, the search in the music space.

A last extra functionality that we would like to implement is automated reverse engineering. All the presented experiments in genetic engineering involving existing human themes were done manually. The possibility of incorporating them automatically into the system would open a whole spectrum of applications, many already commented. However, its implementation would not be trivial since, in order for it to work as expected, the new module should be able to study the compositional structure of a musical piece and represent it properly taking advantage of the genetic encoding.

## 5.3.2 Study of Melomics and the music space

Apart from the initial analysis and the perceptive experiment presented, there are several extra studies that we would like to address:

- Continue the numerical analysis of Melomics music. Revisiting and evolving the concept of musical distance; using principal component analysis and more advanced procedures; using larger and more sets of pieces; taking advantage of newer MIR tools, such as the updated version of jSymbolic (McKay, Cumming, & Fujinaga, 2018); and comparing Melomics' with other existing computer music.
- Related to the previous point, we could study the definition of styles, substyles and how they restrict the musical space of search. Using the new concept of distance and other findings to characterize it.
- The evolutionary algorithm implemented in Melomics is designed to mimic the creative process of iterating a set of ideas (the gene pool). Apart from that, the system provides themes faster in later iterations of the gene pool, presumably because the gene pool is formed with themes that are variations of some others that already passed the filter at phenotype level. A study of convergence would be very interesting. An initial approach can be a comparison between the current evolutionary process and a random search.
- Additional perceptive tests. We could include more styles, more themes and new experiment designs. For example, provided that we can design experiments where the automated synthesis in its current state would not create a bias, we could propose studies where each participant can get a unique computer-generated composition.

### 5.3.3 Adaptive music applications

As discussed, there are a lot of cases that can benefit from music being produced automatically, especially those that are not the focus of composers. A particular example explored is adaptive music, which we think is a powerful idea and we tried to exploit it with a few initial experiments. However, apart from some studies related to therapy, we have not performed experiments using any of the rest of the adaptive systems developed. Many of these applications could be the subject of formal studies, in particular the ones designed for sports, driving, improving of concentration or any of the other therapy apps.

The potential of generative music that complies with human standards also opens the possibility of tailored music, which would consider personal preferences, as well as particular needs or goals, such as responding in real time to the evolution of physiological signals. We would like to explore the concept of adaptive music further, to what we call "personal music", which refers to a way of providing a flow of music that (a) is automatically responsive to any context, not just to a very concrete use-case, and (b) fits the user's preferences. The music would be completely tailored to the person and the situation. It would be necessary to develop the methods and ideas presented in Chapter 4. We made an initial attempt to provide music for different situations and change the context automatically with the mentioned app @life. To fully develop the personal music idea it would be necessary to work on different fronts, for example (a) find methods to accurately detect the present user's context or scenario; (b) improve and develop methods to characterize the current state for each scenario; (c) study the music appropriate for each use-case considered and circumstances; (d) ideally, improve the hypersong concept, especially the music versioning principles and the mechanisms to control the flow of music; (e) find an adequate procedure to set up and update users music preferences, at the same time that is able to propose new music to them; and (f) develop the way the music is specified into the system, in order to reach new genres and be able to satisfy a wider spectrum of tastes.

# Appendix A

# Implementation details

## A.1 Genome in the atonal model

The genome of a composition is implemented as a text file identified with the theme's name and the extension ".gen". This text file is organized in two levels, the first line corresponds to three general parameters separated by blank spaces: (1) the initial iteration value for the grammar; (2) the initial duration, which also serves as duration of reference; and (3) the duration step length, which is the amount of change applied to the current duration when the corresponding genes appear while interpreting the resulting string. These values are read by MATLAB as of type double. The second level is formed by the subsequent lines, each of them containing five track parameters and a production rule. The parameters are separated by blank spaces and read as doubles. They are (1) the iterativity ($r_i$), indicating the way the production rule will stop being applied (as explained in section 2.2.2.1 Genome); (2) the duration factor, affecting the note durations of the corresponding track; (3) the instrument ID, to fetch the musical instrument in charge of performing that track; (4) the musical scale to be applied in the interpretation into musical notes, encoded as an integer; and (5) tessitura, or range of allowed MIDI pitches for the track, expressed as two integers. Production rules are only represented by their right side, since the left sides contain one symbol only and they appear just once, because of being a deterministic grammar. We code these symbols with the reserved character "#" plus an identification number (#0, #1, #2...). The production rules are placed sorted by this ID in ascending order.

Next, we introduce the different symbols that can appear in a production rule and their respective meaning during the interpretation of the final string. These symbols or genes are encoded as one reserved character followed by a numeric ID. It was done this way because there were not enough single characters in the keyboard to represent them all. At genome level, the musical effects are expressed and dealt with using a generic identifier. During the interpretation they are instantiated depending on the actual musical instrument.

| # | | Used to represent structural units or notes (see section 2.2.2 Model for atonal music). No combination is reserved, but as a standard procedure we use #0 for the zygote-rule, #1 to represent the musical rest and #2 to introduce the global structure of the composition. In the interpretation of the resulting string, these genetic symbols produce a musical note in the associated track. They acquire the current values of each of the musical properties that are being computed. |
|---|---|---|
| $ | ID | Used to code the music operators (see section 2.2.2 Model for atonal music). |
| | 1 | Increase pitch value one step in the scale. |
| | 2 | Decrease pitch value one step in the scale. |
| | 3 | Increase duration one step (according to the global parameter). |
| | 4 | Decrease duration one step (according to the global parameter). |
| | 5 | Push current pitch and duration values in the *pitch stack*. |
| | 6 | Pop pitch and duration in the *pitch stack*. |
| | 7 | Push current time position in the *time stack*. |
| | 8 | Pop time position saved the *time stack*. |
| | 9 | Apply the articulation legato to the next compositional element. |
| | 10 | Apply the articulation legato until indicated or the end. |
| | 11 | Terminate the application of legato. |
| | 12 | Apply the articulation portato to the next compositional element. |
| | 13 | Apply the articulation portato until indicated or the end. |
| | 14 | Terminate the application of portato. |
| | 15 | Apply the articulation staccato to the next compositional element. |
| | 16 | Apply the articulation staccato until indicated or the end. |
| | 17 | Terminate the application of staccato. |
| | 18 | Apply the articulation accento to the next compositional element. |
| | 19 | Apply the articulation accento until indicated or the end. |
| | 20 | Terminate the application of accento. |
| | 21 | Apply the articulation tenuto to the next compositional element. |
| | 22 | Apply the articulation tenuto until indicated or the end. |
| | 23 | Terminate the application of tenuto. |
| | 30 | Apply effect1 to the next compositional element. |
| | 31 | Apply effect1 until indicated or the end. |
| | 32 | Terminate the application of effect1. |
| | 33 | Apply effect2 to the next compositional element. |

| | |
|---|---|
| **34** | Apply effect2 until indicated or the end. |
| **35** | Terminate the application of effect2. |
| **36** | Apply effect3 to the next compositional element. |
| **37** | Apply effect3 until indicated or the end. |
| **38** | Terminate the application of effect3. |
| **39** | Apply effect4 to the next compositional element. |
| **40** | Apply effect4 until indicated or the end. |
| **41** | Terminate the application of effect4. |
| **42** | Apply effect5 to the next compositional element. |
| **43** | Apply effect5 until indicated or the end. |
| **44** | Terminate the application of effect5. |
| **45** | Apply effect6 to the next compositional element. |
| **46** | Apply effect6 until indicated or the end. |
| **47** | Terminate the application of effect6. |
| **61** | Apply modulator of dynamics crescendo to the next compositional element. |
| **62** | Apply crescendo until indicated or the end. |
| **63** | Terminate the application of crescendo. |
| **64** | Apply modulator of dynamics diminuendo to the next compositional element. |
| **65** | Apply diminuendo until indicated or the end. |
| **66** | Terminate the application of diminuendo. |
| **67** | Apply fermata to the next compositional element. |
| **68** | Apply fermata to all following notes until indicated or the end. |
| **69** | Terminate the application of fermata. |
| **70** | Apply the property chord to the next compositional element. Instead of a single note, the corresponding instrument performs a chord, built depending on instrumental parameters and the current value of the "property argument" (see meaning of "@" below). |
| **71** | Apply the property chord to all notes until indicated or the end. |
| **72** | Terminate the application of chord. |
| **73** | Apply the property grace to the next compositional element. Instead of a single note, the corresponding instrument performs an additional grace note, built depending on instrumental parameters and the current value of the "property argument" (see meaning of "@" below). |
| **74** | Apply the property grace to all notes until indicated or the end. |
| **75** | Terminate the application of grace. |
| **92** | Apply dynamic *pianississimo*. |
| **93** | Apply dynamic *pianissimo*. |
| **94** | Apply dynamic *piano*. |
| **95** | Apply dynamic *mezzo-piano*. |
| **96** | Apply dynamic *mezzo-forte*. |
| **97** | Apply dynamic *forte*. |
| **98** | Apply dynamic *fortissimo*. |
| **99** | Apply dynamic *fortississimo*. |
| **100** | Push value of pitch and duration in the *global stack*. |
| **101** | Pop value of pitch and duration saved in the *global stack*. |
| **102** | Apply the tempo indicated with the current value of the "property argument" (see meaning of "@" below). |
| **110** | Apply the property arpeggio to the next compositional element. Instead of a |

| | | |
|---|---|---|
| | | single note, the corresponding instrument performs an arpeggio, built depending on instrumental parameters and the current value of the "property argument" (see meaning of "@" below). |
| | 111 | Apply the property arpeggio to all notes until indicated or the end. |
| | 112 | Terminate the application of arpeggio. |
| | 113 | Apply mordent to the next compositional element. |
| | 114 | Apply mordent to all following notes until indicated or the end. |
| | 115 | Terminate the application of mordent. |
| | 116 | Apply inverted mordent to the next compositional element. |
| | 117 | Apply inverted mordent to all following notes until indicated or the end. |
| | 118 | Terminate the application of inverted mordent. |
| | 120 | Apply pitch alteration to the next compositional element, to the position and with a value codified with "property argument" (see meaning of "@" below). |
| @ | | The numeric string following this character is considered as the current value of the "property argument", which serves as a parameter for some musical properties. |

**Table A.1.** Reserved symbols in the atonal genomic model.

# A.2 Genome in the tonal model

The genome of a composition is implemented with a text file identified with a theme's name and the extension ".evg". This text file is organized in different levels by the separator "|" and each level with sections divided using the character ";". The first section of the first level contains the global parameters: initial dynamic, initial tempo, initial chord, initial mode, base duration, duration step length and musical style, read in MATLAB as doubles. Each of the next sections in the first level corresponds to a different musical track, containing all the needed parameters: MIDI ID, initial pitch and tessitura, range of velocities, role ID and instrument ID. The following levels correspond to the production rules of the grammar. Each level comprises as many production rules (represented only by their right side) as different non-terminal symbols there are in all the production rules of the previous level (the first grammar level has only one rule). Symbols are matched with rules in order of appearance.

Next, we introduce the different reserved symbols that can appear in the production rules, followed by their respective meaning when the interpretation of the final string takes place.

| | |
|---|---|
| [ | Push in a stack the current value of pitch, harmonic root and duration. |
| ] | Pop from the stack the last value of pitch, harmonic root and duration. |
| < | Push the current time position, value of pitch, harmonic root and duration in a stack. |
| > | Pop from the stack the last saved time position, value of pitch, harmonic root and duration. |
| *N* | Increase the counters pitch and harmonic root in one unit. |
| *n* | Decrease the counters pitch and harmonic root in one unit. |
| *T* | Increase the current tone shift in one unit. |
| *t* | Decrease the current tone shift in one unit. |
| *R* | Increase current duration one step (according to the global parameter). |
| *r* | Decrease current duration one step (according to the global parameter). |
| *M index, notes, shift, behavior* | Instead of the current pitch, the next instrument symbol interprets the chord resulting from applying the properties of the current harmony and the parameters coming within this operator. *index* fetches a specific chord defined in the musical style; *notes* encodes the actual notes of the selected chord to be played, *shift* indicates a possible additional (local) tonal shift and *behavior* encodes extra directions to compute the final notes. |
| *v accent* | Apply the indicated *accent* to the next note or chord. |
| *W origin, target* | If *target* is 0, apply dynamic indicated by *origin* until indicated or the end. If *target* is a positive value, apply progressive dynamic modulation from *origin* to *target* until indicated or the end. |
| *C role, position, type* | Apply a rhythmic alteration indicated by *type*, in the track indicated by *role* and note given by *position*. |
| *V role, position, type* | Transform a note into rest or vice versa in the track indicated by *role* and note given by *position*. *type* encodes directions to build a note if the fetched position was a rest. |
| *p role, position, n, t* | Apply pitch alteration to the note fetched by *role* and *position*, with a variation in the scale indicated by *n* and a tonal shift indicated by *t*. |
| *w start, end* | When rewriting the next non-terminal, only the symbols in the production rule enclosed by *start* and *end* will appear. |
| *X role* | Prevent the emergence of the track indicated by *role* in the next compositional element (the rewriting of the next non-terminal). |
| *Y durTotal, durAnacrusis* | Produces a musical anacrusis, deleting the required |

| | previous elements in the string, which is computed by using the given arguments. |
|---|---|
| $ *code*, *args* | Apply a predefined musical operation coded in the parameter *code* with the arguments given by *args*. |

**Table A.2.** Reserved symbols in the tonal genomic model.

In this model, to make genome reading easier, we use keyboard characters (not reserved for operations) to code both the non-terminal symbols and the terminal symbols corresponding to the musical tracks. We can dispose of enough keyboard characters because the additional musical properties and the effects, which appear less frequently, are coded using the operator "$". Also, because we imposed an explicit hierarchy in the grammar with no backtracking, so these non-reserved symbols can be reused from one level to another.

# Appendix B

# Computers used

## B.1 Workstation

We used a high-end workstation described in the table below for daily work. This included bibliographic research, software development with MATLAB and Python and testing the music generation system. The computer is connected through a high-speed internet connection provided by the university, which facilitates moving the high amount of musical data between the different computers and clusters. We used Windows, as the main operating system, and a Debian GNU/Linux inside a virtual environment.

| | |
|---|---|
| **Processor** | Intel® Xeon® ES5645 2.4GHz, 6 cores, 2 threads per core |
| **Memory** | DDR3 Triple channel, 12GB |
| **GPU** | NVIDIA® Quadro® 4000, GDDR5 2GB |
| **Hard Drive** | Main system: 500GB<br>Work directory: 2x2TB with a RAID 1 |
| **Network** | Gigabit Ethernet |

**Table B.1.** Workstation specifications.

## B.2 Computer clusters

We designed and acquired two different computer clusters that provided the computational power required to produce the music content present in the web repository, converted and saved in different formats (tens of thousands of them are available in the web repository and even more are stored as a backup repository and for testing purposes); and to carry out many other compositional experiments.

Iamus is a cluster formed by 11 nodes that add up to 352 cores in total, 704GB of main memory and 70TB of storage. This cluster was mostly used to run experiments with the atonal system. It is connected through the university network and accessible via a SSH interface. It runs a Debian GNU/Linux on all its nodes and uses TORQUE and MAUI to manage the resources and to schedule tasks, mostly consisting of MATLAB and Python processes. For an extended description see Pérez-Bravo's dissertation (Perez-Bravo, 2014).

|  |  | Per node | TOTAL |
|---|---|---|---|
| **Processor** | AMD Opteron™ 6128 2GHz, 8 core | 4 (32 cores) | 44 (352 cores) |
| **Memory** | DDR3 | 64GB | 704GB |
| **Storage** | Master node<br>Worker nodes | 10TB<br>6TB | 70TB |
| **Network** | Gigabit Ethernet<br>Fast Ethernet for IPMI | | |

**Table B.2.** Iamus specifications.

The second cluster, called Melomics109,[52] was mainly designed to run the tonal version of Melomics and it was installed in Centro de Bioinnovación in the Parque Tecnológico de Andalucía,[53] integrated as part of the university supercomputing network. It is a cluster with 40 nodes, adding up to 960 processors, 3840 GB of main memory and 320 TB of total storage using a RAID 1+0.

---

[52] https://en.wikipedia.org/wiki/Melomics109 (accessed on May 16, 2021)
[53] http://www.scbi.uma.es/ (accessed on May 16, 2021)

|  | | Per node | TOTAL |
|---|---|---|---|
| **Processor** | AMD Opteron™ 6176 2.3GHz, 12 core | 2 (24 cores) | 80 (960 cores) |
| **Memory** | DDR3 PC3-10600R-9 RDIMM | 96GB | 3840GB |
| **Storage** | 2TB 6G SAS 7.2K 3.5in DP MDL HDD | 4x2TB | 320TB |
| **Network** | Gigabit Ethernet, 3 ports | | |

**Table B.3.** Melomics109 specifications.

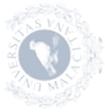

# Appendix C

# Execution statistics

## C.1 Atonal system

The musical style *CtClarinet* (for Contemporary Clarinet) is used in this test to show the quickest possible run, in terms of computational cost, with Melomics atonal system. This configuration gives place to compositions with a duration ranging 160 to 195 seconds and having only one track played by the clarinet. The other style used in the test, representing the most computationally expensive style in the atonal model, is *CtSymphonicOrchestra* (for Contemporary Symphonic Orchestra), which gives place to orchestral pieces with a duration ranging 120 to 320 seconds and between 25 to 43 instruments playing.

In this section we show the times spent by the two computer configurations (workstation and cluster) to produce one composition, using a single core and one thread. Measures for the cluster configuration were taken only from Iamus since this and Melomics109 count on a similar architecture. We provide the times to reach the different compositional stages and produce the corresponding output. In each case the system produces only the information required to generate that particular output. The dependencies are listed below:

- GEN is the first form of compositional representation.

- INT refers to the internal representation. It can be stored as a MATLAB or a Python variable. It only requires the genome to exist in advance.
- XML is the MusicXML file, which can be loaded from any musical editor supporting this standard. It requires the internal representation.
- PDF is the printed score that can be directly read by musicians. It is produced from a MusicXML file, but this is different from the one cited above provided as the standard MusicXML output. That being the case, it only depends on the internal format.
- MID is the MIDI format of symbolic notation; it depends on the internal representation.
- WAV is the earliest audio format that can be synthesized within the system. It uses the MIDI standard as an intermediate step, but these files are different from the MID cited above, so it strictly depends on the internal representation.
- MP3 is the compressed audio obtained from the WAV file.

|  |  | GEN | INT | XML | PDF | MID | WAV | MP3 |
|---|---|---|---|---|---|---|---|---|
|  | AVG | 114.87 | 121.40 | 122.50 | 137.68 | 126.76 | 139.01 | 147.27 |
| CtClarinet | MAX | 134.57 | 140.92 | 142.05 | 157.51 | 146.30 | 158.64 | 166.91 |
|  | MIN | 93.13 | 99.59 | 100.64 | 117.30 | 104.77 | 116.84 | 125.19 |
|  | AVG | 256.20 | 394.45 | 419.91 | 534.70 | 462.42 | 953.01 | 1033.67 |
| CtSymphonic Orchestra | MAX | 562.72 | 743.17 | 775.28 | 912.86 | 825.26 | 1341.09 | 1633.25 |
|  | MIN | 17.46 | 125.03 | 139.37 | 221.97 | 166.03 | 559.81 | 597.12 |

**Table C.1.** Time spent by the atonal system in a workstation. Average, maximum and minimum values (expressed in seconds) were taken after 50 runs.

|  |  | GEN | INT | XML | PDF | MID | WAV | MP3 |
|---|---|---|---|---|---|---|---|---|
|  | AVG | 124.89 | 130.16 | 131.79 | 151.58 | 136.35 | 160.30 | 167.96 |
| CtClarinet | MAX | 144.54 | 151.16 | 151.26 | 170.38 | 158.69 | 178.78 | 186.97 |
|  | MIN | 99.59 | 106.57 | 107.89 | 127.66 | 112.56 | 133.57 | 143.16 |
|  | AVG | 282.50 | 427.53 | 453.22 | 576.07 | 492.21 | 1011.11 | 1092.35 |
| CtSymphonic Orchestra | MAX | 605.16 | 781.26 | 818.05 | 959.71 | 870.49 | 1422.04 | 1716.48 |
|  | MIN | 23.11 | 136.24 | 150.54 | 241.47 | 180.62 | 598.90 | 635.71 |

**Table C.2.** Time spent by the atonal system in a computer cluster. Average, maximum and minimum values (expressed in seconds) were taken after 50 runs.

## C.2 Tonal system

The internal musical style *Relax* is used in this test to show the quickest possible run, in terms of computational cost, with Melomics tonal system. This configuration gives place to compositions with a simple structure, only 5 musical roles and a duration ranging between 100 to 600 seconds. The other style used in the test, representing the most computationally expensive style in the tonal model, is *Disco02*, which gives place to compositions in a disco style, comprising 19 to 22 different roles, a duration between 90 to 210 seconds and a rather complex compositional architecture.

### C.2.1 Generation of one composition

In this section we show the times spent by the two computer configurations (workstation and cluster) to produce one composition, using single core and thread. Measures for the cluster configuration were taken only from Iamus, since this and Melomics109 count on a similar architecture.

We provide the times to reach the different compositional stages and produce the corresponding output. In each case the system produces only the information required to generate that particular output. The dependencies are listed in section C.1 Atonal system. MID_Synth format refers to a MIDI file that is used together with additional information in the internal representation to produce the WAV audio file. Statistics for the audio synthesis are not given in this case, since this component is similar to the atonal synthesis subsystem.

|  |  | GEN | INT | MID | MID_Synth |
|---|---|---|---|---|---|
| Relax | AVG | 0.40 | 24.90 | 31.74 | 31.74 |
|  | MAX | 0.81 | 60.52 | 69.52 | 69.52 |
|  | MIN | 0.23 | 5.44 | 9.72 | 9.72 |
| Disco02 | AVG | 18.59 | 387.73 | 773.01 | 436.30 |
|  | MAX | 25.37 | 1117.00 | 2210.75 | 1204.90 |
|  | MIN | 11.25 | 77.58 | 159.24 | 102.83 |

**Table C.3.** Times spent by the tonal system in a workstation. Average, maximum and minimum values (expressed in seconds) were taken after 50 runs.

|  |  | GEN | INT | MID | MID_Synth |
|---|---|---|---|---|---|
| Relax | AVG | 0.40 | 37.35 | 47.56 | 47.56 |
|  | MAX | 2.26 | 143.65 | 158.66 | 158.66 |
|  | MIN | 0.23 | 6.58 | 13.63 | 13.63 |
| Disco02 | AVG | 17.80 | 387.13 | 796.92 | 427.73 |
|  | MAX | 35.48 | 1249.22 | 2525.67 | 1308.68 |
|  | MIN | 13.85 | 111.00 | 237.88 | 139.38 |

**Table C.4.** Times spent by the tonal system in a computer cluster. Average, maximum and minimum values (expressed in seconds) were taken after 50 runs.

Statistics for MID and MID_Synth in the style *Relax* are the same because in that case they are the exact same file.

## C.2.2 Parallel generation

In this section we show the performance in compositions per hour that the two computer configurations (workstation and cluster) can achieve using parallelism. Measures for the cluster configuration are taken only from Iamus, since this and Melomics109 count on a similar architecture and the differences are a factor of the size of their hardware. Measures taken in separate executions, on each one targeting a different format as the final output and only producing the required previous representation. The dependencies are listed in section C.1 Atonal system.

|  | GEN | INT | MID | MID_Synth |
|---|---|---|---|---|
| Relax | 16951.15 | 593.39 | 414.38 | 414.38 |
| Disco02 | 475 | 40.29 | 28.95 | 29.06 |

**Table C.5.** Parallel generation in a workstation. Average number of compositions per hour using the workstation (8 of its 12 logical cores) in the styles Relax and Disco02.

|  | GEN | INT | MID | MID_Synth |
|---|---|---|---|---|
| Relax | 1708060.70 | 24190.43 | 17495.62 | 17495.62 |
| Disco02 | 43188.28 | 2102.84 | 1235.59 | 1600.72 |

**Table C.6.** Parallel generation in a computer cluster. Average number of compositions per hour using Iamus (352 cores) in the styles Relax and Disco02.

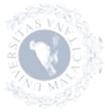

# Appendix D

# Spanish summary and conclusions

En esta sección se proporciona un resumen en español de las líneas desarrolladas en los capítulos principales de la tesis, así como las conclusiones y resultados obtenidos a su finalización.

## D.1 Resumen

### D.1.1 Introducción

La creatividad, o habilidad de producir ideas nuevas y valiosas, comúnmente se asocia al ser humano, pero existen muchos otros ejemplos en la naturaleza donde se puede dar este fenómeno. Inspirados por este hecho, en las ciencias de la computación se han desarrollado numerosos sistemas con propósitos muy diferentes que son capaces de producir resultados mediante este enfoque creativo.

La música, una forma de arte presente durante toda la historia de la humanidad, es el principal campo tratado en esta tesis, intentando aplicar los procedimientos que generan diversidad y creatividad en la naturaleza y en computación.

Proponemos un método de composición basado en búsqueda evolutiva con codificación genética de las soluciones, las cuales son interpretadas durante un complejo proceso de desarrollo, que incluye reglas basadas en conocimiento experto en los distintos niveles de su arquitectura. Este sistema bioinspirado muestra su capacidad creativa y versatilidad en numerosos casos de uso reales y que es indistinguible de la música hecha por humanos, tanto desde una perspectiva analítica, como perceptiva. Este sistema automático también posibilita la aparición de aplicaciones totalmente novedosas, desde herramientas que pueden ayudar a cualquier persona a obtener exactamente la música que desea, a utilidades basadas en nuevos conceptos como la música adaptativa, aplicada a terapia, deporte y muchos otros usos.

## Creatividad computacional

La creatividad es un concepto relativamente reciente. Antes de la edad moderna, esta idea se identificaba más con inspiración divina que con algo de origen exclusivamente humano. Desde entonces, se ha considerado como una de sus características más distintivas. Esto implica por un lado cierto recelo o al menos escepticismo al atribuir creatividad a otras entidades, pero, por otra parte, intentar construir máquinas con esta cualidad representa un gran reto para la ciencia y la ingeniería. En el desarrollo de la informática se ha tratado explícitamente de alcanzar la creatividad en el campo de las artes, normalmente el más asociado a esta habilidad. Sin embargo, también se ha buscado en muchas otras áreas. Un caso conocido es el sistema Watson, de IBM, capaz de procesar lenguaje natural, extraer información, representarla, razonar, aprender sobre ella y producir resultados originales. Este sistema se empleó en el concurso televisivo *Jeopardy!*, donde derrotó a sus oponentes humanos. Posteriormente se usó en otros campos, incluyendo aplicaciones financieras, problemas médicos, consultoría legal o incluso cocina. También se ha observado un creciente interés en el estudio de la creatividad, con algunas propuestas para medirla y también con el fin de distinguir sistemas creativos de otros que simplemente tratan de aparentarlo (Colton, López de Mantaras, & Stock, 2009) (Bringsjord, Bello, & Ferrucci, 2001) (Colton, 2008) (Pease & Colton, 2011).

El sistema que proponemos incorpora diferentes técnicas dentro del campo de la inteligencia artificial para lograr creatividad: (a) Sistemas basados en conocimiento, normalmente implementados en forma de reglas para un determinado dominio. Este enfoque se ha usado previamente en otros trabajos en música (Schaffer & McGee, 1997). (b) Gramáticas formales; un vocabulario de símbolos, con un proceso de reescritura que conduce a un producto final. Este tipo de técnicas se han empleado frecuentemente, por ejemplo, en diseño (Stiny, 1980) (Shea & Fenves, 1997). (c) Métodos de búsqueda, que incluyen multitud de algoritmos de diferente naturaleza y en muchos casos contemplando el uso de heurísticos (Russell & Norvig, 2009) (Campbell, Hoane, & Hsu, 2002).

Nuestro sistema implementa una búsqueda evolutiva, un algoritmo genético en particular. Las composiciones, codificadas como genomas en forma de gramática formal, se desarrollan y después se evalúan usando reglas musicales generales y basadas en estilos. Los genomas que no pasan el test se descartan, pero los que pasan se guardan para ser usados en el futuro.

## Música por ordenador y composición algorítmica

El término música por ordenador se usa para referirse a toda la música que ha sido producida empleando medios informáticos. En los últimos tiempos han surgido nuevas herramientas para ayudar en el proceso de composición: para grabar y reproducir sonidos, mezclar y editar el audio, escribir partituras, sintetizar audio, etc. También existen herramientas para ayudar en los procesos creativos de composición, llamadas CAAC (Computer-Aided Algorithmic Composition), como SuperCollider (McCartney, 2002) o MAX/MSP (Puckette, 2002). Se han inventado nuevos instrumentos musicales, tecnologías de instrumentos virtuales y formatos para almacenar la música de forma simbólica (MIDI o XML) o de forma explícita como onda de audio (WAV, FLAC o MP3). Además, han aparecido nuevos estilos musicales y también se ha desarrollado el campo de la composición automática.

Algunos trabajos formales en composición algorítmica empezaron a aparecer en la década de 1950, por ejemplo un trabajo no publicado de Caplin y Prinz en 1955 (Ariza, 2011), con un generador de líneas melódicas; la conocida y citada Suite de Hiller e Isaacson (Hiller Jr & Isaacson, 1958), basada en cadenas de Markov y un sistema de reglas; MUSICOMP de Baker (Ames, 1987), los algoritmos estocásticos de Xenakis (Ames, 1987), el ordenador dedicado capaz de componer nuevas melodías relacionadas con otras que se proporcionan como entrada, empleando procesos de Markov (Olson & Belar, 1961); el algoritmo de Gill en 1963, que empleaba técnicas de IA clásica como búsqueda jerárquica con backtracking (Gill, 1963); PROJECT1 de Koenig (Ames, 1987), que usaba diversas técnicas, como la composición serial; o el framework de Padber, que empleaba técnicas procedurales (Padberg, 1964). Las máquinas empezaron a ser cada vez más económicas y potentes, por ello el campo de la composición algorítmica ha crecido; sin embargo, ha sido un proceso lento, con poco contacto entre científicos y artistas, con la mayoría de las iniciativas llevadas a cabo por estos últimos y con muy poca continuidad en las investigaciones. Salvo pocas excepciones, como CHORAL (Ebcioglu, 1988), EMI (Cope, 1992) o the Continuator (Pachet F. , 2002), la composición por ordenador no ha tenido gran relevancia y su uso tampoco se ha extendido en la sociedad. Melomics, desarrollado en esta tesis, puede ser considerado uno de estos sistemas que ha alcanzado más allá del entorno de la investigación (La UMA diseña un ordenador que compone música clásica, 2012) (Peckham, 2013). Ha sido usado no sólo para crear un repositorio web musical, dos álbumes originales y servido como fuente de inspiración para compositores profesionales (Diaz-Jerez, 2011), sino que también es la base de diversas aplicaciones novedosas, como la música adaptativa, que se detallan más adelante.

## Inspiración en la biología

La biomimética es un campo de la ingeniería consistente en desarrollar soluciones con inspiración en estructuras o comportamientos observados en la naturaleza. A lo largo de los años, este enfoque ha mostrado un gran potencial para resolver problemas informáticos complejos en multitud de campos, como la inversión financiera, la visión por computador, el tráfico en redes informáticas, el tráfico de la vía pública, los sistemas de control en electrodomésticos, el diseño industrial creativo o el arte. Algunos ejemplos conocidos con enfoque bioinspirado son las redes neuronales artificiales, la lógica difusa o la computación evolutiva. Los algoritmos bioinspirados se usan frecuentemente porque suelen producir soluciones creativas o imprevistas, que pueden no ser complemente óptimas, pero resuelven razonablemente bien el problema.

La evolución es uno de los principales mecanismos que ha contribuido a la diversidad y complejidad observada en la naturaleza. Estos mecanismos han inspirado la aparición de los algoritmos evolutivos, una metodología basada en una población de soluciones que se refina mediante un proceso iterativo consistente en evaluación de los individuos, selección de las mejores soluciones y producción de la siguiente generación de la población. En biología, el proceso de transformación de la primera célula de un organismo, llamada cigoto, en un organismo multicelular completo mediante un proceso de desarrollo regulado por la expresión selectiva de genes (Carroll, Grenier, & Weatherbee, 2004) (Forgács & Newman, 2005) (Mayr, 1961), juega un papel importante en el proceso de evolución (Marcus, 2003) (Lewontin, 2000) (Müller G. B., 2007) (Bateson, 2001) (Borenstein & Krakauer, 2008). La biología evolutiva del desarrollo (conocida como evo-devo) (Carroll S. B., 2005) estudia la evolución de los procesos de desarrollo, que en ciencias de la computación ha inspirado métodos análogos. En los algoritmos evolutivos, si se usa una codificación indirecta adecuada (Stanley & Miikkulainen, 2003), un pequeño genotipo puede desarrollarse en un fenotipo grande y complejo, y pequeñas variaciones en el código genético podrían conducir a grandes variaciones en el fenotipo. Esto supone un enfoque más escalable y robusto que una codificación directa y además cuenta con un gran potencial para proporcionar soluciones más originales. Algunos ejemplos encontrados que han seguido este tipo de metodologías pueden ser el diseño de antenas de satélites de la NASA (Hornby, Lohn, & Linden, 2011), el diseño de fibra óptica microestructurada (Manos, Large, & Poladian, 2007), la generación automática de juegos de mesa (Browne, 2008) o nuevas técnicas de animación de personajes (Lobo, Fernández, & Vico, 2012). El desarrollo embrionario se puede tratar, no solo como un proceso aislado dirigido por el genoma, sino en el que también puede influir un entorno natural (Forgács & Newman, 2005). De esta forma, la aparición de diversidad en las soluciones también dependerá de la interacción con el contexto físico durante el proceso de desarrollo. En ciencias de la computación se han desarrollado algunos trabajos inspirados en ello (Lobo D. , 2010) (Fernández, 2012) (Sánchez-Quintana, 2014).

Para desarrollar el sistema de composición musical propuesto, se toman ideas de esta disciplina bioinspirada, incluyendo representación genética para las composiciones con una codificación indirecta o implícita, que sufrirán un proceso complejo de desarrollo, afectado por un contexto físico, para dar lugar a la composición musical expresada de forma explícita.

## Recuperación de información musical (MIR)

La recuperación de información musical (MIR, por sus siglas en inglés) es un campo interdisciplinar que consiste en obtener información de diferentes tipos de datos musicales. El ejemplo más usual es la obtención de información acerca del estilo, género, instrumentos o voces participantes, nivel de activación, valencia, ritmos, etc. a partir de archivos de audio. No obstante, las fuentes de datos sujetas al estudio con MIR comprenden cualquier tipo de formato, como los archivos de música simbólicos (por ejemplo, en forma de partitura) o incluso información cultural relativa a las piezas musicales, que se puede encontrar en sitios web o redes sociales. Algunos de los propósitos más conocidos de MIR son: la construcción de sistemas de recomendación de música; sistemas de transcripción, para convertir audio en notación de partitura; herramientas de reconocimiento de canciones, como el sistema de Shazam (Wang, 2003); o herramientas de categorización, que la mayoría de servicios musicales en la actualidad utilizan para organizar canciones de acuerdo a sus propiedades musicales. También existe un creciente interés por la materia en investigación, donde se cuenta con ISMIR, una organización internacional que coordina conferencias y está dedicada al estudio de este campo.

## D.1.2 Método de composición algorítmica

Melomics es un sistema de composición musical que basa su funcionamiento en una representación genética para las composiciones y un proceso de desarrollo encargado de interpretar estos códigos y expresar la música de una forma explícita en diversos formatos. Asimismo, el sistema incorpora multitud de reglas que permiten evaluar la bondad de los resultados en distintos niveles del proceso de desarrollo, permitiendo filtrar aquellos resultados que no se ajusten a lo esperado y almacenar estructuras genéticas para su posterior uso, que sí cumplan de forma adecuada los requisitos.

El sistema se ha diseñado con dos versiones, una destinada a producir música atonal, caracterizada por la ausencia de reglas compositivas estrictas, especialmente reglas de armonía; y una segunda versión derivada, orientada a crear composiciones tonales, priorizando el tratamiento de la armonía, mediante las reglas empleadas en la teoría musical actual. En ambas versiones, la representación genética tiene estructura de gramática formal y poseen un flujo de ejecución similar, donde las principales diferencias consisten en la forma de evaluar la bondad de los resultados producidos. De forma general, ambos sistemas se han diseñado siguiendo los mismos principios, ya que las expectativas son similares, se persiguen soluciones originales, útiles, mutables, etc.

Para dar soporte al desarrollo resultante de un genotipo dentro el sistema, fue necesaria la definición de un formato de representación interna capaz de almacenar toda la información tanto compositiva como de ejecución musical, contando, por tanto, con mayor capacidad expresiva que los formatos regulares de música simbólica (MIDI, MusicXML, etc.). El formato interno posee, de forma general, una estructura matricial de tres dimensiones: pista instrumental, tiempo y propiedad musical de nota (inicio, duración, tono, dinámica, efectos...). Adicionalmente se incluye información global (nombre de la composición, fecha de creación, estilo que se usó para producirla, parámetros de mezcla...) e información de pista (ID de rol, nombre del instrumento, parámetros de mezcla...). A partir de este formato se pueden generar los distintos archivos de salida requeridos; en representación simbólica, como una partitura; o en forma de audio, como un archivo WAV.

Melomics, una tecnología que ha estado en continuo desarrollo durante más de seis años, puede considerarse como una contribución importante en el campo de la composición algorítmica, habiendo alcanzado una gran cantidad de hitos durante este tiempo que, de forma resumida y en orden cronológico, enumeramos a continuación: en 2010 la Universidad de Málaga concede un premio spin-off a una propuesta de start-up basada en una aplicación de Melomics; la primera obra musical, *Opus #1* es creada; el Ministerio de Ciencia e Innovación financia el proyecto de investigación MELOMICS; en 2011 se estrena en un concierto en directo *Hello World!*, una obra completa para orquesta de cámara, en el festival Keroxen de Tenerife; Gustavo Díaz-Jerez publica un artículo sobre la composición con el sistema Melomics en Leonardo Music Journal (Diaz-Jerez, 2011); en 2012 músicos reconocidos mundialmente, entre ellos la Orquesta Sinfónica de Londres graban distintas obras de cámara y sinfónicas que formarán parte del álbum Iamus; se celebra un concierto en directo en la ETSI Informática de la Universidad de Málaga, interpretando música de Melomics, como conmemoración del centenario del nacimiento de Alan Turing; la tecnología despierta gran atención de los medios, apareciendo en Discover Magazine entre los 100 hitos científicos del año (Berger, 2013); en 2013 se celebra el Melomics Hackathon en la Universidad de Berkeley; Melomics se presenta en Googleplex durante la sesión de SciFoo music-as-it-could-be; se emplea la tecnología en aplicaciones y experimentos y la aplicación eMTCP es recomendada en el portal de la American Chronic Pain Association; en 2014 se estrena el segundo álbum, 0music, con música perteneciente a géneros populares; se ejecuta un ensayo enfrentando la música de Melomics con música compuesta por humanos en el Museo Interactivo de la Música de Málaga y en el Conservatorio Superior de Música de Málaga; en 2015 se presenta Melomics en Seúl, en el evento SDF2015: Conscious Curiosity; y en marzo de 2016 se estrenan cuatro nuevas obras en MaerzMusic, un festival de música celebrado en Berlin.

## Modelo atonal

La primera versión del sistema Melomics se diseñó fundamentalmente para la producción de música atonal, concretamente en el estilo clásico contemporáneo. Para su construcción se tuvieron en cuenta los elementos de la teoría musical occidental moderna, evitando sin embargo construcciones convencionales de cualquier estilo concreto, de forma que las características rítmicas, armónicas o estructurales no quedaran restringidas de partida por la música producida hasta la fecha.

El código genético en este sistema, almacenado en un archivo de texto plano con extensión ".gen", está basado en una gramática formal de tipo L-system, varios parámetros adicionales que sirven para guiar el proceso de reescritura de la gramática y otros valores que se emplearán en el proceso de traducción de la cadena resultante en elementos musicales. Entre ellos se encuentran la duración de referencia, la escala musical para cada pista, el identificador de instrumento y su tesitura. En este modelo, los símbolos que no poseen una regla de reescritura explícita, se consideran los genes operadores, cuya misión será modificar los valores de un parámetro musical (tono, duración, tiempo de inicio de nota, efectos, volumen, tempo...) durante el proceso de traducción de la cadena de símbolos a su forma más explícita en forma de matriz numérica. Por otra parte, los símbolos con reglas de producción explícitas pueden representar tanto unidades estructurales musicales, como las notas musicales que finalmente serán ejecutadas. Esto dependerá de cómo suceda el proceso de reescritura, según los parámetros de iteración para la gramática y la función que se ha definido para computar las iteraciones restantes de cada regla. En cualquier caso, estos últimos símbolos descritos tendrán un identificador de instrumento asociado.

La cadena de símbolos resultante de la reescritura es objeto de un proceso de estabilización adicional antes de ser interpretada en música. Algunas de estas operaciones consisten en la eliminación de código genético superfluo, como la aparición consecutiva de dos genes operadores idempotentes. Otros reajustes son provocados por el contexto físico, por ejemplo, dependiendo del estilo musical en el que se desarrolle la composición o las características de los instrumentos elegidos, puede ser necesario reescribir subcadenas para satisfacer restricciones impuestas por estos, tales como máximo número de notas consecutivas o la aparición o supresión de determinados efectos musicales.

La última etapa del proceso morfogenético consiste en interpretar la cadena de caracteres en una estructura con forma de matriz numérica que contendrá los valores musicales explícitos de cada nota de cada canal. El proceso se basa en una lectura de la cadena de forma secuencial de izquierda a derecha, donde el comportamiento general consiste en que cada vez que aparece un gen operador, desplazará el valor actual de la variable musical que tiene asociada, partiendo de un valor inicial por defecto o indicado en el código genético. Otros operadores simplemente establecerán un valor específico para las variables. Cuando un gen instrumental aparece, se genera una nota musical adquiriendo el valor actual de cada uno de los parámetros musicales. A partir de la información que quede en esta estructura, será posible obtener distintos formatos de representación de música estándares tanto simbólicos, como sintetizados (De Vicente Milans, 2013).

El conocimiento musical experto es empleado, por una parte, como guía en el proceso de creación de genotipos, limitando las opciones en los procesos estocásticos, por ejemplo, con restricciones según la física de un instrumento, impidiendo que se generen más notas simultáneas de las que es capaz de ejecutar. Por otra parte, el conocimiento se usa para la construcción de las funciones de fitness de este sistema evolutivo, normalmente incluyendo reglas estéticas poco restrictivas, relativas a la rítmica, la armonía o los contornos melódicos.

Bien para probar la capacidad del sistema o bien para producir resultados con un fin específico, hasta la fecha se han efectuado numerosos experimentos con el sistema compositivo, donde podemos destacar los siguientes hitos:

Iamus Opus #1. Esta obra puede considerase la primera pieza profesional de música clásica contemporánea compuesta por un ordenador en su propio estilo. Fue creada en 2010, tras finalizar el desarrollo de la versión primitiva del sistema atonal, en colaboración con Gustavo Díaz Jerez.

Hello World!. Estrenada en 2011, en el festival Keroxen en Santa Cruz de Tenerife, es considerada la primera composición completa totalmente creada por un ordenador, usando notación musical convencional. En esta segunda iteración de Melomics, se disponía de mayor capacidad para gestionar instrumentos orquestales, se disponía de nuevas reglas compositivas, la escritura a formato MusicXML estaba completa y se había mejorado el sistema para producir composiciones de mayor duración y conjunto instrumental.

Ingeniería genética: Nokia tune. Con el fin de probar algunas de las ventajas que puede proporcionar el disponer de una composición en su representación genética, se llevó a cabo un ejercicio de ingeniería inversa, diseñando un genotipo del tema musical conocido como Nokia tune. A continuación, se ejecutaron una serie de mutaciones sobre el genoma. Esto sirvió para estudiar las relaciones entre los diferentes descendientes y tema original.

Repositorio web y álbum Iamus. El sistema usado para producir *Hello World!* fue mejorado para gestionar una mayor diversidad instrumental y se implementó una primera versión de un interfaz de definición de estilos musicales. Como resultado se produjeron decenas de miles de obras, con multitud de configuraciones de los parámetros de entrada, que fueron usadas para poblar un repositorio musical accesible a través de la web. De esta generación se seleccionaron algunas obras y se grabaron en estudio, para dar lugar al álbum Iamus que se lanzó en 2012.

Música para terapias de relajación. La herramienta de especificación de estilos para el sistema atonal se usó para producir distintos estilos de música, concretamente con las configuraciones requeridas para las aplicaciones propuestas de música adaptativa.

## Modelo tonal

El modelo diseñado para generar música atonal permite representar cualquier tipo de música, sin embargo, para restringir la producción de composiciones a géneros más populares y obtenerlas en un tiempo de ejecución menor, fue necesario adaptar el sistema con el fin de permitir una representación más simple de los elementos usuales en música tonal (modos y progresiones armónicas, modos y patrones rítmicos, relaciones instrumentales estrictas, etc.). Se conservó el diseño global del sistema. Se añadieron nuevos símbolos en el código genético, principalmente para la gestión de la armonía y la estructura compositiva, haciéndose esta última corresponder con una organización jerárquica, inspirados por el estudio de estructuras musicales (Bent & Pople, 2010). La interpretación de los nuevos símbolos consta de un mayor nivel de abstracción, siendo por tanto el proceso de traducción más complejo. Por último, se añadieron nuevos parámetros al código genético para guiar el proceso de desarrollo de la gramática y la traducción de la cadena final en una matriz musical

Para facilitar el uso del sistema de música tonal y permitir la especificación sencilla de reglas y dar lugar a estilos musicales concretos, desarrollamos una herramienta con forma de cuestionario sobre elementos musicales, expresados en un lenguaje de composición de un alto nivel de abstracción. Se programó un componente para traducir esta información en restricciones para la generación de genotipos, restricciones durante el desarrollo y reglas de evaluación del ajuste. De esta forma, sin necesidad de conocer el funcionamiento del sistema, simplemente disponiendo de conocimientos sobre composición, es posible lograr que el sistema produzca contenido musical de una forma dirigida.

El modelo genético para el sistema tonal puede ser descrito como una gramática formal determinista de contexto libre, donde los símbolos no terminales se identifican con elementos estructurales de una composición musical, como la composición en sí, períodos, frases o ideas; y los elementos terminales se corresponden con manipuladores de las propiedades musicales como el pitch, duración, acorde actual, etc. y con las notas musicales en sí, que toman esas propiedades.

La estructuración jerárquica que se suele usar para construir genomas musicales con este modelo consta de cinco niveles: (1) composición, que está compuesta de una secuencia de (2) períodos, los cuales se desarrollan en un conjunto de (3) frases, que están formadas por (4) ideas, las cuales constan finalmente de una secuencia de (5) notas. A pesar de esta restricción, se ha procurado mantener, e incluso mejorar algunas de las ventajas que aporta el modelo genético gramatical, como la legibilidad, la flexibilidad, la gran capacidad expresiva y la robustez ante alteración de símbolos y reglas de producción.

La primera etapa del proceso de transcripción es el desarrollo de la gramática de contexto libre hasta sustituir todos los símbolos no terminales en terminales, normalmente tras cuatro iteraciones, correspondientes a la jerarquía descrita. Al igual que en el modelo atonal, la cadena resultante sufre ajustes adicionales, algunos para eliminar elementos superfluos y otros consistentes en operaciones determinadas por reguladores genéticos, como la adición, supresión o alteración de determinados símbolos genéticos o una secuencia de ellos, que pueden depender del estilo musical.

La segunda etapa del desarrollo de un genoma musical es la interpretación de la cadena para dar lugar a una estructura matricial con los valores explícitos de cada nota. Igual que en el modelo atonal, el proceso consiste en una lectura secuencial de izquierda a derecha con un funcionamiento general basado en desplazar los valores musicales actuales de los distintos parámetros, mediante los operadores correspondientes y cada vez que aparece un símbolo terminal asociado a un instrumento, se crea una nota que adquiere las propiedades actuales. Estos mecanismos son más abstractos que en el sistema atonal y muchos de ellos necesitan del estilo musical para ser instanciados. Por ejemplo, para computar el pitch concreto de una nota musical, es necesario tener en cuenta diversos parámetros como el acorde actual, la nota base del acorde, desplazamientos previos de tono o dentro de la escala; e incluso los acordes permitidos para el rol instrumental en cuestión.

Algunas de las pruebas realizadas con música tonal para probar la capacidad del sistema o para producir resultados específicos incluyen:

**Ingeniería genética: SBS Logo**. Se hizo un experimento análogo al realizado con el Nokia tune en el sistema atonal. El sistema Melomics se presentó en el evento SDF2015 en Seúl, que fue retransmitido por el canal privado SBS. Por ello, se tomó el tono corporativo de la empresa; se efectuó ingeniería inversa con la ayuda de un experto para obtener un código genético en el sistema tonal; y se produjo un árbol genealógico, siendo el logo original el familiar central. De esta forma se ilustró (1) un posible proceso evolutivo desde un genoma muy simple que podría haber dado lugar a este tono, (2) diferentes mutaciones del logo y (3) contaminación del código genético de un tema musical independiente con material genético de este tono, preservando la identidad del tema original, pero siendo posible identificar el SBS Logo como parte integrante de él.

**Aplicación para dolor crónico**. Se continuó con el trabajo iniciado con el modelo anterior de producir música para relajación. La principal novedad consistió en el empleo de hypersongs, formados por un conjunto determinado de versiones de una misma composición, que son usadas de forma selectiva dentro del sistema de música adaptativa, detallado más adelante.

**Repositorio web y álbum 0music**. A inicios de 2014 se implementaron numerosas configuraciones de estilos para el sistema tonal, dando lugar a otra colección en el repositorio de música, esta vez en estilos más populares. Una vez más, se seleccionaron algunos de estos temas y se publicaron en forma de álbum, conocido este como 0music.

**Estilos DocumentaryPop y DocumentarySymphonic**. El sistema tonal se configuró a través de la herramienta de estilos para generar música de dos tipos, que se usarían en documentales y otro tipo de producciones. En 2014 se produjeron 200 composiciones en estos estilos que fueron adquiridos por una empresa de distribución musical.

## D.1.3 Propiedades de la música generada

La música generada por Melomics se ha analizado desde una perspectiva analítica, empleando herramientas MIR, a partir de archivos de audio y de notación simbólica. También se ha hecho un estudio desde una perspectiva perceptiva, enfrentando la música de Melomics con música creada por un músico profesional, siendo juzgad por evaluadores humanos.

Análisis con MIR

Este tipo de estudio tuvo como objetivo, por una parte comprobar la relación entre la codificación genética de las composiciones y la música resultante, concretamente cómo afectan los cambios en el código genético en los valores musicales finales, o si códigos genéticos producidos con una configuración de estilo concreta, realmente dan lugar a composiciones agrupadas en un estilo musical y además estas difieren de las producidas usando otro estilo. Por otra parte, se trató de comprobar analíticamente las características de la música producida con Melomics en relación a la música existente compuesta por humanos y en relación a otros tipos de sonidos.

En un primer análisis se usó la herramienta jSymbolic para extraer, de cada composición, los valores de hasta 111 características musicales diferentes, incluyendo parámetros relacionados con el ritmo, textura, dinámica, pitch, melodía y armonía. El primer conjunto analizado fue una colección con aproximadamente 1000 variaciones del Nokia tune, producidas con distintos operadores genéticos implementados, donde se observó que la distancia, medida dentro de este espacio de parámetros con respecto al tono original, se incrementaba con la disrupción genética introducida. En el modelo tonal se llevó a cabo un experimento similar con el árbol filogenético del SBS Logo.

Empleando la misma herramienta, se analizaron 656 obras de cámara del sistema atonal, con distintos conjuntos de instrumentos, en el estilo clásico contemporáneo. Tras calcularse el centroide de este grupo, se analizaron nuevas piezas pertenecientes a la misma configuración de estilo, las cuales se ubicaron, sin sorpresa, esparcidas entre las obras del estilo de referencia; a continuación se analizaron piezas pertenecientes a un etilo etiquetado como "Modern Classical" de una colección de composiciones en MIDI creadas por humanos, denominado *Bodhidharma*, y estas aparecieron en su mayoría ubicadas entre las composiciones del estilo clásico contemporáneo de Melomics, dando a entender que existe solapamiento entre ambos estilos musicales; por último se analizaron piezas producidas por el sistema tonal en el estilo *Disco02*, siendo el grupo de composiciones que apareció más alejado del estilo de referencia, en el espacio de parámetros musicales definido. En un estudio similar con el sistema tonal, se comprobó que el parecido musical entre los resultados de los estilos configurados *DancePop* y un estilo del que evoluciona, *Disco02*, es mayor que entre *DancePop* y *DocumentarySymphonic,* un estilo que es independiente del primero.

Por último, se trató de poner en contexto la música de diferentes estilos de Melomics, empleando las herramientas que proporciona The Echo Nest. Entre sus denominados atributos acústicos, "Valence" pareció aportar información útil para distinguir contenido musical de otro tipo de sonidos e incluso diferenciar la música más popular, de la música contemporánea, que en general carece de configuraciones armónicas comunes. Posiblemente debido a que incluya un análisis de la armonía, para puntuar la emoción evocada.

## Estudio perceptivo

Se llevó a cabo un estudio propuesto en colaboración con el Dr. Alfredo Raglio y efectuado en el Museo Interactivo de la Música de Málaga y en el Conservatorio Superior de Música de Málaga, siguiendo la idea del *Musical Output Toy Test* (MOtT) (Ariza, 2009), inspirada en el Test de Turing, pero sustituyendo el papel del interrogador por cuestionarios para que los participantes o críticos, emitan su juicio sobre las piezas que se les presentan. El objetivo del experimento fue medir las emociones y representaciones mentales que evocaban la música de Melomics en comparación a la humana y, en última instancia, comprobar si los sujetos eran capaces de discernir una de la otra. Para ello se crearon dos piezas musicales con las mismas especificaciones de estilo, balada de guitarra, una por Melomics y otra por un compositor humano. Posteriormente ambas piezas fueron interpretadas por músicos humanos y por el sistema de síntesis de Melomics. Se incluyeron las cuatro piezas resultantes y una grabación de duración similar que contenía sonidos naturales (The sounds of Nature Collection (Jungle River, Jungle birdsong and Showers)) combinados con sonidos de animales (Animals/Insects, n.d.). Los participantes del experimento se clasificaron en dos grupos, de acuerdo a su experiencia en música: con más de cinco años se consideraron como músicos y en otro caso como no-músicos. Todos debían responder el mismo cuestionario sobre las muestras de audio que se les reproducían, las cuales eran diferentes según el grupo asignado. Tras el análisis de los datos aportados por los 251 sujetos totales, no se observó sesgo en cuanto a la atribución del origen compositivo de las piezas musicales. Esta imposibilidad de discriminar se dio en ambos grupos por separado, músicos y no-músicos.

## D.1.4 Composición musical adaptativa

Disponer de un sistema compositivo automático, además de posibilitar la producción rápida de composiciones, tiene otra serie de ventajas como composición de temas sin desviación de las expectativas definidas o el versionado de piezas musicales. Una aplicación original de esta tecnología es la música adaptativa, que consiste en proveer música que cambia en tiempo real, en respuesta al valor detectado de determinadas variables del entorno o el oyente. En la literatura se han encontrado diversas aplicaciones de este concepto. Sin embargo, observamos que se usan bien composiciones previamente producidas por artistas humanos que, al cambiar la entrada detectada, se hace alternar la reproducción entre unas piezas y otras (en ocasiones con un simple efecto crossfade); o se usan señales de audio generadas en directo, pero que distan de ser composiciones musicales reales. Nuestra contribución en esta área es un método capaz de proveer con un flujo de música genuina que se adapta de forma continua y lo hace mediante la manipulación de los parámetros musicales de una composición en los distintos niveles de abstracción; incluyendo variables como el ritmo, la evolución de la armonía, de la densidad instrumental, relación entre roles instrumentales o la estructura compositiva.

Sistema de generación basado en realimentación. Hypersong

Se ha implementado un sistema estandarizado para dar soporte al concepto de composición adaptativa en tiempo real para las aplicaciones propuestas, el cual requirió los siguientes elementos:

- En cada aplicación, un componente para evaluar el estado del entorno, por ejemplo, el estado de ansiedad de un sujeto mediante el análisis de su frecuencia cardíaca, que será la forma de definir cómo debe ser el flujo de música que se proporciona.
- Un mecanismo para traducir la información del análisis del entorno, en los parámetros para seleccionar la música apropiada de salida, implementado mediante un autómata de Moore.
- El sistema capaz de generar el flujo de música adecuado. Para establecer la música a proporcionar en cada aplicación y estado, se definió e implementó el concepto de hypersong, una estructura de audio musical de dos dimensiones: versión de la composición y tiempo. Cada versión está dividida en fragmentos de audio seccionados según la estructura compositiva. De esta forma, dada la información del autómata y el instante actual de reproducción, el sistema es capaz de indexar el fragmento temporal y versión de las hypersongs que ha de añadir a continuación en el flujo de reproducción normal.

## Aplicaciones

Haciendo uso de la tecnología musical desarrollada, se diseñaron una gran variedad de aplicaciones, entre ellas destacamos:

eMTSD.  Una aplicación para dispositivos móviles, para facilitar la iniciación del sueño. Funciona situando el dispositivo bajo la almohada, de forma que sea posible caracterizar el estado de somnolencia en el usuario a través del análisis de la actividad detectada con el acelerómetro. Esta información se usa para alimentar un autómata de cuatro estados, que establece la reproducción de una versión de las composiciones progresivamente más suave, según se detecte un estado de somnolencia mayor, hasta detener la reproducción completamente cuando el sujeto está dormido. Posteriormente se desarrollaron versiones de esta aplicación, con funcionamiento y música ligeramente diferentes para su uso durante la siesta y para ayudar a la conciliación del sueño de niños pequeños.

**PsMelody**. Una aplicación de escritorio, implementada en Windows, para favorecer el rendimiento en el trabajo y al mismo tiempo tratar de disminuir los niveles de estrés. PsMelody analiza las pulsaciones del teclado y el movimiento del ratón para caracterizar el tipo de actividad que se está realizando y la energía que se emplea. A partir de ello, se reproducen composiciones que cambian de versión para favorecer el trabajo, pero evitando caer en situación de ansiedad.

**@car.** Una aplicación móvil para favorecer la concentración durante la conducción y tratar de reducir la tensión, especialmente bajo condiciones irregulares, como tiempo atmosférico extremo o tráfico saturado. Esta aplicación toma como entrada los datos proporcionados por el acelerómetro, el GPS, el magnetómetro y el fotómetro. A partir de esta información computa el estado actual de la conducción, distinguiendo cuatro niveles de fatiga o dificultad, desde conducción con tráfico saturado y condiciones climáticas adversas, hasta conducción fluida por vía interurbana a alta velocidad. La versión de la música reproducida es más suave cuando la dificultad de la conducción es mayor.

**@memory**. Esta aplicación móvil hace uso de la música de Melomics para ejercitar la memoria asociativa, proponiendo un juego basado en identificar melodías con figuras que se muestran en pantalla previamente (la música no es adaptativa en este caso) y en cada fase la dificultad va incrementando. Tanto el funcionamiento como las especificaciones musicales de esta aplicación sirven como base para futuras propuestas para tratar enfermedades relacionadas con la memoria.

**@rhythm**. Es una aplicación para ayudar a mejorar la concentración y la coordinación. Presenta pequeños trozos de música y se debe tratar de identificar el ritmo predominante mediante pulsaciones sobre la pantalla táctil del dispositivo en un lugar concreto. Igual que la anterior aplicación, esta aplicación hace uso del repositorio de melodías automáticamente generadas, versionadas y distribuidas por Melomics.

**eMTCP**. Es una aplicación móvil dirigida a disminuir los niveles de dolor en pacientes que sufren de dolor crónico, aprovechando el efecto distractor de la música. Se implementó para llevar a cabo un experimento clínico cuyas hipótesis eran (a) que la música adaptativa es capaz de mejorar la calidad del sueño en pacientes con dolor crónico, correspondiéndose con una reducción de la percepción del dolor; y (b) que la música adaptativa reduce la percepción subjetiva del dolor por parte del sujeto. Se implementó una máquina de estados que distinguía entre ocho estados de ansiedad, transitándose de uno a otro según el grado de dolor indicado por el usuario de forma interactiva.

STM. El Sistema de Terapia Musical para hospitales fue el primer sistema de música adaptativa implementado. Se desarrolló en lenguaje MATLAB para funcionar en pequeños ordenadores portátiles y cuenta con una interfaz gráfica de usuario sencilla. Se empleó para llevar a cabo experimentos de terapias de relajación con diferentes configuraciones en pacientes de clínicas y hospitales. El primero de los ensayos propuestos consistió en un experimento doble ciego, donde se pretendía comprobar la mejora en la tasa de lactancia de recién nacidos, cuando se proporcionaba música relajante adaptativa a las madres para disminuir su nivel de ansiedad. El parámetro de entrada en este caso, para caracterizar el estado de agitación, es la frecuencia cardíaca, la cual se analiza teniendo en cuenta su desviación con respecto a la frecuencia basal, previamente calculada.

@jog. Una aplicación móvil que efectúa podometría usando el acelerómetro, para detectar el ritmo dominante actual de un corredor y reproducir música de acuerdo con esta información. El smartphone se sitúa en el brazo del corredor o en un bolsillo bien ajustado. La interfaz de usuario es sencilla, para facilitar una posible manipulación durante el ejercicio y la aplicación posee distintas configuraciones y niveles de dificultad, diseñadas para mejorar el rendimiento deportivo. Las hypersongs empleadas en esta aplicación están formadas por 25 versiones, con distinto beats per minute (BPM) como principal variación entre ellas, aunque no la única. El estilo de base es música disco, pero con una gran apertura de los parámetros, para abarcar una mayor cantidad de preferencias musicales, sin perder la capacidad de animar al usuario a mantener el ritmo. @bike es una versión de la aplicación @jog, para ayudar a mejorar el rendimiento en el ciclismo (de interior o exterior). Cuenta con una configuración de parámetros distinta para la detección del ritmo, en este caso el dispositivo ha de ir bien fijado en el muslo o en los gemelos del deportista. El estilo musical empleado es el mismo que en @jog, pero se aumenta el rango de BPM, para satisfacer la mayor cantidad de ritmos que pueden darse usando una bicicleta.

## D.2 Conclusiones

En esta tesis hemos presentado el sistema de composición Melomics, describiendo el diseño de sus dos versiones, ambas basadas en métodos bioinspirados y detallando los principales mecanismos implicados y la forma en la que se trata la información generada a lo largo del proceso. Hemos evaluado los resultados que se obtienen con este sistema, empleando diferentes enfoques. Por último, hemos expuesto una nueva forma de proveer música, aprovechando las características de la composición automatizada.

Para construir el sistema se han usado distintos métodos dentro del campo de la inteligencia computacional. Para la definición de la música y los estilos, hemos usado técnicas basadas en conocimiento, para proporcionar soluciones válidas hemos empleado búsqueda evolutiva y, como soporte para representar y manipular los elementos musicales siguiendo este enfoque, se ha usado una representación genética implícita con un proceso de desarrollo asociado. Como resultado, hemos construido un método capaz de crear música genuina y manipularla, que cuenta con las siguientes ventajas:

- El producto es innovador, ya que no se emplean métodos basados en imitación y se dispone de libertad para explorar el espacio de búsqueda, definido por reglas más o menos restrictivas. Estas actúan como un mecanismo de control para que las piezas producidas cumplan con los requisitos especificados.
- La sintaxis para representar los genomas y, por tanto, la música es (a) altamente expresiva, ya que permite la representación de cualquier pieza musical que pueda ser escrita en notación musical estándar; (b) flexible, cualquier secuencia puede ser escrita de infinitas formas; (c) compacta, a pesar de incluir información compositiva y de interpretación, consume entre la mitad y un tercio del espacio del equivalente en formato MIDI, entre un tercio y un cuarto de lo que correspondería a MusicXML y definitivamente menos que cualquier formato de audio; y (d) robusta, lo que significa que si un genoma es alterado, no sólo sigue siendo capaz de producir una pieza válida, sino que además esta compartirá elementos en común con el tema original, siendo este una mutación del primero.

- El sistema es asequible: (a) de inicializar, ya que no se necesita disponer de una fuente de ejemplos previos; (b) en relación al consumo de memoria, ya que no se necesita mover o almacenar una gran cantidad de datos para entrenar; (c) aunque necesita la intervención de un experto para configurar las reglas para converger a un estilo particular, una vez hecho, la ejecución no es muy costosa computacionalmente. Usando una un hilo de ejecución en una CPU actual, tanto el sistema atonal como el tonal producen una composición genuina en su estilo más complejo en 6,5 minutos; y es posible la ejecución en paralelo.

Usando el sistema, se ha generado un repositorio de música web, composiciones relevantes como *Opus #1* y *Hello World!*, así como dos álbumes musicales: Iamus álbum y 0music.

Se ha desarrollado una herramienta que permite especificar un estilo musical rellenando un cuestionario de especificaciones. Este incluye valores globales, guías para la estructura y parámetros específicos, los cuales Melomics puede usar para producir genotipos y para conducir el proceso compositivo. La forma en que se ha diseñado hace posible que persona con poco conocimiento musical pueda proporcionar instrucciones al sistema para crear música. Hemos confirmado que cuando personas diferentes especifican un mismo estilo, las instrucciones dadas a través del cuestionario fueron muy similares y por tanto así la música obtenida. La herramienta se exportó a un formulario web, donde las preguntas se transcriben directamente a un archivo de entrada que puede ser interpretado por el sistema. Este podría ser el primer paso para una futura herramienta a disposición de la comunidad de usuarios.

También hemos presentado el concepto hypersong, una estructura musical que comprende un tema musical y un conjunto de versiones, todas seccionadas en el tiempo, dando lugar a una matriz bidimensional de fragmentos musicales. Este concepto nos ha permitido desarrollar diferentes aplicaciones basadas en proporcionar un flujo de música que se adapta según las condiciones, lo cual sólo es práctico si este flujo puede ser creado automáticamente. Esta tecnología fue puesta a disposición del público en forma de API, alimentada por un repositorio de música estándar y de una colección de hypersongs, haciendo posible el desarrollo de nuevas aplicaciones por parte de la comunidad.

Que las piezas producidas por Melomics puedan ser consideradas como auténticas piezas musicales es algo que tiene que ser determinado, en última instancia, por el juicio de la gente y de los expertos, como ocurre siempre en cualquier forma de arte. El Test de Turing se diseñó para identificar trazas de pensamiento en el procesamiento de una máquina, pero su naturaleza interactiva hace difícil su adaptación al ámbito musical. Sin embargo, los principios subyacentes son inspiración válida para nuevos experimentos para medir cómo de parecidas son la música artificial y la humana.

Melomics se configura, si desea, mediante un conjunto de parámetros abstractos, como la cantidad de disonancia, la repetitividad, los instrumentos, la duración de la composición, etc. Estos son traducidos a forma de genoma y participan en un proceso evolutivo. No se requiere de una entrada creativa por parte del usuario, ni de una interacción continua (función de fitness interactiva). La entrada inicial actúa meramente como directrices musicales, exactamente las que se le darían aun músico para crear una composición.

Sin embargo, normalmente la composición de música se entiende como un proceso creativo donde los humanos pueden expresar sensaciones. Entonces, ¿cómo podría llegar a considerarse la música artificial música auténtica? La definición de música puede ser complicada, incluso si sólo se considera la instrumental. Por ejemplo, si se considera como un conjunto ordenado de sonidos, el concepto sería demasiado amplio, ya que incluiría sonidos, que si bien están organizados, normalmente no entendemos como música y, al mismo tiempo, corre el riesgo de excluir composiciones creadas por Melomics, ya que hasta que son sintetizadas (o interpretadas), son simplemente partituras. Dos propiedades que normalmente se requieren en la definición de música son tonalidad, o la presencia de ciertas características musicales, y una atención a propiedades estéticas. La primera puede ser garantizada por Melomics a través de la función de fitness. La segunda es más compleja de respetar, ya que no se puede atribuir ninguna intención estética al software. Sin embargo, si nos centramos en el oyente, en lugar de en el compositor, el hecho de que el compositor sea una máquina deja de ser relevante. El juicio positivo de Peter

Russell (Ball, Artificial music: The computers that create melodies, 2014) resultó interesante, ya que, al no conocer el origen de la composición, no expresó ninguna duda sobre la naturaleza musical de la pieza. Esto nos animó a seguir este enfoque para establecer si lo que el sistema produce es realmente música o no, atendiendo a la opinión de la crítica y del público general. Si ellos consideraban el resultado final como música, entonces podría ser considerado como música.

En contraste con anteriores trabajos, el experimento presentado ilustra una metodología controlada y rigurosa y es llevado a cabo sobre una gran muestra de participantes. La primera pregunta del cuestionario ("Dirías que lo que acabas de escuchar es música?") pretendía medir potenciales diferencias percibidas entre la composición creada por el músico humano y el resto de muestras. La muestra que contenía sonidos naturales fue clasificada como música por el 41,7% de los músicos profesionales. En contraste, ambas piezas musicales fueron clasificadas como música por la mayoría de los sujetos (por encima del 90%). Con respecto a la capacidad para evocar imágenes mentales, no existió diferencia significativa al estudiar las muestras musicales, evocando imágenes para alrededor del 50% de sujetos, mientras que esta cifra se situó en el 90% para los sonidos naturales. El estudio de datos cualitativos mostró que la mayoría de los términos usados en las descripciones encajaban en la categoría "naturaleza", a diferencia de las piezas musicales, que no favorecían a ninguna de las categorías definidas. Estos resultados resaltan la diferencia entre los sonidos naturales y la música presentada, que genera un conjunto de imágenes mentales más amplio y complejo, con independencia del entrenamiento musical del oyente o de quién compuso o interpretó las piezas. En relación a los estados emocionales, uno de los resultados más reveladores fue que, para los sonidos naturales, las descripciones asignadas al estado "calmado" fueron del 89%, mientras que para las piezas musicales este porcentaje fue del 47.3% como máximo, a pesar de que el estilo musical podía considerarse como calmado. Igual que en el caso de las imágenes mentales, todas las piezas musicales parecieron provocar un amplio rango de emociones, independientemente del oyente, el compositor o el intérprete. El segundo resultado interesante vino del estudio de la valencia. Resultaron ser significativamente positivas ($p < 0.001$) al describir sonidos naturales, mientras que la música, con independencia del grupo de estudio, también provocaron sentimientos negativos.

La última parte del experimento confirmó que los sujetos eran incapaces de distinguir el origen de la composición. A pesar de que se usaron sólo dos piezas musicales, la muestra era amplia y el hecho de que más de la mitad eran músicos profesionales, es otro indicador de la robustez de las conclusiones. Estas confirman la hipótesis planteada, sugiriendo que las composiciones creadas por el ordenador podrían ser consideradas como "música auténtica". Esto es, por supuesto, un primer resultado en la línea de evaluaciones de calidad y este modelo puede ser extendido a diferentes estilos musicales.

Debido a la naturaleza multidisciplinar de este proyecto, existen multitud de caminos que podrían desarrollarse a continuación, tanto desde un punto de vista empresarial, como líneas de investigación.

Un ejemplo considerado es la obtención de nuevos estilos musicales y géneros. A pesar de que Melomics puede representar cualquier contenido musical, en la práctica sólo produce música en los estilos especificados. Sería necesario definir una forma adecuada de explorar el resto del espacio de búsqueda, para llegar a esos estilos que Melomics no ha alcanzado aun y también para alcanzar estilos que todavía no se han descubierto. Sin embargo, algunos de estos introducirían problemas mayores, como la inclusión de voz. Se necesitaría construir la letra, lo cual es una línea de investigación en curso (Pudaruth, Amourdon, & Anseline, 2014) (Malmi, Takala, Toivonen, Raiko, & Gionis, 2016), y desarrollar un sistema de síntesis vocal, de mayor complejidad que la síntesis de voz normal.

Otra línea de trabajo es la "música personal". Esto sería una evolución del concepto de música adaptativa. En este caso el flujo de música tendría que (a) ser automáticamente responsivo a cualquier contexto, no simplemente a un caso de uso concreto, y (b) satisfacer las preferencias cada usuario. La música sería completamente a medida de la persona y la situación. Habría que desarrollar las ideas y métodos presentados en el Capítulo 4. Ahí se comenta que ya se ha hecho un intento preliminar de proporcionar música para diferentes situaciones y cambiar de contexto automáticamente (aplicación @life). Para desarrollar completamente la idea de música personal, sería necesario trabajar en diferentes frentes, por ejemplo (a) encontrar métodos para detectar el contexto o escenario actual del usuario; (b) mejorar y desarrollar métodos de caracterización del estado actual para cada escenario; (c) estudiar la música apropiada para cada caso de uso considerado y sus circunstancias; (d) idealmente, mejorar el concepto de hypersong, especialmente los principios de versionado musical y los mecanismos de control del flujo musical; (e) encontrar un procedimiento adecuado para configurar y actualizar las preferencias musicales de cada usuario; y (f) desarrollar la forma en que la música se especifica al sistema, para alcanzar más géneros musicales y ser capaz de satisfacer un mayor rango de gustos.

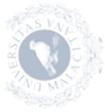

# Bibliography

Abdallah, S., Gold, N., & Marsden, A. (2016). Analysing symbolic music with probabilistic grammars. In D. Meredith (Ed.), *Computational Music Analysis* (pp. 157-189). Cham, Switzerland: Springer.

Albarracín-Molina, D. D. (2010). Diseño e implementación de una estrategia para un juego de conexión. M.A. Thesis. Málaga, Spain: Universidad de Málaga. Retrieved May 16, 2021, from https://riuma.uma.es/xmlui/handle/10630/6770

Albarracin-Molina, D. D., Moya, J. C., & Vico, F. J. (2016). An evo-devo system for algorithmic composition that actually works. *Proceedings of the Companion Publication of GECCO 2016.* Denver, Colorado, USA: ACM.

Albarracín-Molina, D. D., Raglio, A., Rivas-Ruiz, F., & Vico, F. V. (2021). Using Formal Grammars as Musical Genome. *Applied Sciences, 11*(9), 4151. doi:doi.org/10.3390/app11094151

Ames, C. (1987). Automated composition in retrospect: 1956-1986. *Leonardo, 20*, 169-185.

*Animals/Insects.* (n.d.). Retrieved April 30, 2021, from freeSFX: http://www.freesfx.co.uk/soundeffects/animals_insects/

Ariza, C. (2009). The interrogator as critic: The turing test and the evaluation of generative music systems. *Computer Music Journal, 33*(2), 48-70.

Ariza, C. (2011). Two pioneering projects from the early history of computer-aided algorithmic composition. *Computer Music Journal, 35*(3), 40-56.

Ball, P. (2012, July 1). Iamus, classical music's computer composer, live from Malaga. *The Guardian*. Retrieved April 30, 2021, from http://www.theguardian.com/music/2012/jul/01/iamus-computer-composes-classical-music

Ball, P. (2014, August 8). Artificial music: The computers that create melodies. *BBC*. Retrieved April 30, 2021, from http://www.bbc.com/future/story/20140808-music-like-never-heard-before/

Barceló, A., Barbé, F., Llompart, E., De la Peña, M., Durán-Cantolla, J., Ladaria, A., . . . G., A. A. (2005). Neuropeptide Y and leptin in patients with obstructive sleep apnea syndrome: role of obesity. *American journal of respiratory and critical care medicine, 171*(2), 183-187.

Bateson, P. (2001). Where does our behaviour come from? *Journal of biosciences, 26*(5), 561-570.

Bent, I. D., & Pople, A. (2010). Analysis. In *Grove Music Online*. Oxford, United Kingdom: Oxford University Press. doi:doi:10.1093/gmo/9781561592630.article.41862

Bentley, P. J. (1999). Is evolution creative. *Proceedings of the AISB. 99*, pp. 28-34. Edinburgh, United Kingdom: Society for the Study of Artificial Intelligence and Simulation of Behaviour.

Bentley, P. J., & Corne, D. W. (2001). An Introduction to Creative Evolutionary Systems. In D. W. Corne, & P. J. Bentley (Eds.), *Creative evolutionary systems* (pp. 1-75). Elsevier Science.

Berger, K. (2013, January 25). 100 Top Stories of 2012. Digital Composer Records with London Symphony Orchestra. *Discover Magazine*. Retrieved May 16, 2021, from https://www.discovermagazine.com/technology/70-digital-composer-records-with-london-symphony-orchestra

Biles, J. (1994). Genjam: A genetic algorithm for generating jazz solos. *Proceedings of the 1994 International Computer Music Conference*, *94*, pp. 131-137. Aarhus, Denmark.

Boden, M. A. (2009). Computer models of creativity. *AI Magazine, 30*(3), 23.

Borenstein, E., & Krakauer, D. C. (2008). An end to endless forms: epistasis, phenotype distribution bias, and nonuniform evolution. *PLoS Comput Biol, 4*(10), e1000202-e1000202.

Boulanger, R. C. (2000). *The Csound book: perspectives in software synthesis, sound design, signal processing, and programming.* MIT press.

Boulanger-Lewandowski, N., Bengio, Y., & Vincent, P. (2012). Modeling Temporal Dependencies in High-Dimensional Sequences: Application to Polyphonic Music Generation and Transcription. *Proceedings of the 29th International Coference on International Conference on Machine Learning* (pp. 1881-1888). Edinburgh, Scotland: ACM.

Bringsjord, S., Bello, P., & Ferrucci, D. (2001). Creativity, the Turing test, and the (Better) Lovelace test. *Minds and Machines, 11*, 3-27.

Briot, J. P., Hadjeres, G., & Pachet, F. D. (2017). *Deep learning techniques for music generation -- A survey.* Ithaca, NY, United States: arXiv, Cornell University. Retrieved from https://arxiv.org/abs/1709.01620

Browne, C. (2008). Automatic generation and evaluation of recombination games. Ph.D. Thesis. Brisbane, Australia: Queensland University of Technology.

Buttram, T. (2003). *DirectX 9 Audio Exposed: Interactive Audio Development, chap. Beyond Games: Bringing DirectMusic into the Living Room.* Wordware Publishing Inc.

Campbell, M., Hoane, A. J., & Hsu, F. H. (2002). Deep blue. *Artificial Intelligence, 134*(1-2), 57-83.

Carroll, S. B. (2005). *Endless forms most beautiful: The new science of evo devo and the making of the animal kingdom.* WW Norton & Company.

Carroll, S., Grenier, J., & Weatherbee, S. (2004). *From DNA to Diversity: Molecular Genetics and the Evolution of Animal Design* (2 ed.). Wiley-Blackwell.

Castellaro, M. (2011). El concepto de representación mental como fundamento epistemológico de la psicología. *Límite, 24*, 55-68.

Chambel, T., Correia, L., Manzolli, J., Miguel, G. D., Henriques, N. A., & Correia, N. (2007). Creating video art with evolutionary algorithms. *Computers & Graphics, 31*(6), 837-847.

Christensen, T. (Ed.). (2002). *The Cambridge history of Western music theory.* Cambridge University Press.

Colombo, F., Seeholzer, A., & Gerstner, W. (2017). Deep artificial composer: A creative neural network model for automated melody generation. *Proceedings of the International Conference on Evolutionary and Biologically Inspired Music and Art* (pp. 81-96). Amsterdam, The Netherlands: Springer.

Colton, S. (2008). Creativity Versus the Perception of Creativity in Computational Systems. *AAAI Spring Symposium: Creative Intelligent Systems* (pp. 14-20). Palo Alto, California: The AAAI Press.

Colton, S., López de Mantaras, R., & Stock, O. (2009). Computational creativity: Coming of age. *AI Magazine, 30*(3), 11-14.

Cook, M., & Colton, S. (2011). Automated collage generation–with more intent. *Proceedings of the Second International Conference on Computational Creativity*, (pp. 1-3). México City, México.

Cope, D. (1992). Computer modeling of musical intelligence in EMI. *Computer Music Journal, 16*(2), 69-83.

Crocker, R. L. (1966). *A history of musical style.* Courier Corporation.

Dacey, J. (1999). Concepts of Creativity: A History. In M. A. Runco, & S. R. Pritzer (Eds.), *Encyclopedia of Creativity* (Vol. 1). Elsevier.

Davidson, E. H. (2010). *The regulatory genome: gene regulatory networks in development and evolution.* Academic Press.

Davidson, E. H., & Erwin, D. H. (2006). Gene regulatory networks and the evolution of animal body plans. *Science, 311*(5762), 796-800.

De Smedt, T. D. (2013). *Modeling Creativity: Case Studies in Python.* University Press Antwerp.

De Vicente Milans, R. (2013). Síntesis musical humanizada con notación musical simbólica. Master thesis. Málaga, Spain: Universidad de Málaga.

Delgado, M., Fajardo, W., & Molina-Solana, M. (2009). Inmamusys: Intelligent multiagent music system. *Expert Systems with Applications, 36*(3), 4574-4580.

Díaz, J. L., & Enrique, F. (2001). La estructura de la emoción humana: Un modelo cromático del sistema afectivo. *Salud Mental, 24*, 20-35.

Diaz-Jerez, G. (2011). Composing with Melomics: Delving into the computational world for musical inspiration. *Leonardo Music Journal, 21*, 13-14.

Ebcioglu, K. (1988). An expert system for harmonizing four-part chorales. *Computer Music Journal, 12*(3), 43-51.

Eck, D., & Schmidhuber, J. (2002). Finding temporal structure in music: blues improvisation with LSTM recurrent networks. *Proceedings of the 2002 12th IEEE Workshop on Neural Networks for Signal Processing* (pp. 747-756). Martigny, Switzerland: IEEE.

Eerola, T., & Toiviainen, P. (2004). MIR In Matlab: The MIDI Toolbox. *Proceedings of the ISMIR 2004, 5th International Conference on Music Information Retrieval,* (pp. 22-27). Barcelona, Spain.

Fernández, J. D. (2012). The Evolution of Diversity in the Structure and Function of Artificial Organisms. Ph.D. Thesis. Málaga, Spain: Universidad de Málaga.

Fernandez, J. D., & Vico, F. J. (2013). AI Methods in Algorithmic Composition: A Comprehensive Survey. *Journal of Artificial Intelligence Research, 48*, 513-582.

Ferrucci, D., Brown, E., Chu-Carroll, J., Fan, J., Gondek, D., Kalyanpur, A. A., . . . Welty, C. (2010). Building Watson: An Overview of the DeepQA Project. *AI magazine, 31*(3), 59-79.

Ferrucci, D., Levas, A., Bagchi, S., Gondek, D., & Mueller, E. T. (2013). Watson: beyond jeopardy! *Artificial Intelligence, 199*, 93-105.

Fiebelkorn, T. (2003). *Patent No. WO2003018097 A1*.

Forgács, G., & Newman, S. (2005). *Biological physics of the developing embryo.* Cambridge University Press.

Garner, T., Grimshaw, M., & Nabi, D. A. (2010). A preliminary experiment to assess the fear value of preselected sound parameters in a survival horror game. *Proceedings of the 5th Audio Mostly Conference: A Conference on Interaction with Sound* (p. 10). Piteå, Sweden: ACM.

Gero, J. S. (2000). Computational models of innovative and creative design processes. *Technological Forecasting and Social Change, 64*(2-3), 183-196.

Gero, J. S., & Maher, M. L. (Eds.). (1993). *Modeling Creativity and Knowledge-Base Creative Design.* Psychology Press.

Gervás, P. (2009). Computational Approaches to Storytelling and Creativity. *AI Magazine, 30*(3), 49-62.

Gilbert, É., & Conklin, D. (2007). A probabilistic context-free grammar for melodic reduction. *Proceedings of the International Workshop on Artificial Intelligence and Music, 20th International Joint Conference on Artificial Intelligence* (pp. 83-94). Hyderabad, India: ACM.

Gill, S. (1963). A Technique for the Composition of Music in a Computer. *The Computer Journal, 6*(2), 129-133.

Goodfellow, I., Pouget-Abadie, J., Mirza, M., Xu, B., Warde-Farley, D., Ozair, S., . . . Bengio, Y. (2014). Generative adversarial nets. *Proceedings of the 27th International Conference on Neural Information Processing Systems* (pp. 2672-2680). Montréal, Canada: ACM.

Goodwin, B. (2001). Living form in the making. In B. Goodwin, *How the leopard changed its spots: the evolution of complexity* (pp. 77-114). Princeton University Press.

Gutiérrez-Gutiérrez, J. D. (2013). Aplicación móvil para la reproducción de música adaptada en tiempo real a los distintos momentos, estados y eventos que se producen durante la conducción de un automóvil. Master Thesis. Málaga, Spain: Universidad de Málaga.

Harnad, S. (2000). Minds, machines and Turing: The Indistinguishability of Indistinguishables. *Journal of Logic, Language, and Information, 9*(4), 425-445.

Hiller Jr, L. A., & Isaacson, L. M. (1958). Musical composition with a High-Speed digital computer. *Journal of the Audio Engineering Society, 6*(3), 154-160.

Honda, T. (1998). *Washington, DC: U.S. Patent and Trademark Office Patent No. D403,013.*

Hood, L., & Galas, D. (2003). The digital code of DNA. *Nature, 421*(6921), 444-448.

Hornby, G. S., & Pollack, J. B. (2002). Creating high-level components with a generative representation for body-brain evolution. *Artificial life, 8*(3), 223-246.

Hornby, G., Lohn, J. D., & Linden, D. S. (2011). Computer-Automated Evolution of an X-Band Antenna for NASA's Space Technology Mission. *Evolutionary computation, 19*(1), 1-23.

Huang, A., & Wu, R. (2016). *Deep Learning for Music.* Ithaca, NY, United States: arxiv, Cornell University. Retrieved from https://arxiv.org/abs/1606.04930

Huang, S. T., Good, M., & Zauszniewski, J. A. (2010). The effectiveness of music in relieving pain in cancer patients: a randomized controlled trial. *International journal of nursing studies, 47*(11), 1354-1362.

Huron, D. (2002). Music information processing using the Humdrum toolkit: Concepts, examples, and lessons. *Computer Music Journal, 26*(2), 11-26.

Huzaifah, M., & Wyse, L. (2017). Deep generative models for musical audio synthesis. In E. R. Miranda, *Handbook of Artificial Intelligence for Music*

(pp. 639-678). Springer International Publishing. doi:10.1007/978-3-030-72116-9

Ianigro, S., & Bown, O. (2019). Exploring Transfer Functions in Evolved CTRNNs for Music Generation. *Proceedings of the International Conference on Computational Intelligence in Music, Sound, Art and Design* (pp. 234-248). Leipzig, Germany: Springer.

Institute of Medicine (US). (2011). *Relieving Pain in America: A Blueprint for Transforming Prevention, Care, Education, and Research.* Washington, DC, USA: The National Academies. doi:doi.org/10.17226/13172

Ioja, S. W., & Rennert, N. J. (2012). Relationship between sleep disorders and the risk for developing type 2 diabetes mellitus. *Postgraduate medicine, 124*(4), 119-129.

Juslin, P. N., & Sloboda, J. A. (2011). MEASUREMENT. In *Handbook of Music and Emotion: Theory, Research, Applications* (pp. 187-344). Oxford, United Kingdom: Oxford University Press.

Kania, A. (2007). The philosophy of music. In *Stanford Encyclopedia of Philosophy.* Stanford, CA, USA: Stanford University. Retrieved April 30, 2021, from https://plato.stanford.edu/entries/music/

Keroxen aúna coreografía y música sobre el escenario de El Tanque. (2011, October 14). *EL DÍA.* Retrieved May 16, 2021, from http://web.eldia.es/2011-10-14/CULTURA/6-Keroxen-auna-coreografia-musica-escenario-Tanque.htm

Koelsch, S. (2014). Brain correlates of music-evoked emotions. *Nature Reviews Neuroscience, 15*, 170-180.

Koelsch, S. (2015). Music-evoked emotions: Principles, brain correlates, and implications for therapy. *Annals of the New York Academy of Sciences, 1337*, 193-201.

La UMA diseña un ordenador que compone música clásica. (2012, July 1). *El Mundo.* Retrieved May 16, 2021, from

http://www.elmundo.es/elmundo/2012/07/01/andalucia_malaga/134116
4727.html

Lai, H. L., & Good, M. (2005). Music improves sleep quality in older adults. *Journal of advanced nursing, 49*(3), 234-244.

Lamere, P. (2008). Social tagging and music information retrieval. *Journal of New Music Research, 37*(2), 101-114.

Levine, M., & Tjian, R. (2003). Transcription regulation and animal diversity. *Nature, 424*(6945), 147-151.

Lewontin, R. (2000). *The Triple Helix: Gene, Organism, and Environment*. Harvard University Press.

Lindemann, A., & Lindemann, E. (2018). Musical organisms. *Proceedings of the International Conference on Computational Intelligence in Music, Sound, Art and Design* (pp. 128-144). Parma, Italy: Springer.

Lindenmayer, A. (1968). Mathematical models for cellular interaction in. *Journal of Theoretical Biology, 18*, 280-315.

Lobo, D. (2010). Evolutionary development based on genetic regulatory models for behavior-finding. Ph.D. Thesis. Málaga, Spain: Universidad de Málaga.

Lobo, D., Fernández, J. D., & Vico, F. J. (2012). Behavior-finding: morphogenetic designs shaped by function. In R. Doursat, H. Sayama, & O. Michel (Eds.). Springer Berlin Heidelberg.

Lui, I., & Ramakrishnan, B. (2014). *Bach in 2014: Music Composition with Recurrent Neural Network*. Ithaca, NY, United States: arXiv, Cornell University. Retrieved from https://arxiv.org/abs/1412.3191

MacCallum, R. M., Mauch, M., Burt, A., & Leroi, A. M. (2012). Evolution of music by public choice. *Proceedings of the National Academy of Sciences of the United States of America, 109*(30), 12081-12086.

Malmi, E., Takala, P., Toivonen, H., Raiko, T., & Gionis, A. (2016). Dopelearning: A computational approach to rap lyrics generation. *Proceedings of the*

*22nd ACM SIGKDD International Conference on Knowledge Discovery and Data Mining* (pp. 195--204). San Francisco, California, USA: ACM.

Manos, S., Large, M. C., & Poladian, L. (2007). Evolutionary design of single-mode microstructured polymer optical fibres using an artificial embryogeny representation. *Proceedings of the 9th annual conference companion on Genetic and evolutionary computation* (pp. 2549-2556). London, United Kingdom: ACM.

Marcus, G. (2003). *The birth of the mind: How a tiny number of genes creates the complexities of human thought.* Basic Books.

Markoff, J. (2011, February 16). Computer Wins on 'Jeopardy!': Trivial, It's Not. *The New York Times.* Retrieved May 16, 2021, from https://www.nytimes.com/2011/02/17/science/17jeopardy-watson.html

Mason, S., & Saffle, M. (1994). L-Systems, melodies and musical structure. *Leonardo Music Journal, 4,* 31-38.

Mayr, E. (1961). Cause and effect in biology. *Science, 134*(3489), 1501-1506.

McCartney, J. (2002). Rethinking the computer music language: SuperCollider. *Computer Music Journal, 26*(4), 61-68.

McCormack, J. (2005). A developmental model for generative media. *European Conference on Artificial Life* (pp. 88-97). Canterbury, United Kingdom: Springer.

McKay, C. (2010). Automatic Music Classification with jMIR. Ph.D. Thesis. Montréal, Canada: McGill University.

McKay, C., & Fujinaga, I. (2005). The Bodhidharma system and the results of the MIREX 2005 symbolic genre classification contest. *Proceedings of the ISMIR 2005, 6th International Conference on Music Information Retrieval.* London, United Kingdom.

McKay, C., Cumming, J., & Fujinaga, I. (2018). JSYMBOLIC 2.2: Extracting Features from Symbolic Music for use in Musicological and MIR Research.

*Proceedings of the ISMIR 2018, 19th International Society for Music Information Retrieval Conference*, (pp. 348-354). Paris, France.

Meyer, L. B. (1956). *Emotion and Meaning in Music.* Chicago, IL, USA: University of Chicago Press.

Meyer, L. B. (1989). *Style and music: Theory, history, and ideology.* University of Chicago Press.

Moffat, D. C., & Kelly, M. (2006). An investigation into people's bias against computational creativity in music composition. *Proceedings of the 3rd international joint workshop on computational creativity (ECAI06Workshop)*, (pp. 1-8). Riva del Garda, Italy.

Monteith, K., Martinez, T., & Ventura, D. (2010). Automatic generation of music for inducing emotive response. *Proceedings of the International Conference on Computational Creativity*, (pp. 140-149). Lisbon, Portugal.

Moroni, A., Manzolli, J., von Zuben, F., & Gudwin, R. (2000). Vox populi: An interactive evolutionary system for algorithmic music. *Leonardo Music Journal, 10*, 49-54.

Morreale, F., De Angeli, A., Masu, R., Rota, P., & Conci, N. (2014). Collaborative creativity: The Music Room. *Personal and Ubiquitous Computing, 18*(5), 1187-1199.

Morreale, F., Masu, R., & De Angeli, A. (2013). Robin: an algorithmic composer for interactive scenarios. *Proceedings of the Sound and Music Computing Conference 2013, SMC 2013*, (pp. 207-212). Stockholm, Sweden.

Müller, G. B. (2007). Evo-devo: extending the evolutionary synthesis. *Nature Reviews Genetics, 8*(12), 943-949.

Müller, M. (2002). Computer go. *Artificial Intelligence, 134*(1), 145-179.

Nam, J., Choi, K., Lee, J., Chou, S. Y., & Yang, Y. H. (2018). Deep Learning for Audio-Based Music Classification and Tagging: Teaching Computers to Distinguish Rock from Bach. *IEEE signal processing magazine, 36*(1), 41-51.

Nam, Y. W., & Kim, Y. H. (2019). Automatic jazz melody composition through a learning-based genetic algorithm. *Proceedings of the International Conference on Computational Intelligence in Music, Sound, Art and Design* (pp. 217-233). Leipzig, Germany: Springer.

Newman, S. (2003). From physics to development: the evolution of morphogenetic mechanisms. In G. B. Müller, & S. A. Newman, *Origination of organismal form: beyond the gene in developmental and evolutionary biology* (pp. 221-240).

Ohno, S. (1987). Repetition as the essence of life on this earth: Music and genes. In R. Neth, R. C. Gallo, M. F. Greaves, & H. Kabisch (Eds.), *Modern Trends in Human Leukemia VII* (pp. 511-519). Berlin/Heidelberg, Germany: Springer.

Olson, H. F., & Belar, H. (1961). Aid to music composition employing a random probability system. *The Journal of the Acoustical Society of America, 33*(9), 1163-1170.

Oramas, S., Barbieri, F., Nieto Caballero, O., & Serra, X. (2018). Multimodal deep learning for music genre classification. *Transactions of the International Society for Music Information Retrieval, 1*(1), 4-21.

Pachet, F. (2002). Interacting with a musical learning system: The Continuator. *Proceedings of the International Conference on Music and Artificial Intelligence* (pp. 103-108). Edinburgh, Scotland, UK: Springer.

Pachet, F., & Cazaly, D. (2000). A taxonomy of musical genres. *Proceedings of the Content-Based Multimedia Information Access Conference (RIAO)* (pp. 1238-1245). Paris, France: ACM.

Padberg, H. A. (1964). Computer-composed canon and free-fugue. Ph.D. Thesis. St. Louis, Missouri, USA: Saint Louis University.

Papadopoulos, G., & Wiggins, G. (1999). AI methods for algorithmic composition: A survey, a critical view and future prospects. *Proceedings of the AISB Symposium on Musical Creativity*, (pp. 110-117). Edinburgh, United Kingdom.

Pearce, M., & Wiggins, G. (2001). Towards a framework for the evaluation of machine. *Proceedings of the AISB'01 Symposium on Artificial Intelligence and Creativity in the Arts and Sciences*, (pp. 22-32). York, United Kingdom.

Pease, A., & Colton, S. (2011). On impact and evaluation in computational creativity: A discussion of the Turing test and an alternative proposal. *Proceedings of the AISB symposium on AI and Philosophy* (pp. 15-22). York, United Kingdom: Society for the Study of Artificial Intelligence and Simulation of Behaviour.

Peckham, M. (2013, 01 4). Finally, a Computer that Writes Contemporary Music Without Human Help. *Time*. Retrieved May 16, 2021, from Time: http://techland.time.com/2013/01/04/finally-a-computer-that-writes-contemporary-music-without-human-help/

Peppard, P. E., Young, T., Palta, M., & Skatrud, J. (2000). Prospective study of the association between sleep-disordered breathing and hypertension. *New England Journal of Medicine, 342*(19), 1378-1384.

Perez-Bravo, J. (2014). Montaje y configuración de un cluster de computación para su uso con Torque y Maui. M.A. Thesis. Málaga, Spain: Universidad de Málaga.

Pinch, T., & Trocco, F. (1998). The social construction of the early electronic music synthesizer. *Icon, 4*, 9-31.

Plutchik, R. (2001). The nature of emotions: Human emotions have deep evolutionary roots, a fact that may explain their complexity and provide tools for clinical practice. *American Scientist, 89*, 344-350.

Prusinkiewicz, P. (1986). Score generation with L-systems. *Proceedings of the International Computer Music Conference* (pp. 455-457). The Hague, Netherlands: Michigan Publishing.

Prusinkiewicz, P., & Lindenmayer, A. (1990). *The algorithmic beauty of plants*. Springer-Verlag New York.

Puckette, M. (2002). Max at seventeen. *Computer Music Journal, 26*(4), 31-43.

Pudaruth, S., Amourdon, S., & Anseline, J. (2014). Automated generation of song lyrics using CFGs. *2014 Seventh International Conference on Contemporary Computing (IC3)* (pp. 613-616). Noida, India: IEEE.

Redacción Creativa. (2011, October 14). Keroxen ofrece este fin de semana una intensa programación de música y danza. *Creativa Canaria*. Retrieved May 16, 2021, from https://web.archive.org/web/20120904172449/http://www.creativacanaria.com/index.php/musica/1582-keroxen-ofrece-este-fin-de-semana-una-intensa-programacion-de-musica-y-danza

Reddin, J., McDermott, J., & O'Neill, M. (2009). Elevated pitch: Automated grammatical evolution of short compositions. *Workshops on Applications of Evolutionary Computing EvoMUSART* (pp. 579-584). Tübingen, Germany: Springer Berlin Heidelberg.

Requena, G., Sánchez, C., Corzo-Higueras, J. L., Reyes-Alvarado, S., Rivas-Ruiz, F., Vico, F., & Raglio, A. (2014). Melomics music medicine (M3) to lessen pain perception during pediatric prick test procedure. *Pediatric Allergy and Immunology, 25*(7), 721-724.

Rer Nat Wiebkin, H. D. (1985). *Patent No. DE3338649 A1*.

Ritchie, G. (2007). Some empirical criteria for attributing creativity to a computer program. *Minds and Machines, 17*(1), 67-99.

Ritter, P. L., González, V. M., Laurent, D. D., & Lorig, K. R. (2006). Measurement of pain using the visual numeric scale. *The Journal of rheumatology, 33*(3), 574-580.

Rodrigues, A., Costa, E., Cardoso, A., Machado, P., & Cruz, T. (2016). Evolving l-systems with musical notes. *Proceedings of the International Conference on Computational Intelligence in Music, Sound, Art and Design* (pp. 186-201). Porto, Portugal: Springer.

Rohrmeier, M. (2011). Towards a generative syntax of tonal harmony. *Journal of Mathematics and Music, 5*, 35-53.

Roig, C., Tardón, L. J., Barbancho, I., & Barbancho, A. M. (2014). Automatic melody composition based on a probabilistic model of music style and harmonic rules. *Knowledge-Based Systems, 71*, 419-434.

Runco, M. A., & Albert, R. S. (2010). Creativity research: a historical view. In J. C. Kaufman, & R. J. Sternberg, *The Cambridge Handbook of Creativity* (pp. 3-19). Cambridge University Press.

Russell, J. A. (1980). A circumplex model of affect. *Journal of personality and social psychology, 39*, 1161-1178.

Russell, S., & Norvig, P. (2009). *Artificial Intelligence: A Modern Approach* (3 ed.). Prentice Hall.

Sachs, C. (2012). *The history of musical instruments*. Courier Corporation.

Sánchez-Quintana, C. A. (2014). Aportaciones y Aplicaciones de Disciplinas Bioinspiradas a la Creatividad Computacional. Ph.D. Thesis. Málaga, Spain: Universidad de Málaga.

Schaffer, J. W., & McGee, D. (1997). *Knowledge-based programming for music research* (Vol. 13). AR Editions, Inc.

Schedl, M., Gómez, E., & Urbano, J. (2014). *Music Information Retrieval: Recent Developments and Applications*. Now Publishers.

Shea, K. C., & Fenves, S. J. (1997). A shape annealing approach to optimal truss design with dynamic grouping of members. *Journal of Mechanical Design, 119*(3), 388-394.

Silver, D., Huang, A., Maddison, C. J., Guez, A., Sifre, L., van den Driessche, G., . . . Hassabis, D. (2016). Mastering the game of Go with deep neural networks and tree search. *Nature, 529*(7587), 484-489.

*Sleep in America Polls*. (n.d.). (National Sleep Foundation) Retrieved May 16, 2021, from https://www.sleepfoundation.org/professionals/sleep-america-polls

SoundFont Technical Specification. (2006, February 3). Retrieved May 16, 2021, from http://www.synthfont.com/sfspec24.pdf

Stanley, K. O., & Miikkulainen, R. (2003). A taxonomy for artificial embryogeny. *Artificial Life, 9*(2), 93-130.

Steinbeis, N., & Koelsch, S. (2009). Understanding the intentions behind man-made products elicits neural activity in areas dedicated to mental state attribution. *Cerebral Cortex, 19*(3), 619-623.

Steinberg, K. (1999). Steinberg Virtual Studio Technology (VST) Plug-in Specification 2.0 Software Development Kit. Hamburg: Steinberg Soft- und Hardware GMBH. Retrieved May 16, 2021, from http://jvstwrapper.sourceforge.net/vst20spec.pdf

Sternberg, R. J., & O'Hara, L. A. (1999). 13 Creativity and Intelligence. In R. J. Sternberg (Ed.), *Handbook of creativity* (pp. 251-272). Cambridge, MA: Cambridge University Press.

Stiny, G. (1980). Introduction to shape and shape grammars. *Environment and planning B, 7*(3), 343-351.

Stranges, S., Tigbe, W., Gómez-Olivé, F. X., Thorogood, M., & Kandala, M. B. (2012). Sleep problems: an emerging global epidemic? *Sleep, 35*(8), 1173-81.

Tatarkiewicz, W. (1980). Creativity: History of the Concept. In W. Tatarkiewicz, *A History of Six Ideas an Essay in Aesthetic* (pp. 244-265). Springer.

*The sounds of Nature Collection (Jungle River, Jungle birdsong and Showers).* (n.d.). Retrieved April 30, 2021, from Gumroad: https://gumroad.com/l/nature

Thywissen, K. (1999). GeNotator: an environment for exploring the application of evolutionary techniques in computer-assisted composition. *Organised Sound, 4*(2), 127-133.

Turing, A. M. (1950). Computing machinery and intelligence. *Mind, 59*, 433-460.

Turner, S. R. (2014). *The creative process: A computer model of storytelling and creativity.* Psychology Press.

Tzanetakis, G., & Cook, P. (2000). MARSYAS: a framework for audio analysis. *Organised Sound, 4*(3), 169-175. doi:https://doi.org/10.1017/S1355771800003071

Ullmann, P. (2009). *Patent No. US20090149699 A1.*

Vico, F. J. (2007). *Selfo: A class of self-organizing connection games.* Málaga, Spain: Universidad de Málaga. Retrieved May 16, 2021, from http://hdl.handle.net/10630/6708

Vico, F. J., Sánchez-Quintana, C. A., & Albarracín, D. D. (2011). MELOMICS Contributions of computer science and biology to receptive music therapy. Evidence for Music Therapy Practice, Research & Education. *Selected Readings & Proceedings of the VIII European Music Therapy Congress* (pp. 521-530). Cádiz, Spain: Granada: Grupo Editorial Universitario.

Wallas, G. (1926). *The art of thought.* New York, Harcourt, Brace and Company.

Wang, A. (2003). An Industrial Strength Audio Search Algorithm. *ISMIR*, (pp. 7-13). Baltimore, Maryland, USA.

Williams, D., Kirke, A., Miranda, E. R., Roesch, E., Daly, I., & Nasuto, S. (2015). Investigating affect in algorithmic composition systems. *Psychology of Music, 43*(6), 831-854.

Wilson, A. J. (2009). A symbolic sonification of L-systems. *Proceedings of the International Computer Music Conference* (pp. 203-206). Montréal, Canada: Michigan Publishing.

Witten, I. H., & Frank, E. (2005). *Data mining: Practical machine learning tools and techniques.* New York: Morgan Kaufman.

Wooller, R. W., & Brown, A. R. (2005). Investigating morphing algorithms for generative music. *Third Iteration: Third International Conference on Generative Systems in the Electronic Arts.* Melbourne, Australia.

Yannakakis, G. N. (2012). Game AI revisited. *Proceedings of the 9th conference on Computing Frontiers* (pp. 285-292). Cagliari, Italy: ACM.

Yin, X., Liu, Z., Wang, D., Pei, X., Yu, B., & Zhou, F. (2015). Bioinspired self-healing organic materials: Chemical mechanisms and fabrications. *Journal of Bionic Engineering, 12*(1), 1-16.

You, J. (2014, November 18). Gecko-inspired adhesives allow people to climb walls. *Science Magazine*. Retrieved May 16, 2021, from https://www.sciencemag.org/news/2014/11/gecko-inspired-adhesives-allow-people-climb-walls

Yu, C., Sasic, S., Liu, K., Salameh, S., Ras, R. H., & van Ommen, J. R. (2020). Nature–Inspired self–cleaning surfaces: Mechanisms, modelling, and manufacturing. *Chemical Engineering Research and Design, 155*, 48-65.

This research could not have been completed the way it is, without an intensive use of Wikipedia®.[54] This multilingual, free-access and free-content internet encyclopedia has become an ideal and essential tool for everyone, to access knowledge of any kind, in any field. During my research and particularly along the development of this thesis, Wikipedia can be considered as the primary bibliographic source. I am grateful to all the skilled and enlightened devoted contributors, who make possible these contents to be instantly updated, fixed and always reachable.

---

[54] https://www.wikipedia.org/ (accessed on May 16, 2021)

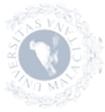